%% file: paper.tex
\documentclass[a4paper,fleqn,usenatbib]{mnras}

\usepackage{graphicx}	
\usepackage{amsmath}	
\usepackage{amssymb}	
\usepackage{multicol}        
\usepackage{bm}		
\usepackage{pdflscape}	
\usepackage{natbib}
\usepackage{amsfonts}
\usepackage{placeins}
\usepackage{array,longtable}		
\usepackage{geometry}
\usepackage{multirow, booktabs} 				
\usepackage{caption}
\usepackage{subcaption}
\usepackage{acronym}
\usepackage{url}
\usepackage{float}
\usepackage[export]{adjustbox}
\usepackage[utf8]{inputenc}

\bibpunct{(}{)}{,}{a}{}{;} 

\newcommand{\kms}{\,km\,s$^{-1}$} 
\newcommand{\Hi}{\textsc{Hi}}

\usepackage[T1]{fontenc}
\usepackage{ae,aecompl}

\usepackage{newtxtext,newtxmath}

\acrodef{ghasp}[\textsc{ghasp}]{Gassendi H-Alpha survey of Spirals}
\acrodef{sdss}[\textsc{sdss}]{Sloan Digital Sky Survey}
\acrodef{ned}[\textsc{ned}]{NASA/IPAC Extragalactic Database}

\title[GHASP Mass Models]{GHASP: an H$\alpha$ kinematics survey of spiral
galaxies - XII. Distribution of luminous and dark matter in spiral and irregular nearby galaxies using R$_c$-band photometry }

\author[M. Korsaga et al.]{M. Korsaga$^{1,2}$\thanks{marie.korsaga@lam.fr},
P. Amram$^{1}$,
C. Carignan$^{2,3}$,
B. Epinat$^{1}$
\\
$^{1}$Aix Marseille Univ, CNRS, CNES, LAM, Marseille, France\\
$^{2}$Department of Astronomy, University of Cape Town, Private Bag X3, Rondebosch 7701, South Africa\\
$^{3}$Observatoire d'Astrophysique de l'Universit\'{e} de Ouagadougou, BP 7021, Ouagadougou 03, Burkina Faso}

\pubyear{2017}

\begin{document}
\label{firstpage}
\pagerange{\pageref{firstpage}--\pageref{lastpage}}
\maketitle

\begin{abstract}

Mass models of 100 nearby spiral and irregular galaxies, covering morphological types from Sa to Irr, are computed using H$\alpha$ rotation curves and R$_c$-band surface brightness profiles. The kinematics was obtained using a scanning Fabry-Perot interferometer. One of the aims is to compare our results with those from \citet{Korsaga+2018}, which used mid-infrared (MIR) WISE W1 (3.4 $\mu$m) photometric data. For the analysis, the same tools were used for both bands. Pseudo-Isothermal (ISO) core and Navarro-Frenk-White (NFW) cuspy models have been used. We test Best Fit Models (BFM), Maximum Disc Models (MDM) and models for which M/L is fixed using the B - V colors. Similarly to what was found in the MIR 3.4 $\mu$m band, most of the observed rotation curves are better described by a central core density profile (ISO) than a cuspy one (NFW) when using the optical R$_c$-band. In both bands, the dispersion in the (M/L) values is smaller for the fixed M/L fits. As for the W1 photometry, the derived DM halos’ parameters depend on the morphological types. We find similar relations than those in the literature, only when we compare our results for the bulge-poor sub-sample because most of previous results were mainly based on late-type spirals. Because the dispersion in the model parameters is smaller and because stellar masses are better defined in that band, MIR photometry should be preferred, when possible, to the optical bands. It is shown that for high-z galaxies, sensible results can still be obtained without full profile decomposition.

\end{abstract}




\newpage

\section{Introduction}

The $\Lambda$-Cold Dark Matter ($\Lambda$CDM) model has a great deal of success in replicating large-scale structures such as the cosmic microwave background (e.g. Planck's results), baryonic and dark matter (DM) mass distributions in galaxy cluster using weak lensing, or structure growing through the ages \citep[e.g.][]{Springel+2006}. However, the $\Lambda$CDM model still struggles to explain galactic scale structures like, e.g. the cusp-core problem. Indeed, the cusp-core controversy, which consists of observing that inner galaxy rotation curves rise less steeply than expected from pure DM structure formation in $\Lambda$CDM simulations \citep[e.g.][]{Flores+1994}, is far from being closed as a large number of studies show it continuously. In order to illustrate it, we will only mention two very recent works reaching pretty puzzling opposite conclusions. For a recent review, see e.g. \citet{Bullock+2017}. 

On the one hand, \citet{Read+2018} studied the central DM density profile of the Draco dwarf spheroidal galaxy that is supposed to be a prime candidate for hosting a pristine DM cusp, unaffected by stellar feedback during galaxy formation. These authors found that Draco has an inner DM density consistent with a $\Lambda$CDM cusp. On the other hand, \citet{Carleton+2018} concluded that ultra-diffuse Galaxies which suffer a dramatic reduction in surface brightness due to tidal stripping and heating, are remarkably well modelled if they reside in cored halos.  Their semi-analytic simulations show that galaxies in cored dark-matter halos expand significantly in response to tidal stripping and heating in agreement with observations, whereas galaxies embedded in cuspy halos experience limited evolution.

Spiral and Irregular galaxies request a substantial amount of DM  to model their mass distribution in the flat outer parts of their rotation curves (RCs) \citep[e.g.][]{Carignan+1985,VanAlbada+1986}. Because of the higher spatial resolution of H$\alpha$ 2D velocity fields, the kinematics derived from those high spectral and spatial resolution data cubes allow to accurately construct the RCs that are best suited to derive the shape of the DM halo in the central part of the galaxies. Indeed, the rising inner regions of the RC actually constrain the parameters of the mass models \citep{Ouellette+1999}, including the Mass-to-Light ratio (M/L) of their luminous components and the shape and parameters of the DM halo component. Therefore, high resolution RCs derived from 2D velocity fields allow to obtain accurate correlations between the dark halo properties and the optical properties of the galaxies \citep{Ouellette+2001}. The stellar mass distribution is the hardest parameter to constrain, and the ill-defined stellar M/L ratios translate into degeneracies. To tackle this problem, the high resolution RCs observed by the Gassendi H$\alpha$ survey of Spirals (GHASP) will be used \citep{Epinat+2008b,Epinat+2008a}. 

A serious complication comes from the fact that the baryonic matter is not monotonically distributed in a single component. The presence of a bulge component usually makes the rotational velocities rise rapidly thus impacting the solid body shape of the RC.  The latter reaches a plateau in the central regions and one needs to disentangle the contributions of the bulge and of the disc by assigning different spatial distribution (sphere vs disc)  and different M/L ratios to each component since they are gravitationally supported by different mechanisms (velocity dispersion and velocity rotation respectively) and have different stellar populations, the stars being on average older in bulges than in discs and therefore (M/L)$_{bulge}$ > (M/L)$_{disc}$. 
  
One goal of this paper is to test baryonic and DM mass distributions from early to late morphological types to account for galaxies with a bulge. In this study, the distribution of the DM profile is defined by either a pseudo-isothermal sphere (ISO) with a core density profile, as suggested by most of the observations \citep{Begeman+1987,Kravtsov+1998} or by  a cuspy dark halo density profile, as suggested by cosmological numerical simulations \citep{Navarro+1996}. Cuspy profiles describe well the DM distribution of steeply rising RCs while the core profiles give a better representation of slowly rising RCs \citep{Ouellette+2001, Blok+2001,Swaters+2003,Gentile+2004}. Most previous work based on late type and low surface brightness galaxies mentioned that the constant core density profile rather than the cuspy profile describes better the mass distribution in the inner parts of the galaxies \citep{Gentile+2004,Spano+2008,Bottema+2015}. 

To represent the stellar luminous mass distribution, prior to the development of good IR detectors that were developed to be used in space, earlier studies on galaxy mass distribution have been using optical bands \citep[B-band or preferentially R-band, see e.g.][]{Kormendy+2004} while most of the recent studies used MIR data, either the 3.6$\mu$m band of Spitzer \citep[see e.g.][]{Lelli+2016} or the 3.4$\mu$m band of WISE \citep[see e.g.][]{Korsaga+2018}. One of the main reason to revisit the R$_c$-band in this paper is to study if the derived DM halo parameters are very different when using an optical band compared to a MIR-band, which sample different stellar populations. If this is the case, one would have to forget about most of the earlier results and concentrate only on recent studies.

In a previous study by \citet{Korsaga+2018}, the mass distribution was derived using a sample of 121 galaxies by combining the optical H$\alpha$ kinematical and the mid-infrared (MIR) WISE photometric data. Two different techniques were used for each model; the best fit model and the one with the M/L calculated from the MIR color index (W1-W2); pseudo-isothermal maximum disc models were also derived. \citet{Korsaga+2018} found that (i) most rotation curves were better described by core rather than cuspy profiles; (ii) the relation between the DM parameters and the luminosity of galaxies depends on morphological types and (iii) the value of the M/L provided by the maximum disc models is $\sim$4 times higher and the one given by the (W1-W2) color is $\sim$3 higher than the M/L for the best fit model.

The same fitting procedure to construct the mass models was used in both photometric bands, such that we should be able to see how the DM parameters vary when using optical photometry instead of MIR for a similar sample of galaxies. In addition, this work gives us an excellent opportunity to compare the core and cuspy density profiles for both early and late type galaxies since our sample covers all morphological types and not only late-types, which was often the case in earlier studies. Our main objective is thus to trace the relation between the DM parameters (central density and the scaling radius) and the luminosity of galaxies covering all morphological types in order to study if the R$_c$ results are comparable with the MIR results found in \citet{Korsaga+2018}. We also want to study the distribution of the scale radius and of the central halo density as a function of the optical disc scale length,  see how the M/L is distributed when using the optical R$_c$- and MIR W1-bands and test which band provides correlations with the smallest dispersion.

This paper is organized as follows. In section \ref{sect:sample} we present the sample and the selection criteria for the H$\alpha$ RCs. Section \ref{sect:photometry} describes the decomposition of the R$_c$-band luminosity profiles, analyses the R$_c$-band scale relations and compares them with the MIR ones. In Section \ref{sect:massmodel}, we describe the formalism of the different mass models and present the results. In Section \ref{sect:rcw1}, we compare the results obtained using optical R$_c$ and MIR W1 photometry. A summary and the main conclusions are given in Section \ref{sect:conclusion}. Appendix \ref{appendixA} lists the global properties and mass models parameters for a subsample of 8 galaxies. The complete catalogue is available online. In Appendix \ref{appendixB}, we present the mass models of two galaxies. The remaining sample is also available online. We use a Hubble constant H$_0 = 75$~km~s$^{-1}$~Mpc$^{-1}$ throughout this paper.

\section{Sample selection}
\label{sect:sample}
We use a sample of 100 spiral and irregular nearby galaxies with high resolution H$\alpha$ rotation curves (RCs), derived from 2D velocity fields, and the optical R$_c$-band photometry: 73 from surface brightness photometry obtained at the Observatoire de Haute Provence (OHP) \citep{Barbosa+2015} and 27 from SDSS DR7 \citep{Abazajian+2009} archival data. The RCs of this sample are selected from the GHASP survey, which contains 203 galaxies observed with the 1.93 m telescope of the OHP using a Fabry-Perot  interferometer scanning around the H$\alpha$ emission line with high spectral and spatial resolutions. The spectral resolution is $\sim 10000$ and the spatial resolution is $\sim 2$ arcsec over a large field of view of $\sim 6\times 6$ arcmin$^2$. The GHASP objects are mostly isolated to reduce the perturbative effects of neighbours. The GHASP data have been published in several previous papers \citep{Garrido+2002, Garrido+2003, Garrido+2004, Garrido+2005,Epinat+2008b,Epinat+2008a}.

The present study combines the H$\alpha$ rotation curves and the optical R$_c$-band photometry. We apply the following selection criteria to define the sample of 100 galaxies. Firstly, we  began  with 124 galaxies, which have both H$\alpha$ rotation curves and optical R$_c$-band photometric data. Secondly, we assigned a quality flag to the rotation curve of each galaxy, ranging from 1 to 3 (see column 15 of Table \ref{tab:photometry}): -1- for very high, -2- for high and -3- for low quality. To construct mass models with good constraints, we therefore selected only galaxies with flag -1- and -2- which reduced the sample to 100 galaxies. We discarded 24 galaxies which were attributed a flag 3 for various reasons: too high inclination (>75$^o$), presence of a strong bar, signs of galaxy interaction, strong lopsidedness or very low SNR data.

\begin{figure*}
	\hspace*{-0.85cm} \includegraphics[width=5.87cm]{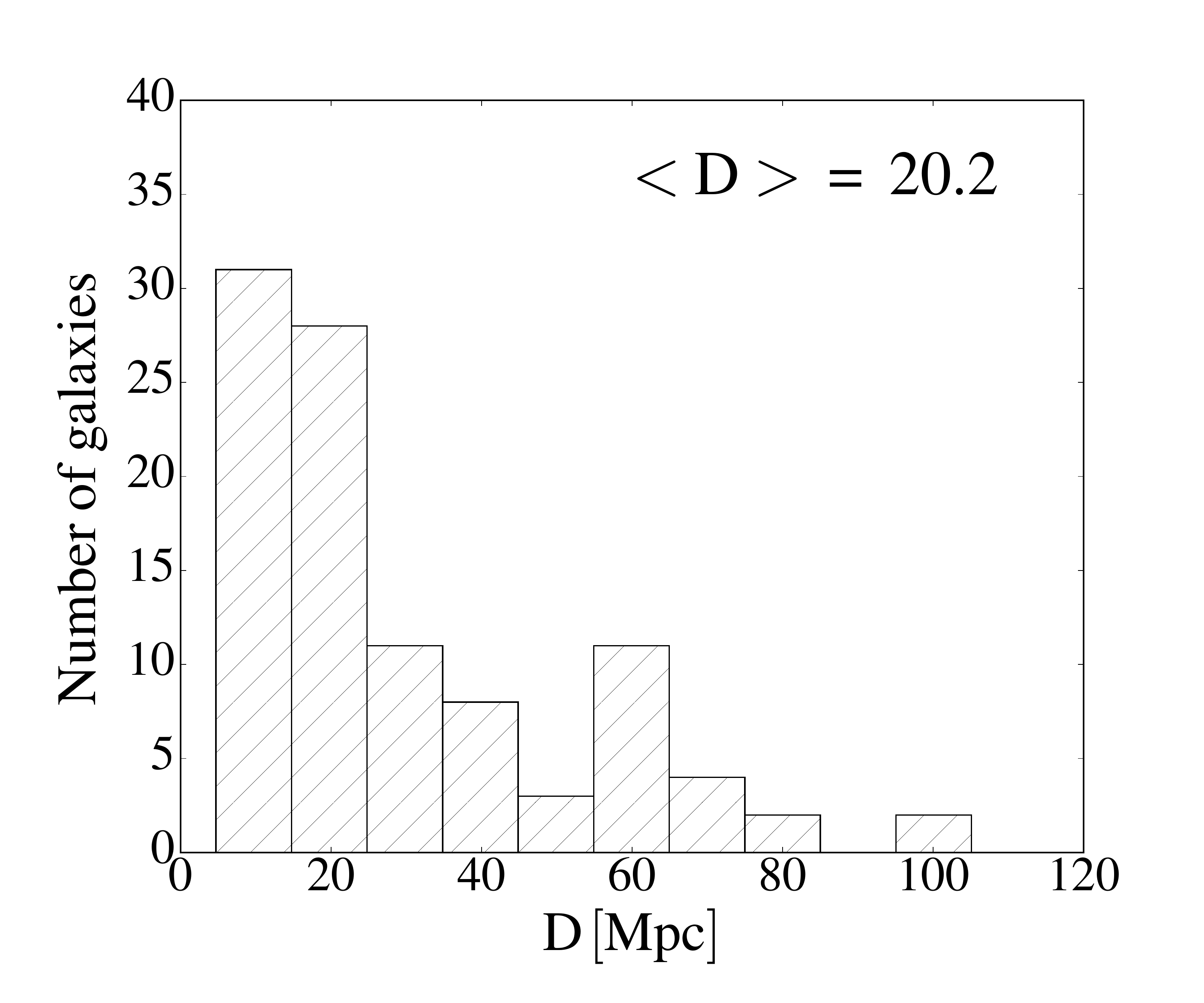}
	\hspace*{-0.35cm} \includegraphics[width=5.87cm]{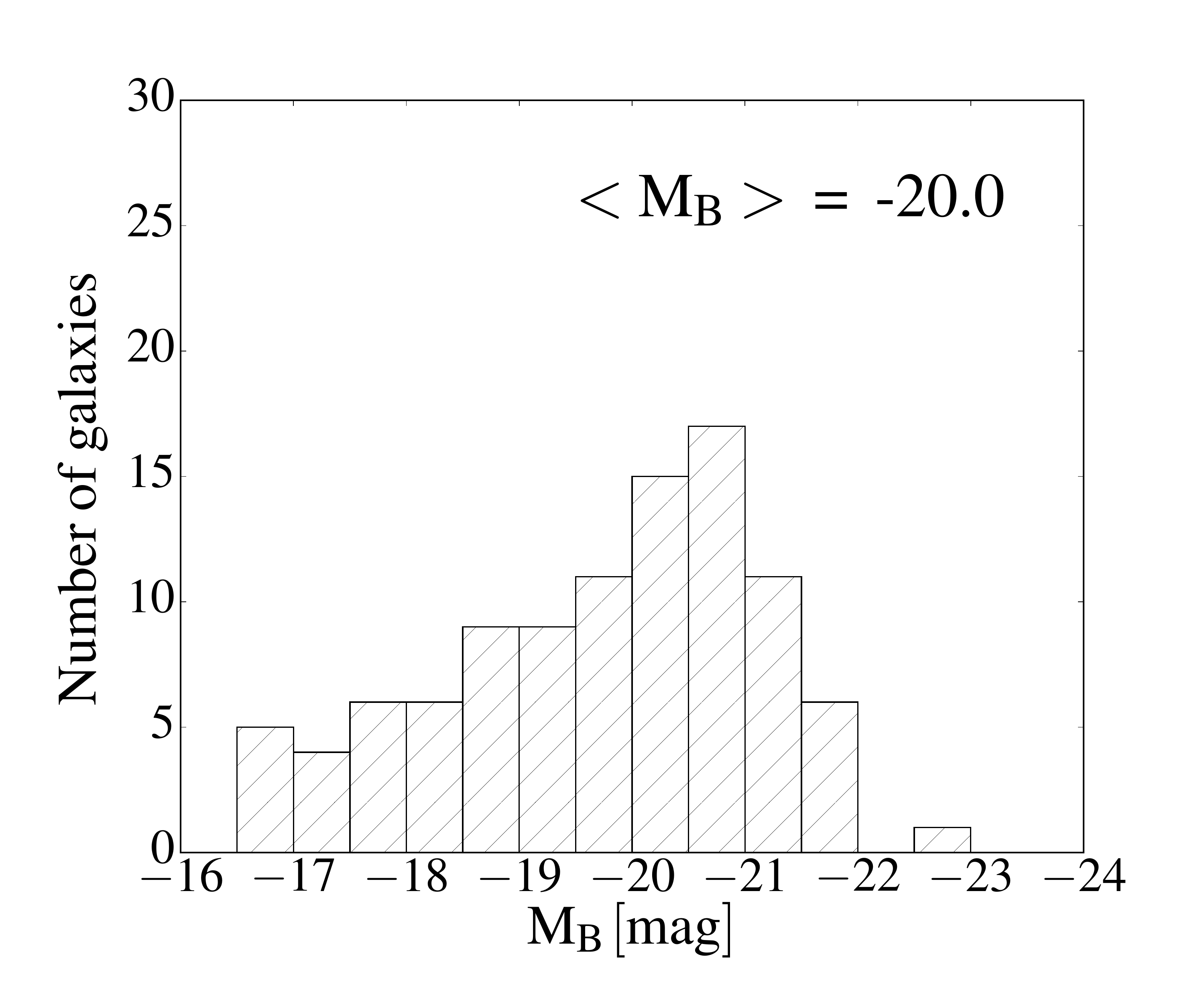}
	\hspace*{-0.35cm} \includegraphics[width=5.87cm]{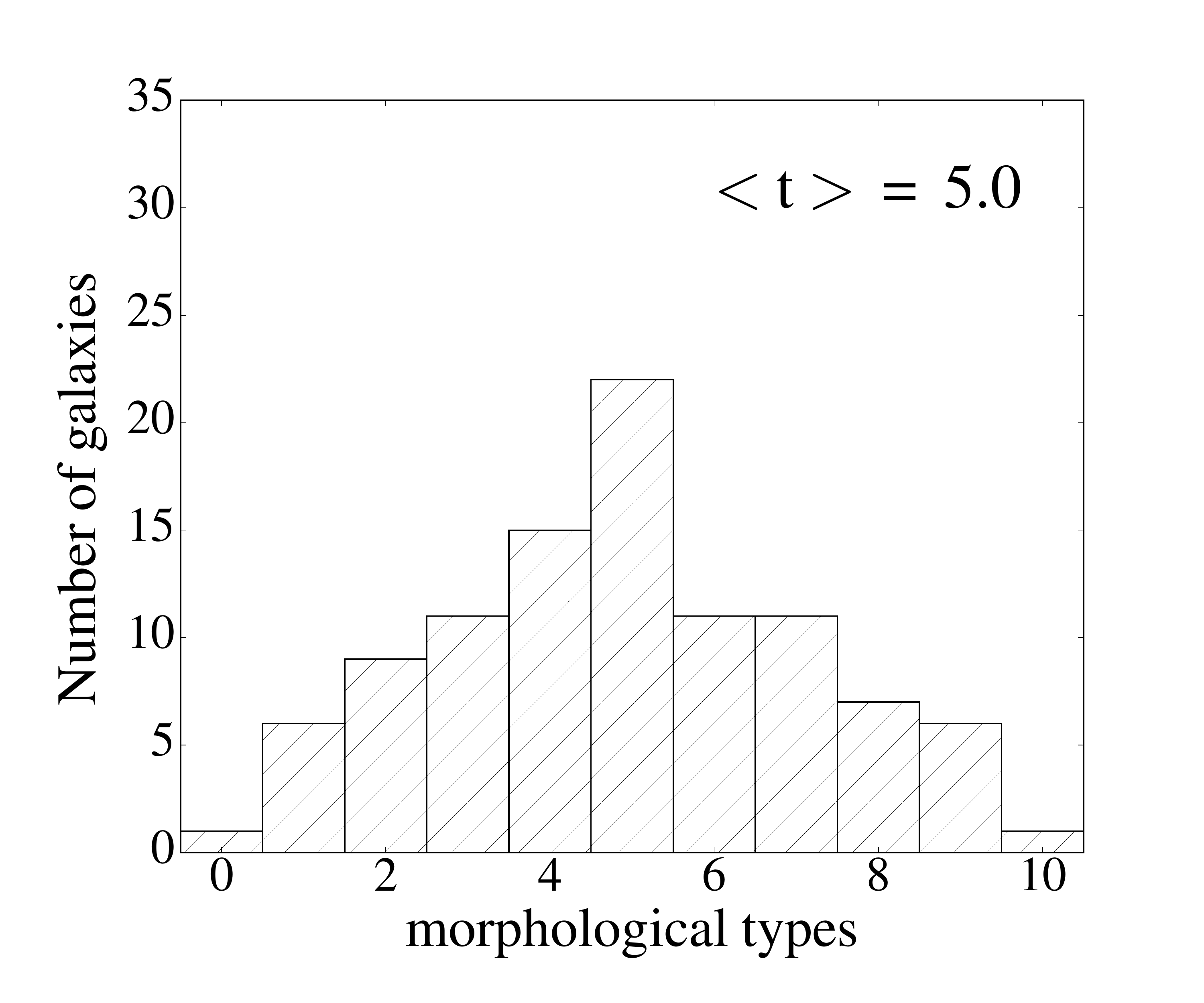}
	\hspace*{-0.85cm} \includegraphics[width=5.87cm]{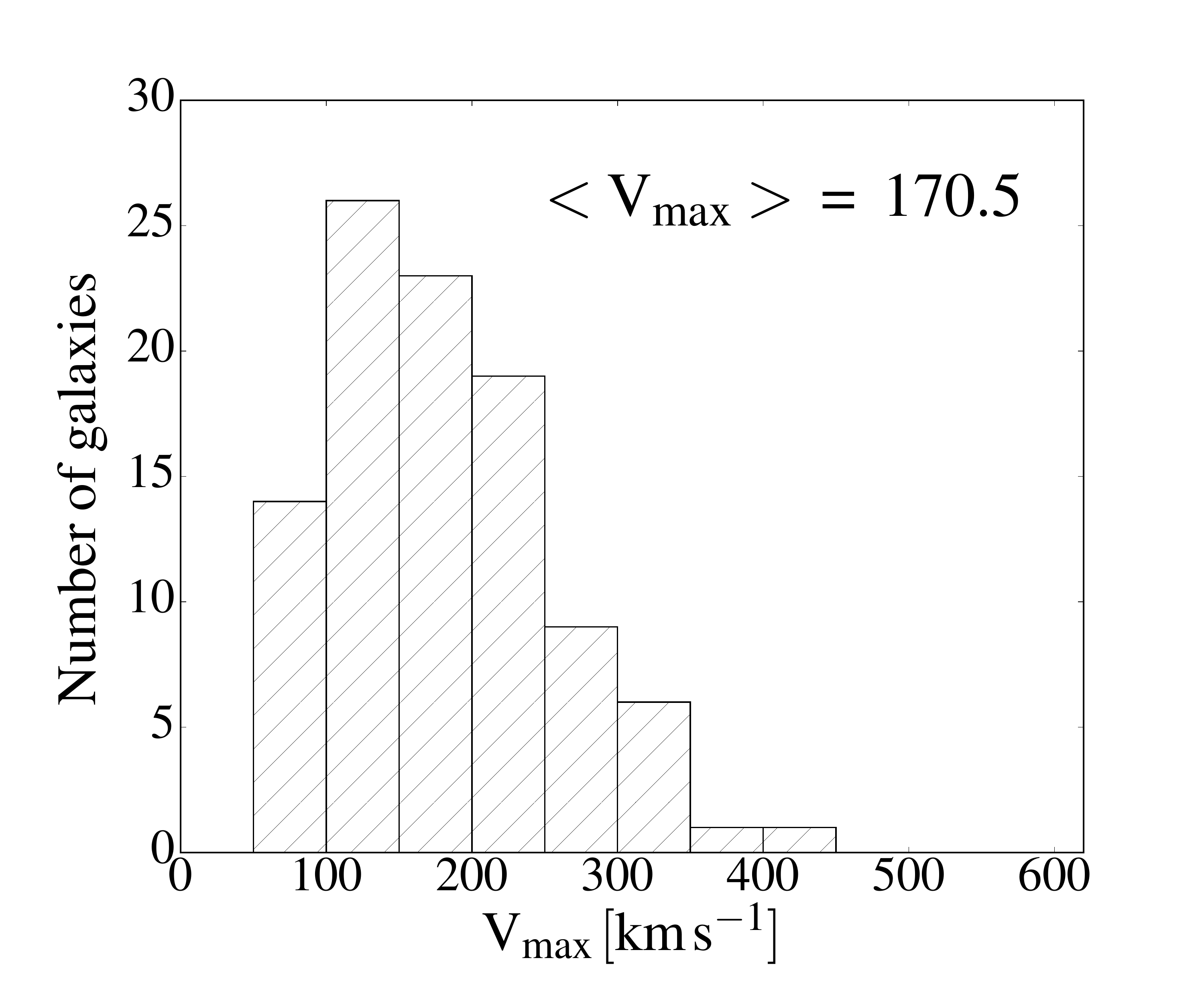}
	\hspace*{-0.35cm} \includegraphics[width=5.87cm]{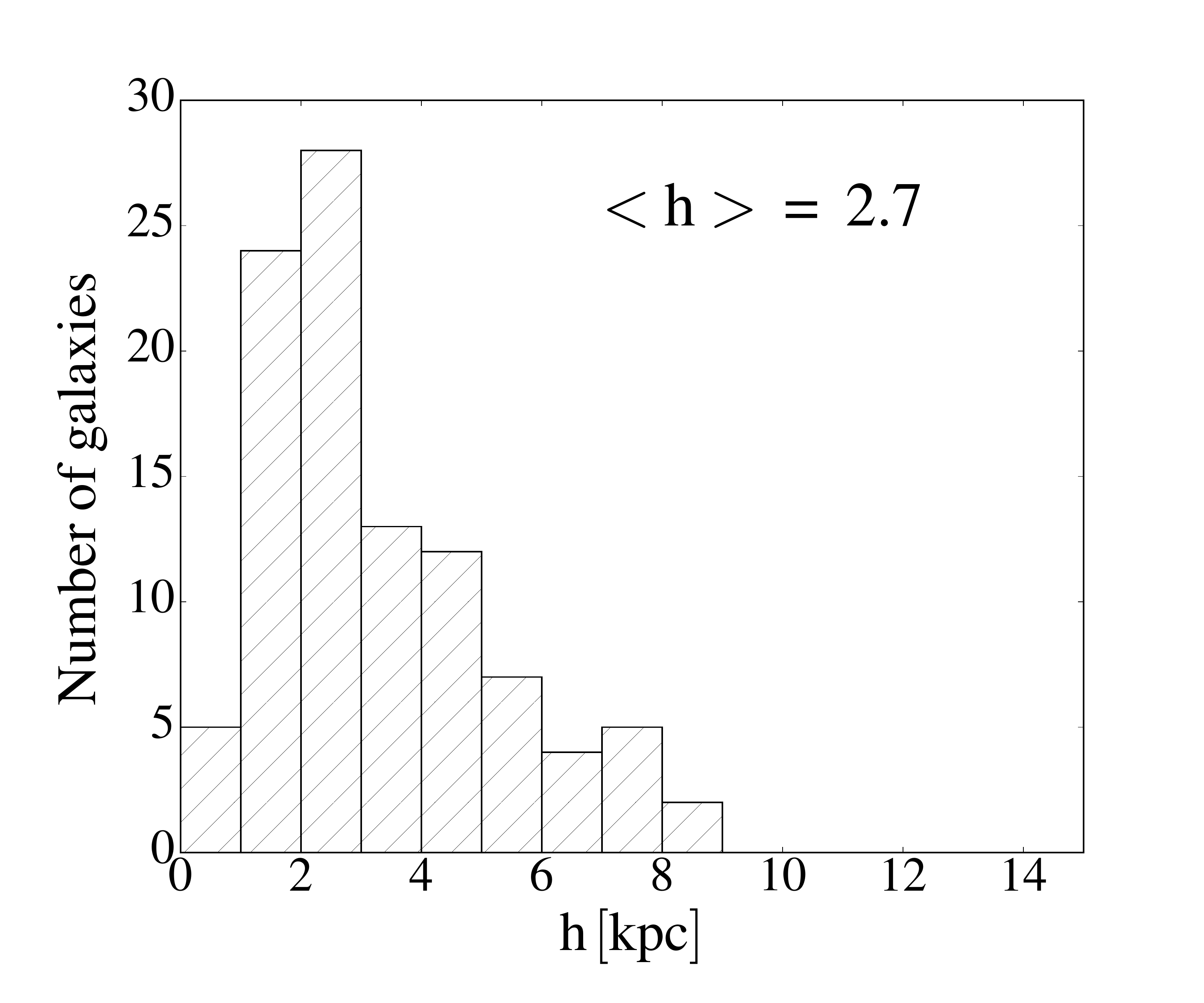}
	\hspace*{-0.35cm} \includegraphics[width=5.87cm]{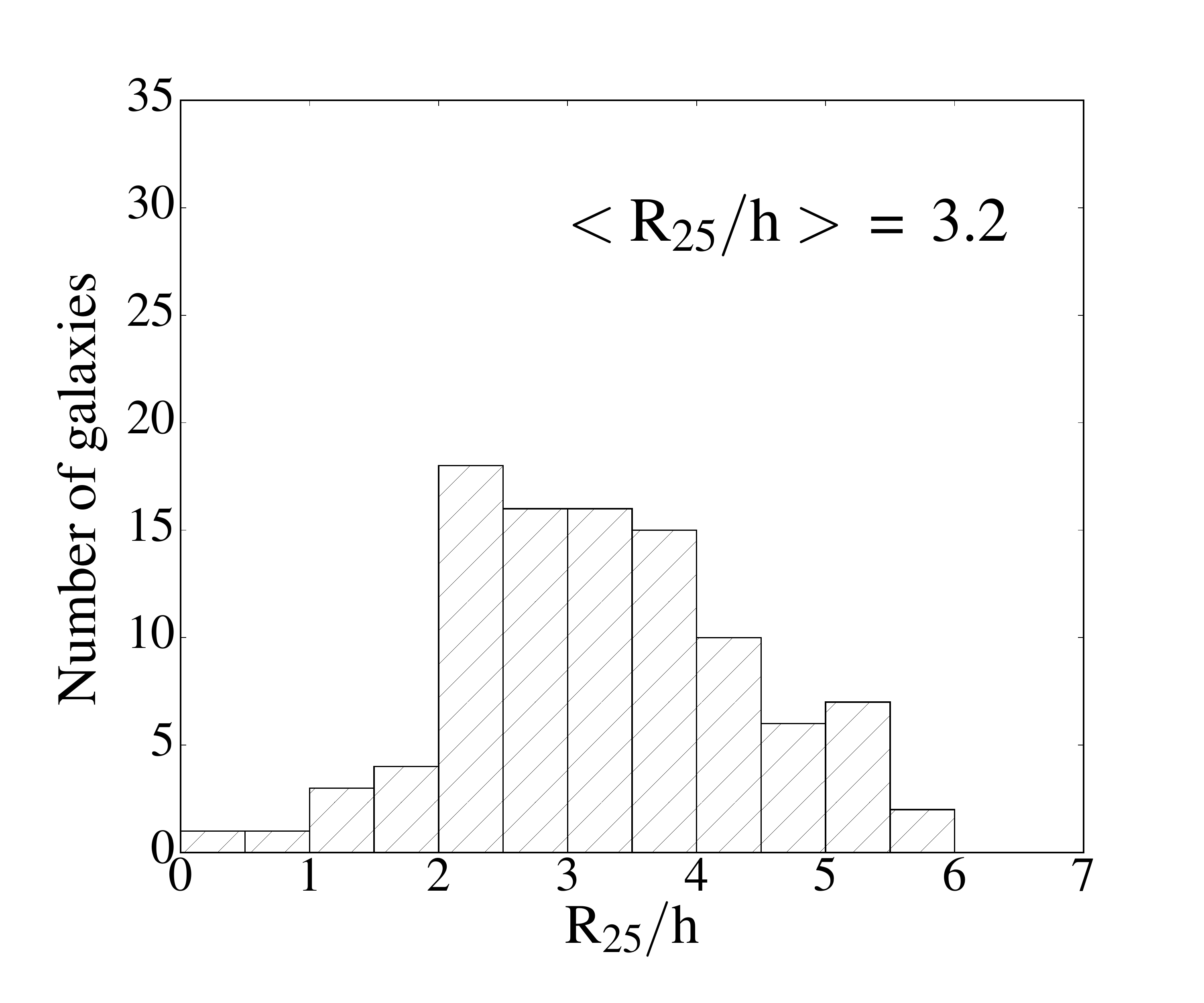}
\caption{Properties of the sample. First line: from left to right, distribution of the galaxies' distances, of the absolute magnitudes and of the morphological types, respectively. Second line: from left to right, distribution of the maximum velocities, of the disc scale lengths and of the ratios of the isophotal radius R$_{25}$ to the disc scale length, respectively. The median of each parameter is shown in the respective panel.}
\label{fig:histogram}
\end{figure*}

The overall properties of our sample are shown in Fig. \ref{fig:histogram}. The sample spans a range in (i) distances (D) from 5 to 100 Mpc with a median value of 20.2 Mpc; (ii) absolute blue magnitudes M$_B$, from -16 to -23 with a median value of -20; (iii) morphological types (t) from t=0 to t=10 with a median value of t=5.0 ($\rm S_c$); (iv) maximum rotation velocities (V$_{max}$), from 50 to 500 \kms with a median value of 170.5 \kms; (v) disc scale lengths (h), from 1 to 8 kpc with a median value of 2.7 kpc and (vi) isophotal radii at the limiting surface brightness of 25 mag arcsec$^{-2}$ (r$_{25}$), such that r$_{25}$/h ranges from 1 to 6 with a median value of 3.2, which is identical to the value of 3.2 found by \citet{Freeman+1970} in the B-band for most spiral galaxies.

\section{R$_c$-band surface photometry}
\label{sect:photometry}

\subsection{The luminosity profile}
It is well known that, within the optical range, the R$_c$-band surface brightness photometry describes better the old stellar population which represents the bulk of the stellar mass than, for instance, the B-band from which a lot of analysis has been based on in the past. Therefore, we used 73 photometric R$_c$-band data obtained on the OHP 1.2 m telescope and completed the sample with homogeneous available data in archives which consists of 27 $g+r+i$-bands profiles derived from the  Sloan Digital Sky Survey (SDSS) archival data. This leads us to a total of 100 galaxies.

The OHP images have a field of view of 11.7' $\times$ 11.7' and were taken with a single 1024 $\times$ 1024 CCD having a pixel size of 0.68 arcsec$^{-1}$. The isophotal level used as reference is 23.5 mag arcsec$^{-2}$. The OHP surface brightness profiles are taken from \citet{Barbosa+2015}. The SDSS data are taken from SDSS DR7 \citep{Abazajian+2009}, which provides imaging and calibration in the $ugriz$ pass bands. The R$_c$ surface brightness profiles were computed by using the multi-band scaling relation  \citep{Barbosa+2015} to transform the SDSS $ugriz$-bands into R$_c$-band: 
\begin{equation}
\mu_R(r)= 0.42\mu_g(r) -0.38 \mu_r(r) +0.96\mu_i(r) -0.16
\label{eq:sdss}
\end{equation}
where $\mu$(r) is the surface brightness profile at radius r. The surface brightness profiles were computed by using the IRAF task ELLIPSE, which gives the parameters that describe the luminosity of the galaxy as a function of the semi-major axis, the position angle, the ellipticity and the curve of growth that quantify the total apparent magnitude present in each isophote. 

\subsection{The light profile decomposition}
\label{sub:decomposition}

The radial profile decomposition of the 73 OHP R$_c$-band data has been done by  \citet{Barbosa+2015}. We utilize exactly the same method for the 27 remaining galaxies for which we used SDSS data. The different luminosity profiles are shown in Appendix \ref{appendixB}.  The surface brightness profiles were decomposed, when needed, into multiple components (disc, bulge, bar, spiral arm, ring, lens, ...) by using a 2D fitting Python routine. Type I discs were modelled by using a simple exponential disc:  
\begin{equation}
I_d( r ) = I_0\,  \exp{\, \left(-\frac{r} {\rm h}\right)}
\label{equation1}
\end{equation}
with $I_0$ being the central intensity of the disc and h its scale length. Type II and Type III discs, which correspond respectively to discs with downward truncations and discs with upward bends,  were modelled using broken exponential disc profiles:
\begin{equation}
I_d( r ) = SI_0\, \exp{\, \left(-\frac{r} {\rm h_i}\right)}\, .\, {\{1+ \exp[\alpha(r - r_b)]}\}^{\frac{1} {\alpha}(\frac{1} {\rm h_i} - \frac{1} {\rm h_0})}
\label{equation2}
\end{equation}
where I$_0$ is the central intensity of the disc, h$_i$ and h$_0$ the inner and outer disc scale length respectively, r$_b$ is the break radius, $\alpha$ is the sharpness of the disc transition between the inner and the outer region, and S is a scaling factor, S$= [1+ \exp(- \alpha r_b)]^{- \frac{1} {\alpha}(\frac{1} {\rm h_i} - \frac{1} {\rm h_0})}$. Bulges, bars, rings, lenses components were determined by using a S\'{e}rsic function given by :
\begin{equation}
I{_b}( r ) = I{_e} \exp{ \left( -b{_n} \left[(r/r_e)^{1/n} - 1\right]\right)}
\end{equation}

\begin{figure*}
	\begin{center}
             \includegraphics[width=8.1cm]{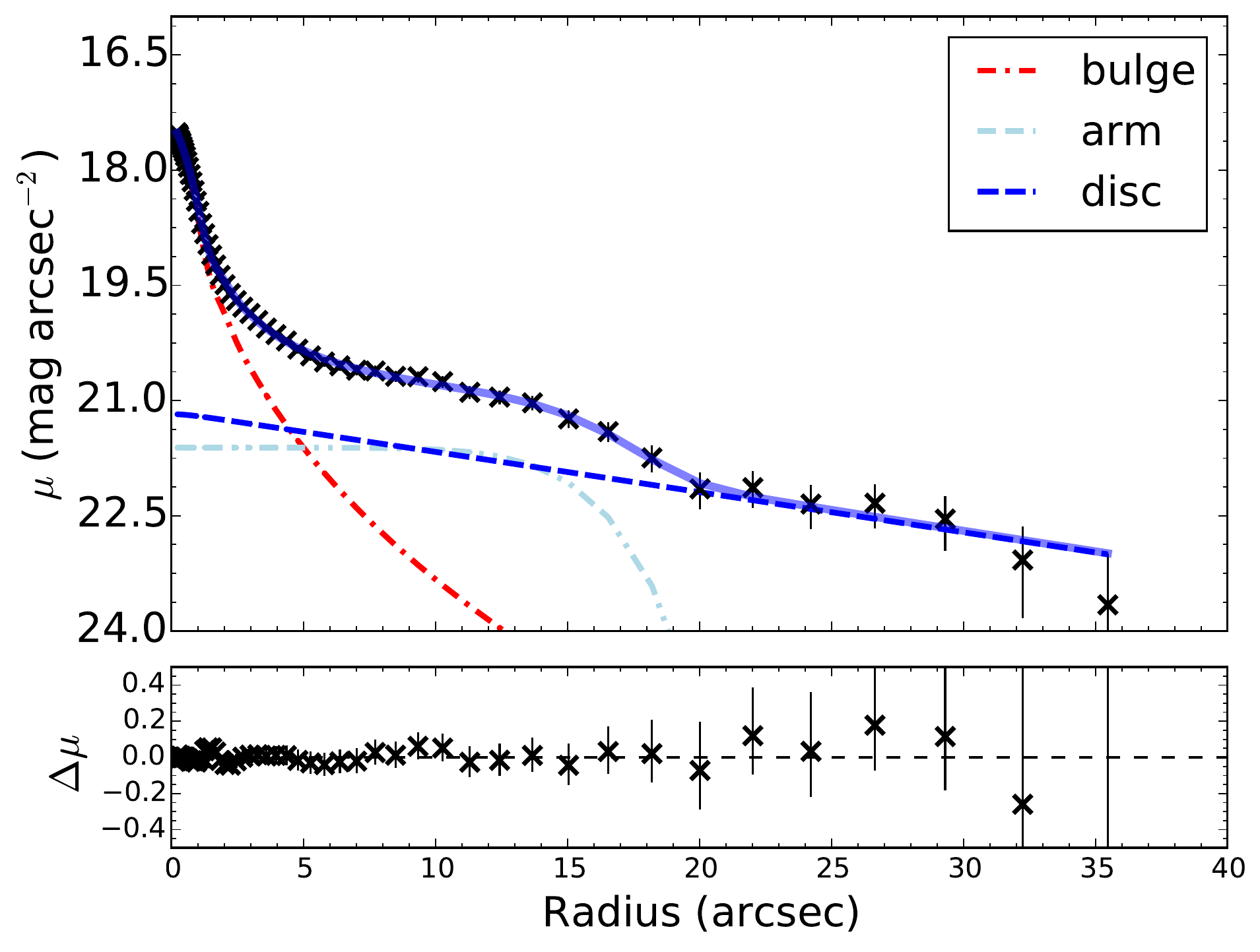}
             \includegraphics[width=8.0cm]{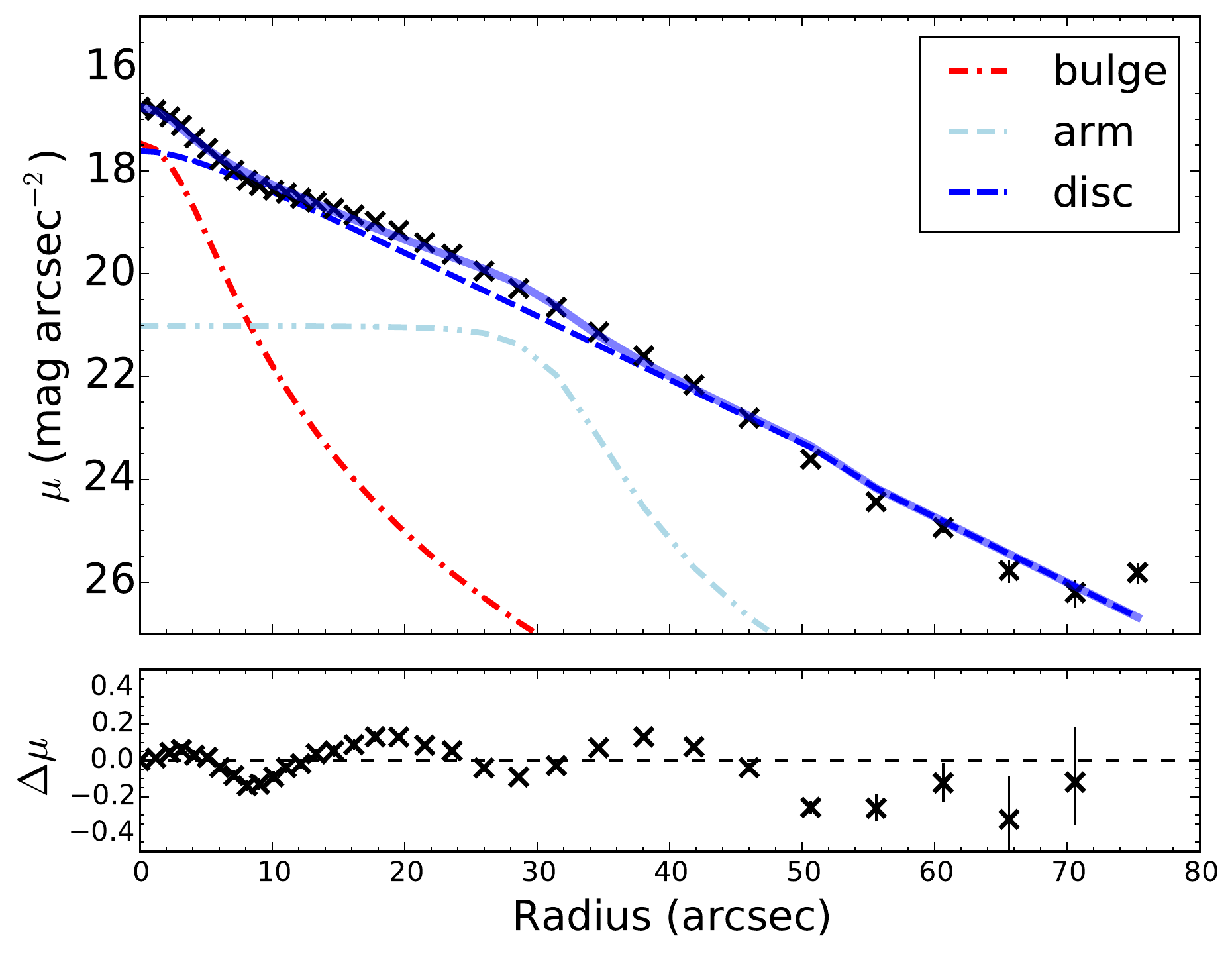}
\caption{Examples of surface brightness decomposition for the galaxy UGC 5045. The bottom panel of each plot represents the difference between the observed surface brightness distribution and the model displayed in light blue in the top panels.   Left panels are for the R$_c$-band and right panels for the W$_1$ band. }
\label{fig2:decomp}
\end{center}
\end{figure*}

We show an example of decomposition in multiple components in Fig. \ref{fig2:decomp}.  This example illustrates two general trends: (i)  MIR-band profiles are more extended than R$_c$-band profiles, the arms components are also detected further away in MIR than in R$_c$; (ii) the bulge is more cuspy in R$_c$ than in MIR, this cuspiness is nevertheless mainly linked to the difference in spatial resolutions (seeing $\sim$2-3" for R$_c$ versus a resolution of 6" for W1 in the case of the space mission WISE). Fig. \ref{fig2:decomp} provides a simple example for which only three components were needed (bulge, exponential disc and spiral arms) but decompositions usually request more components.

Spherical and flat components do not provide, for the same given mass, the same rotational velocity, thus we have to disentangle them. On the other hand, in order to simplify the construction of our mass models, we aim to minimise the number of parameters to fit. Furthermore we assume only two components: a spherical one containing the bulge that we call the bulge component and a planar one containing the disc and eventually other components embedded within the disc such as bar(s), spiral arm(s) and ring(s), that we call disc component. The different parameters of the bulge and disc components for all the sample are shown in Table \ref{tab:photometry}.

When the last radius of the surface brightness profile of the disc is smaller than the one of the rotation curve, we extrapolate the last points of the profile using the decomposition parameters in order to extend the radius to a value larger than the last radius of the rotation curve to avoid creating a truncation bump.
It was never necessary to extend bulge surface brightness profiles since bulges are less bright than discs at large radii because their luminusoty decreases rapidly at small radius.

\subsection{Mass-to-light ratio}
\label{sub:M/L}

\begin{figure}
\begin{center}
		\includegraphics[width=8cm]{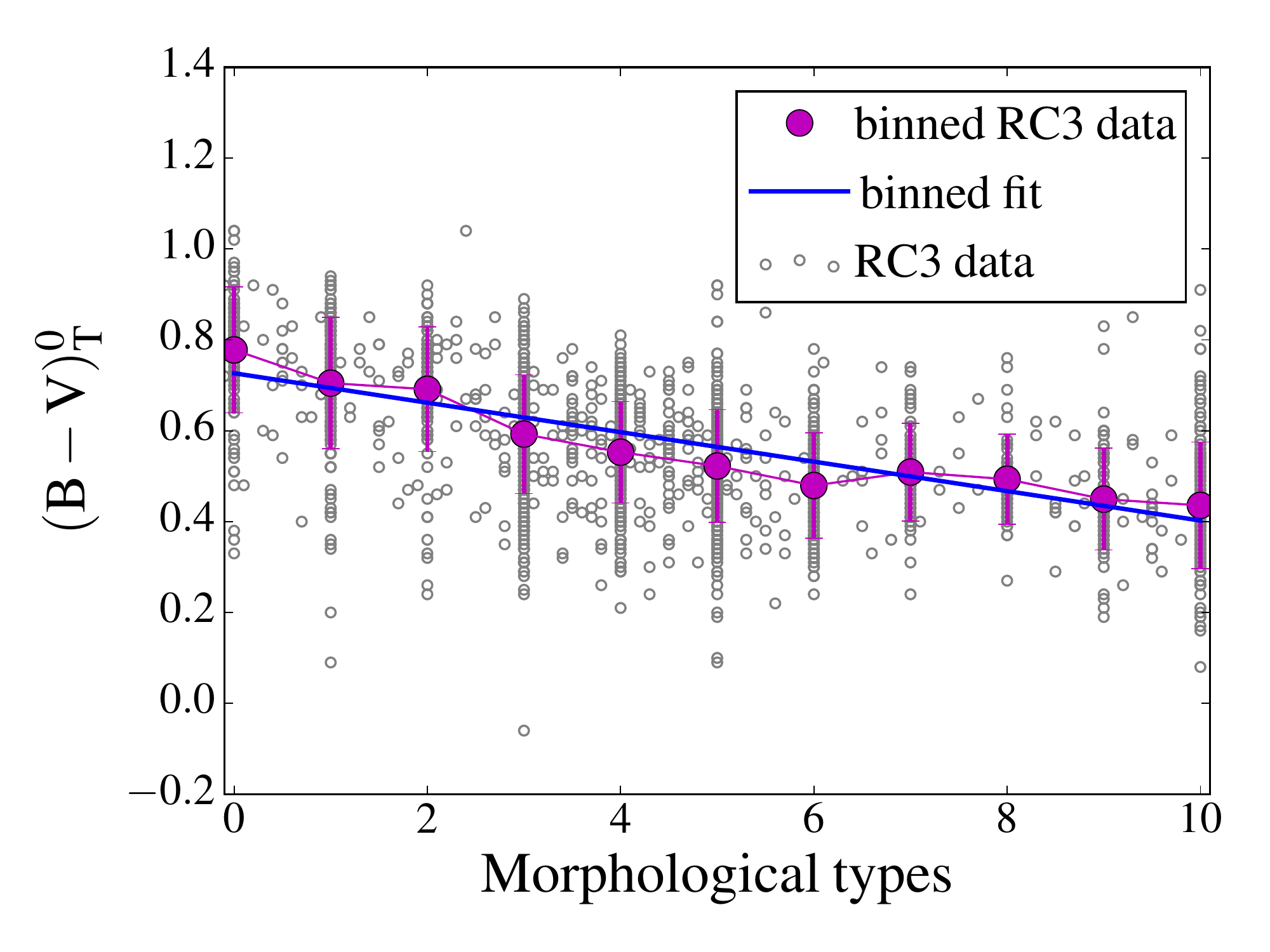}
\caption{$\rm (B - V)^0_T$ colors as a function of the morphological types. The gray dots represent the RC3 data and the magenta dots represent the RC3 binned data which are obtained by binning the data in morphological types with a bin size of 1,  and these points are connected with a magenta line. The blue line shows the fit of the binned data.}
\label{fig:bvfit}
\end{center}
\end{figure}

To obtain the stellar mass density profiles, we used the R$_c$-band luminosity profiles. The transition between the photometry and the dynamics is based on the estimation of the stellar mass-to-light ratio  (M/L). The stellar M/L values can be calculated as a function of their color, based on stellar population models \citep{Jong+2001}. The correlation between the stellar M/L and the optical color is rather tight, especially for galaxies with smooth star formation histories. Those relations between the color index and the M/L do not allow allocating masses to young stellar populations due to the fact that, at low color index, all bands are affected by a degeneracy \citep[e.g.][]{deDenus-Baillargeon+2013}. However, a general trend is expected between color and M/L in the sense that  younger populations have lower M/L and are relatively bluer compared to older populations which have higher M/L ratio. It is thus important to understand the stellar populations distribution. The relation between the M/L in the R$_c$-band and the $\rm (B - V)$ color \citep{Jong+2001} is given by : 
\begin{equation}
\rm log(M/L_{R}) = -0.660 + 1.222\, \rm (B-V)
\label{eq7}
\end{equation}
We used the color corrected for extinction designated as $\rm (B - V)^0_T$ in the RC3 catalogue when available. We found these $\rm (B - V)^0_T$ values for 62 galaxies of our sample. To estimate the values for the remaining 38 galaxies, we selected the 1706 galaxies of the RC3 (out of the 23011 galaxies) having a tabulated $\rm (B - V)^0_T$ value and a morphological types spanning the range from 0 to 10. We plotted  in Fig. \ref{fig:bvfit}, $\rm (B - V)^0_T$ as a function of the morphological types for those galaxies.  As expected $\rm (B - V)^0_T$ decreases with morphological types. To derive a relation between $\rm (B - V)^0_T$ and morphological types, we binned the morphological types (bin size=1) and choose the average value within each bin, these are the magenta dots showed in the figure and the blue line is the fit to those dots taking the error bars into account. We therefore used the relation found when using the fit to derive the remaining $\rm (B - V)^0_T$. This relation is :
\begin{equation}
\rm (B - V)^0_T= (-0.032 \pm 0.004) \, \times \,  \rm t   +(0.73 \pm 0.02)
\label{eq:bvfit}
\end{equation}
where t is the morphological type. The $\rm (B - V)^0_T$ are shown in column (4) of Table \ref{tab:photometry} and those marked with an asterisk correspond to the values derived from equation \ref{eq:bvfit}.

\begin{figure*}
\begin{center}
	\vspace{0.cm}\includegraphics[width=6.0cm]{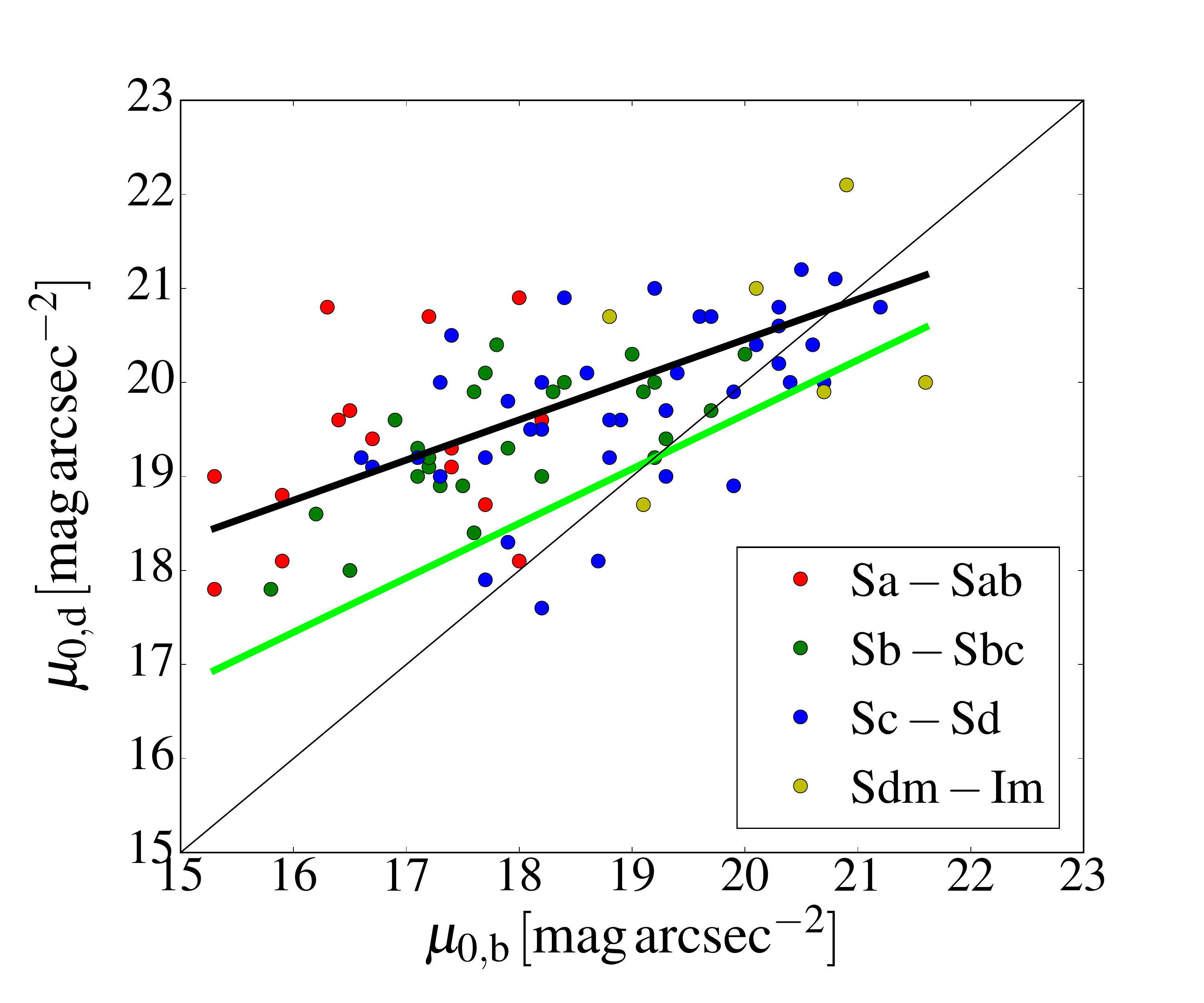}
	\vspace{0.cm}\includegraphics[width=6.0cm]{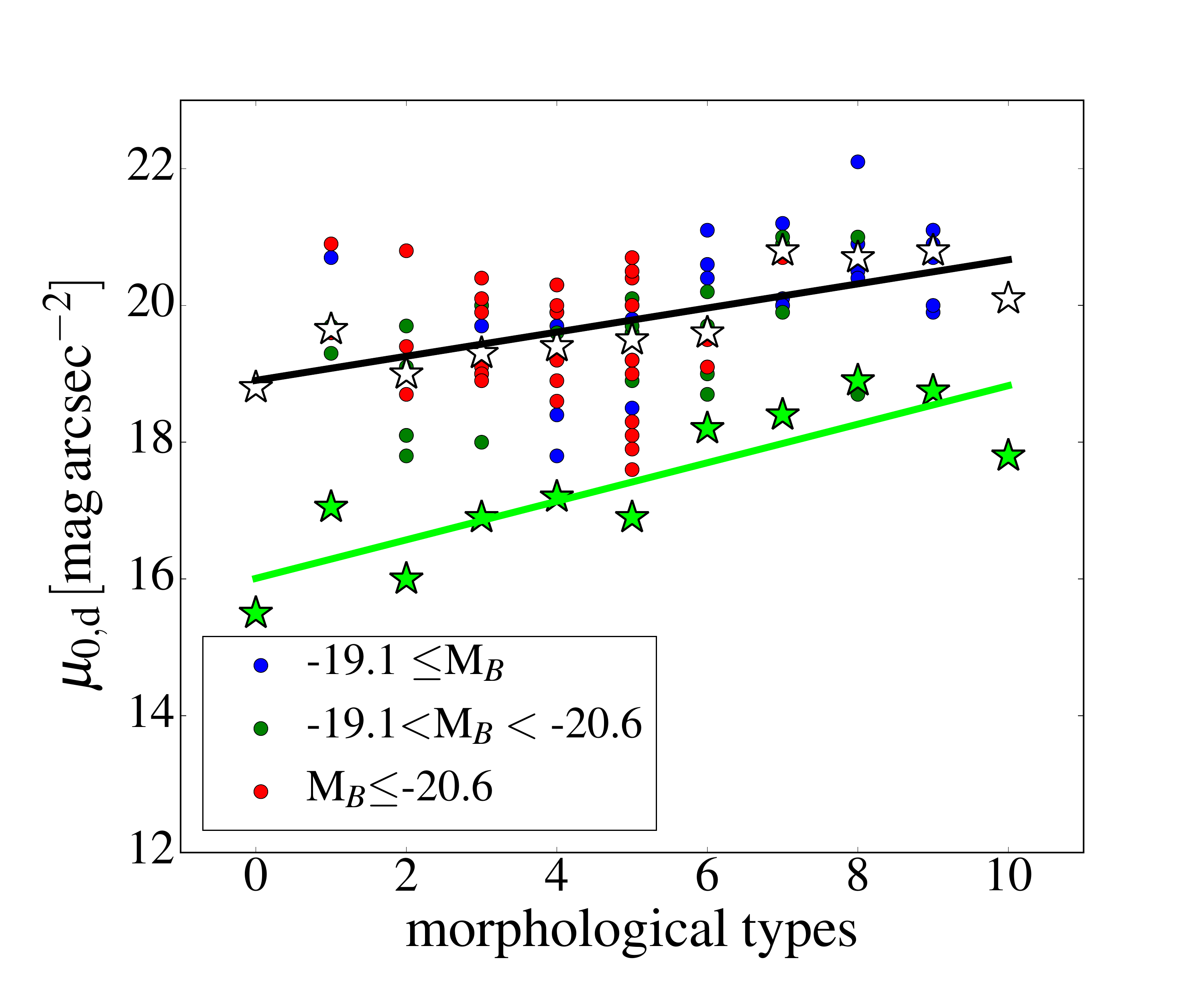}
\end{center}
\caption{The left panel shows the disc central surface brightness versus the bulge central surface brightness by morphological type. The thin black line represents the y=x relation. Thick black and lime lines represent the fit of Rc and W1 bands respectively. The right panel shows  the disc central surface brightness versus morphological type. The open black stars represent the median in morphological types and the thick black line is the fit of the median data for the R$_c$-band; the lime stars show the median in morphological types and the thick lime line is the fit of the median data for the W1-band.}
\label{fig:fig4}
\end{figure*}

\begin{figure*}
	\hspace{-0.2cm}\vspace{-0.cm}\includegraphics[width=6.cm]{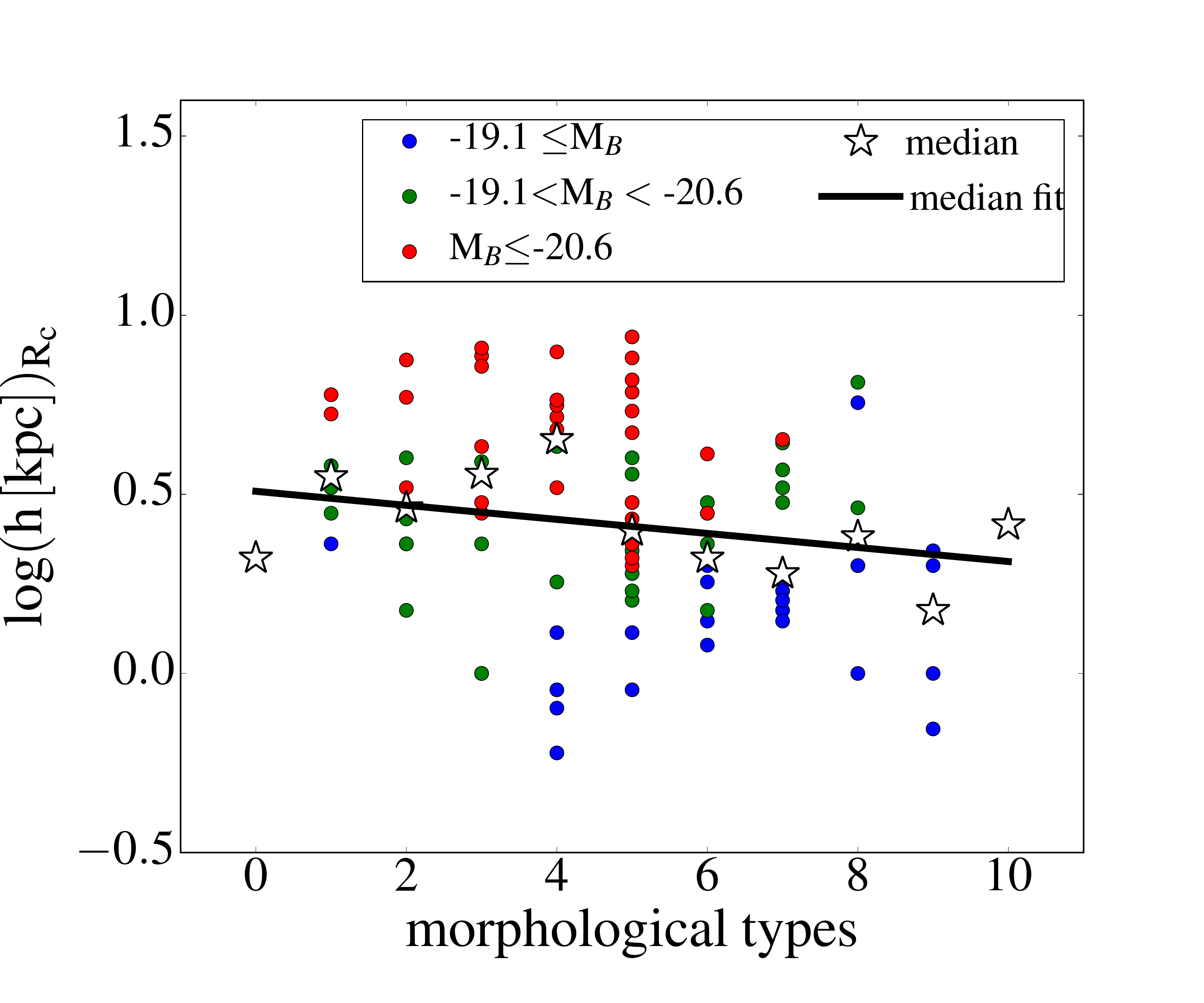}\hspace{-0.25cm}
	\hspace{-0.2cm}\vspace{-0.cm}\includegraphics[width=6.cm]{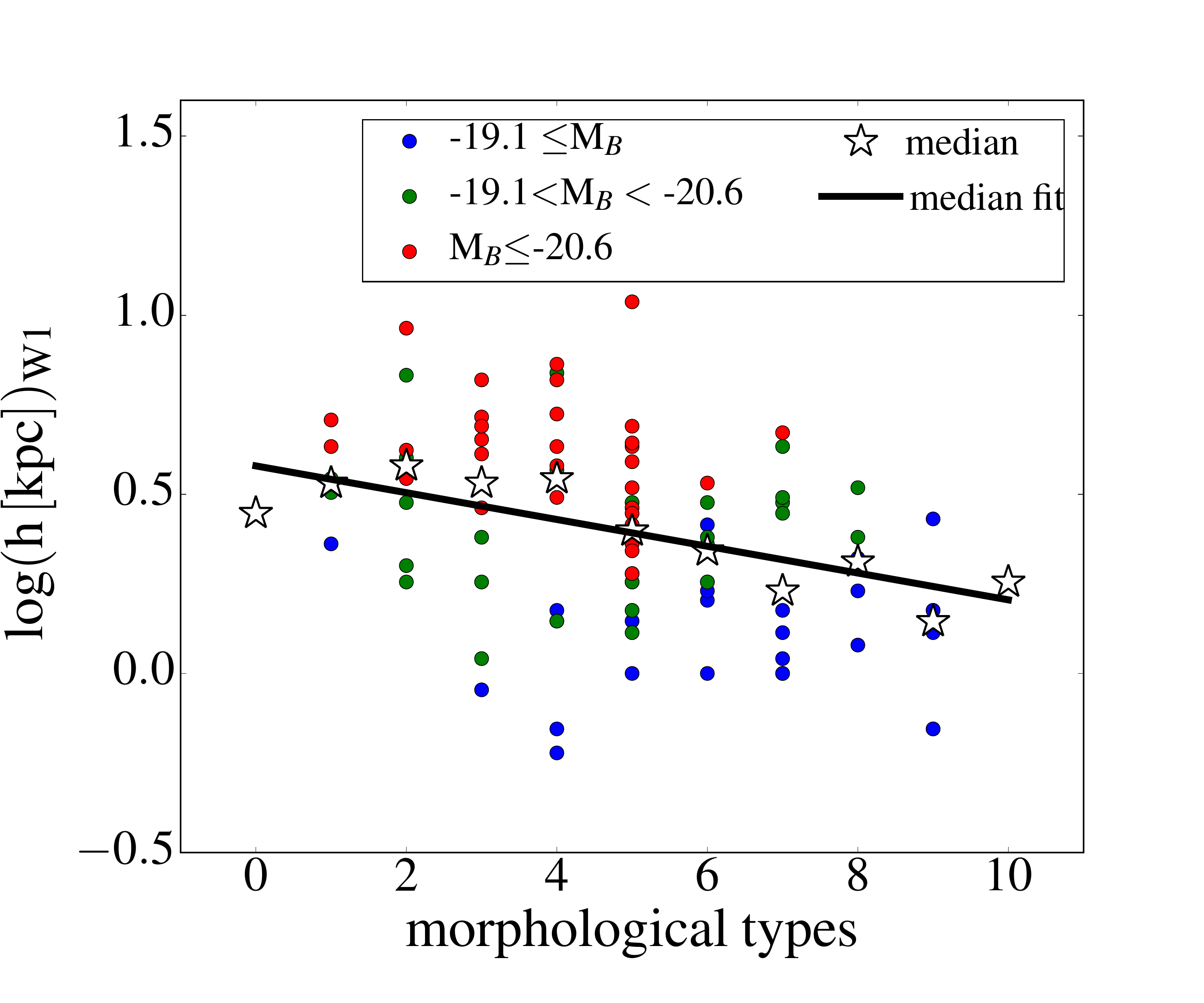}\hspace{-0.25cm}
	\hspace{-0.2cm}\vspace{-0.cm}\includegraphics[width=6.cm]{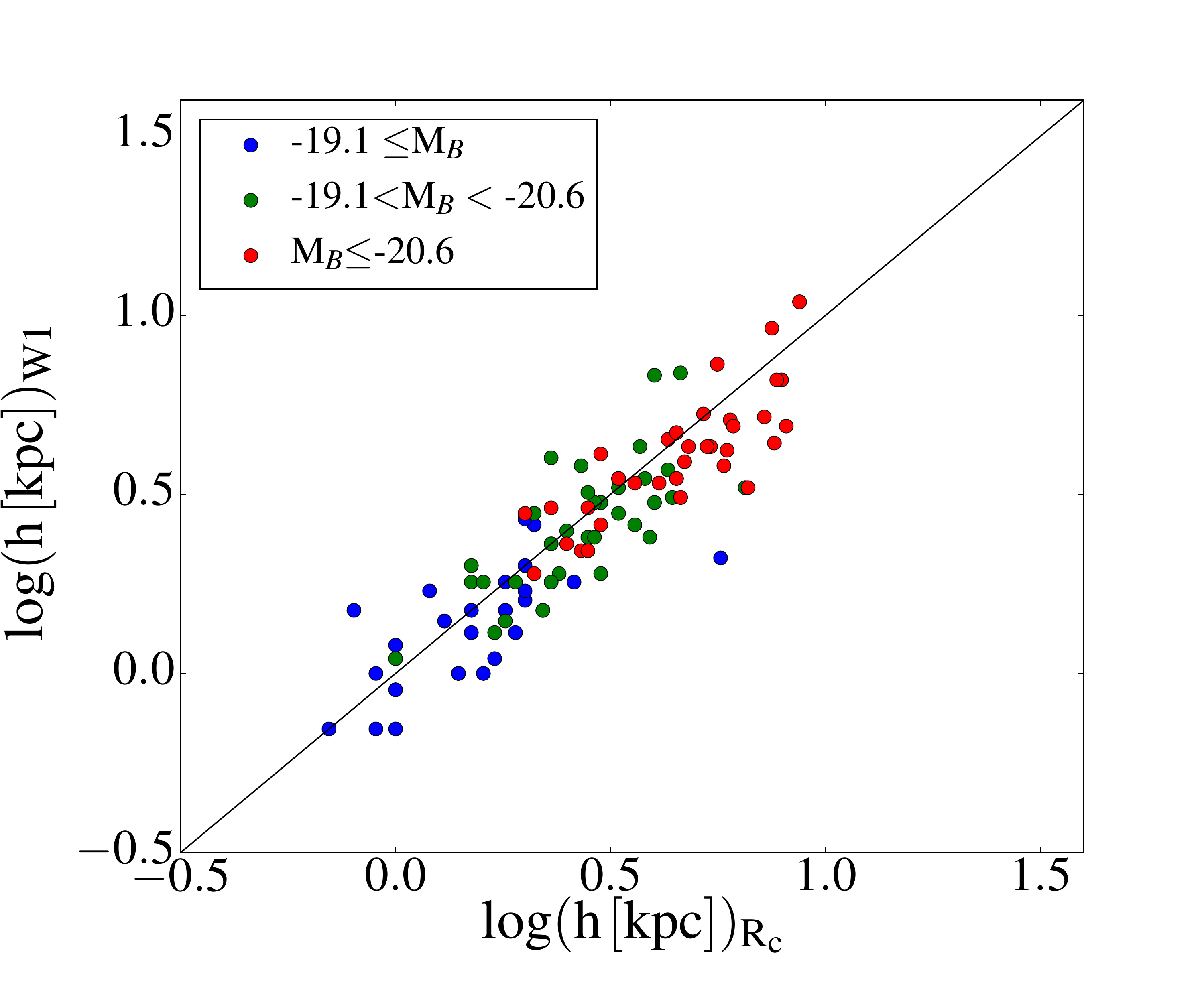}
\caption{The left and middle panels show respectively the disc scale lengths in the R$_c$-band and in the W1-band versus morphological types. The open black stars represent the median in morphological types. The thick black line is the fit of the median data. The right panel represents the disc scale length in the W1-band versus the disc scale length in the R$_c$ band. The thin black line shows the y=x relation.}
\label{fig:fig5}
\end{figure*}

\begin{figure}
\begin{center}
	\vspace{0.cm}\includegraphics[width=6.0cm]{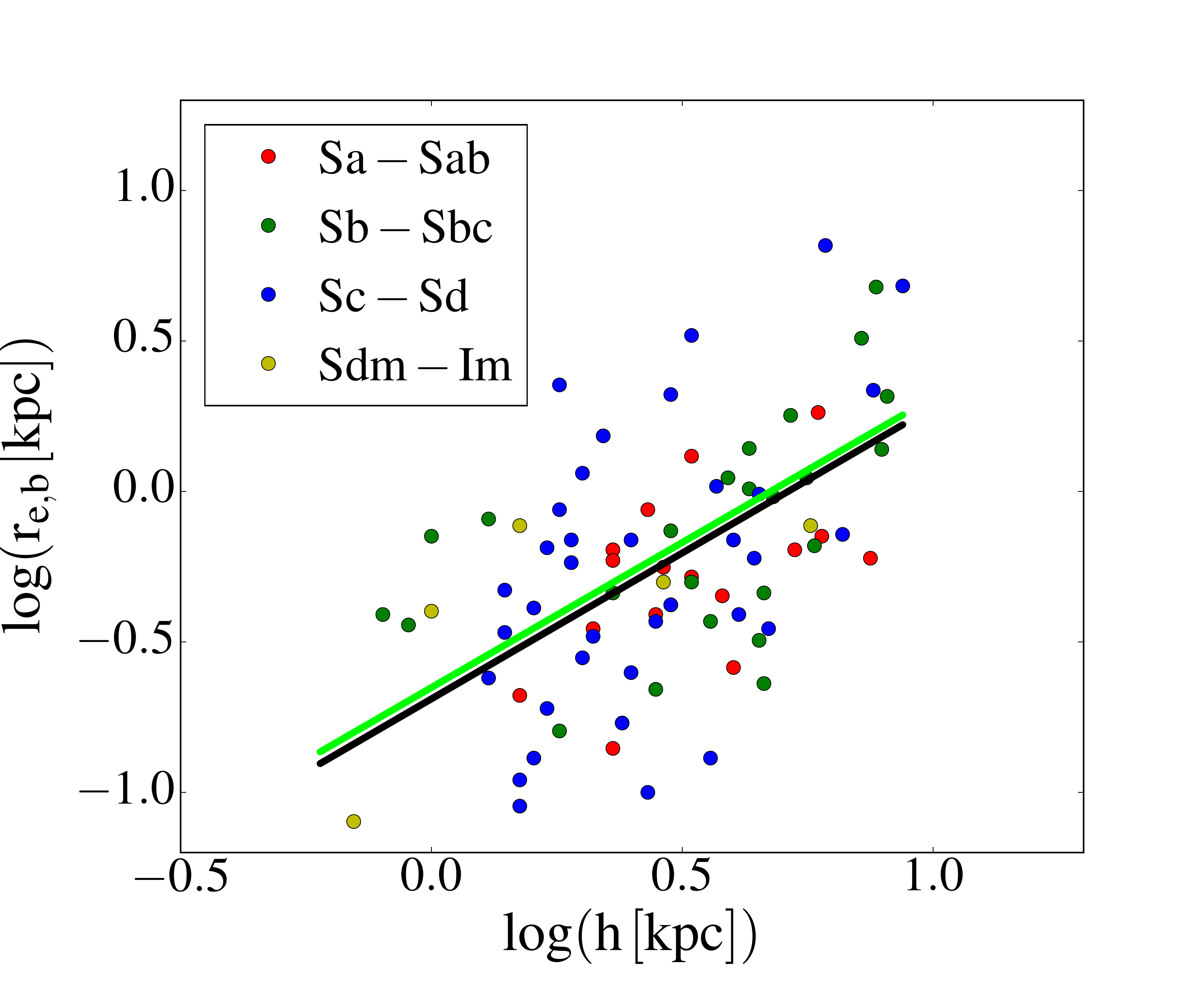}\vspace{-0.55cm}
	\vspace{0.cm}\includegraphics[width=6.0cm]{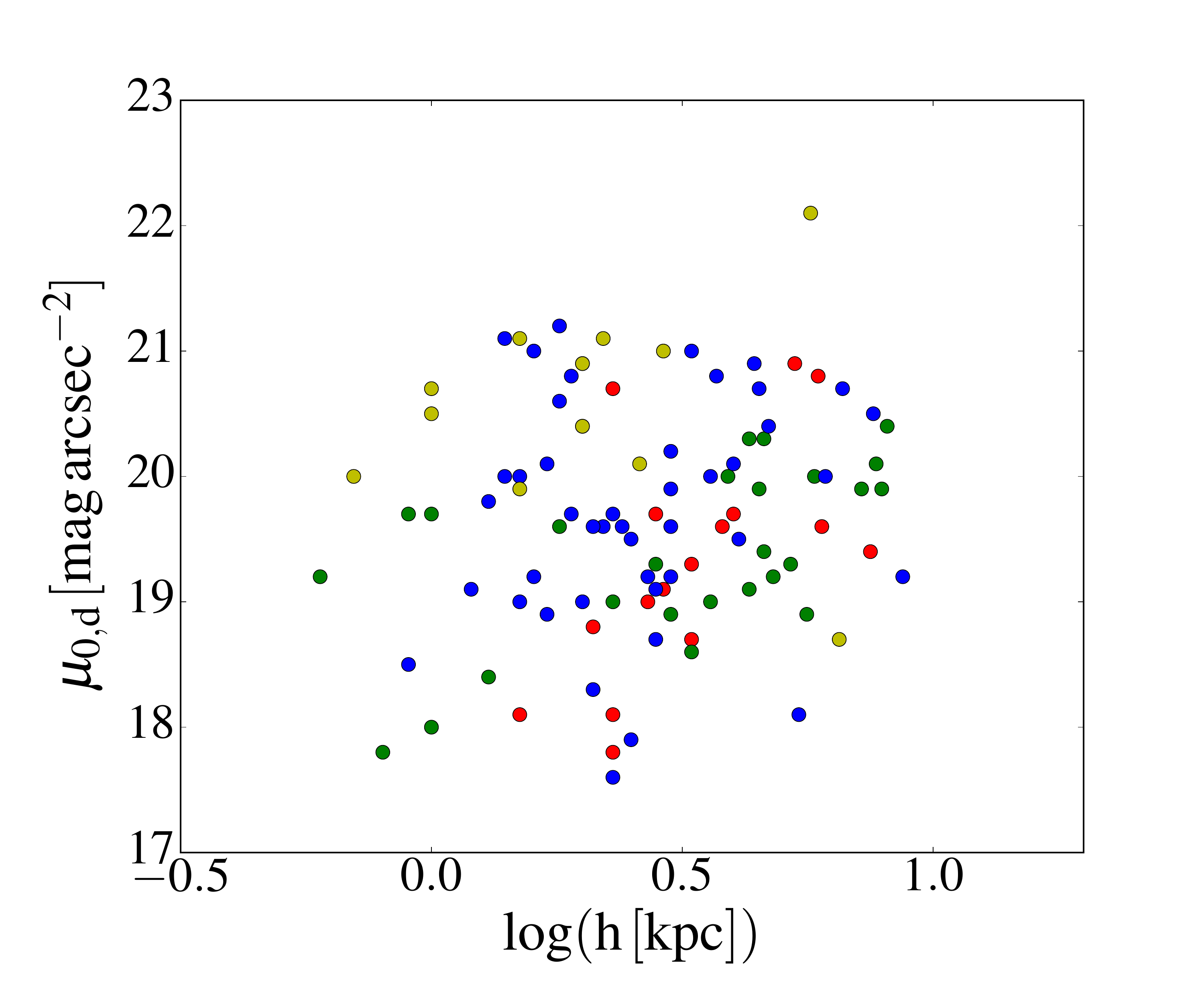}\vspace{-0.55cm}
	\vspace{0.cm}\includegraphics[width=6.0cm]{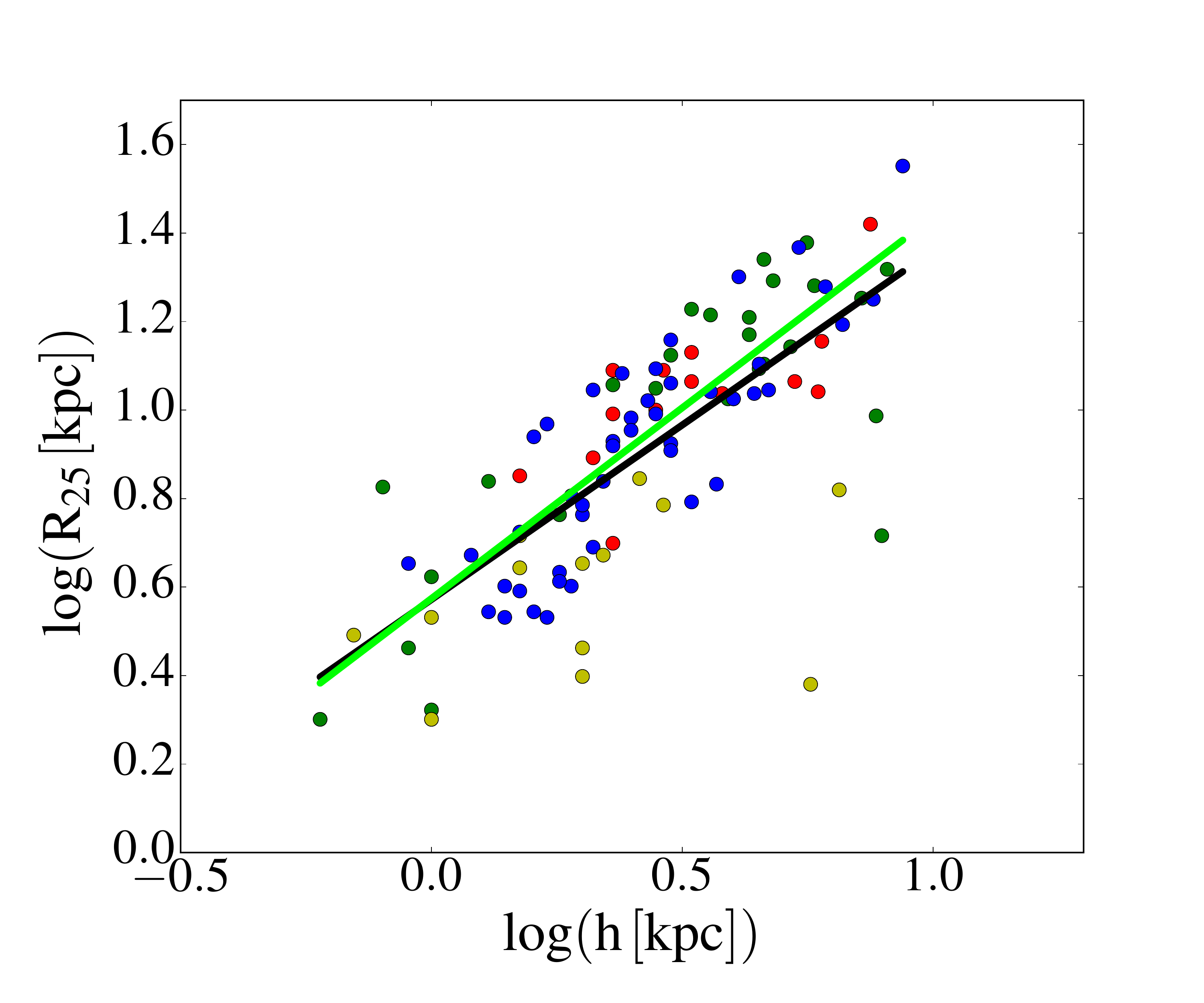}
\end{center}
\caption{Correlations between the parameters derived from the R$_c$ luminosity profiles. From top to bottom: the bulge effective radius, the disc central surface brightness and the isophotal radius R$_{25}$ versus the disc scale length respectively. The thick black and lime lines represent the fits of the Rc and W1 band respectively. The legends for the three panels are shown in the top panel.}
\label{fig:fig6}
\end{figure}

\begin{figure}
\begin{center}
	\vspace*{-0.0cm}\includegraphics[width=6.0cm]{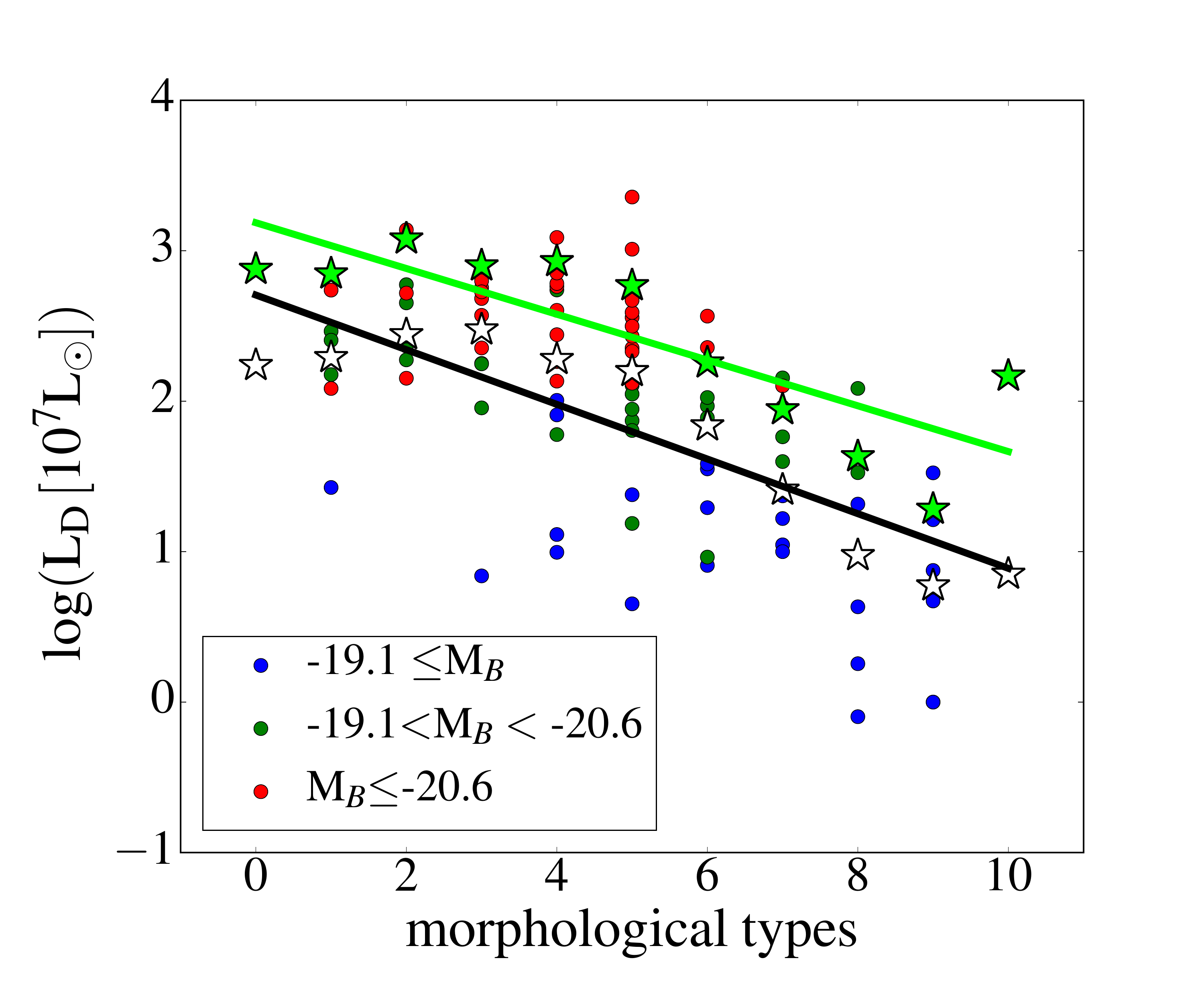}\vspace{-0.55cm}
	\vspace*{-0.0cm}\includegraphics[width=6.0cm]{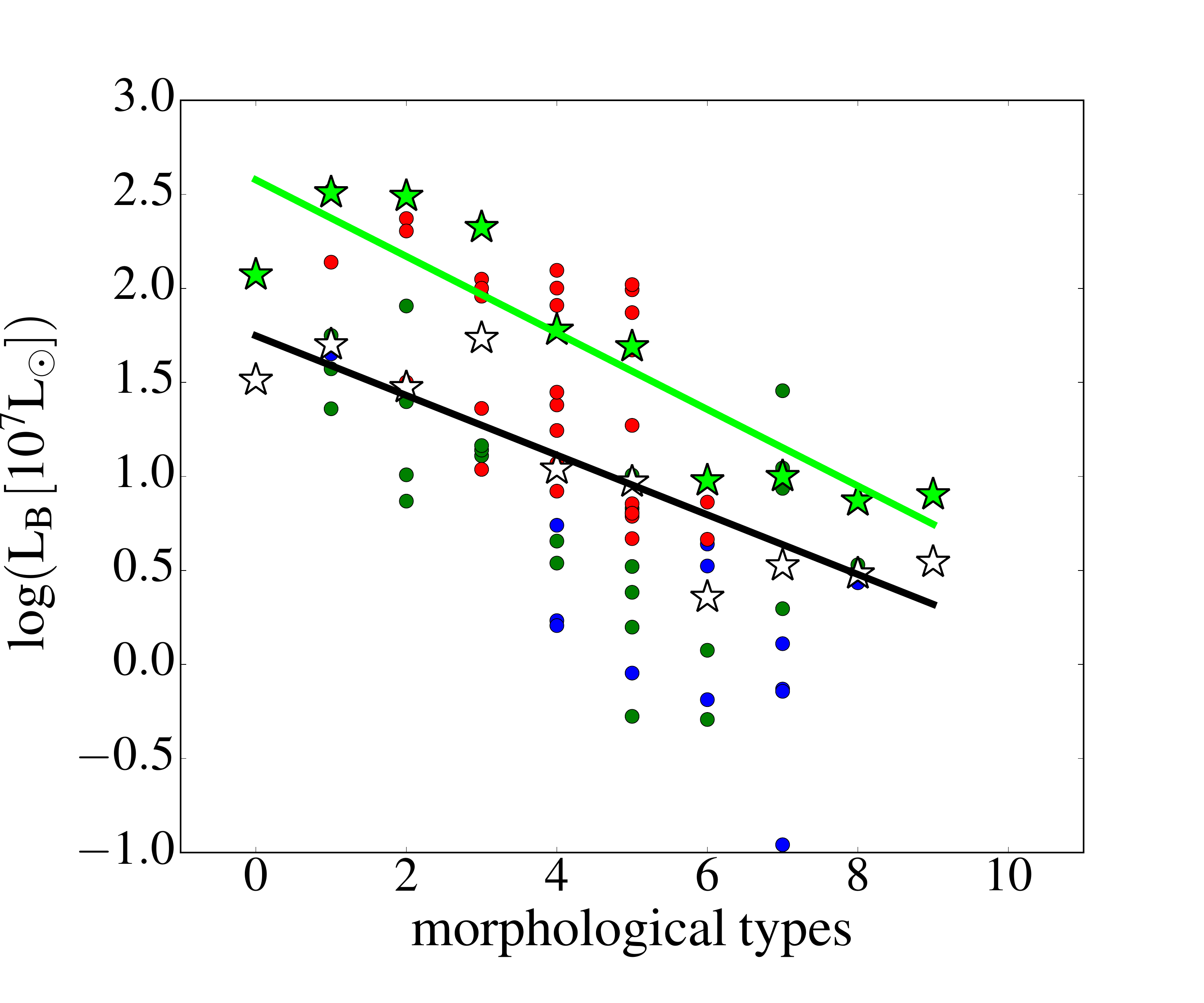}\vspace{-0.55cm}
	\vspace*{-0.0cm}\includegraphics[width=6.0cm]{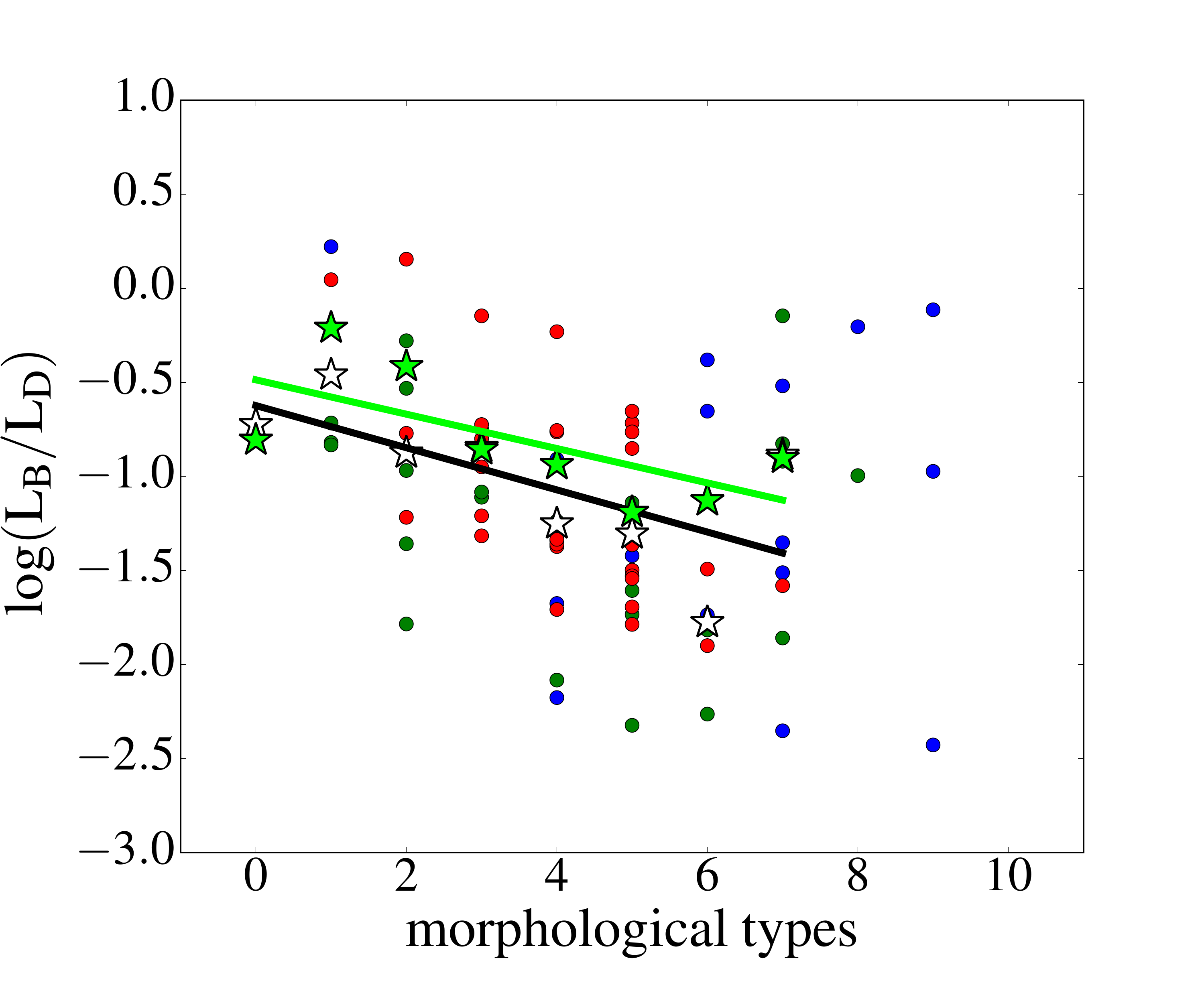}
\end{center}
\caption{From top to bottom: disc luminosity, bulge luminosity and ratio of disc-to-bulge luminosity versus the morphological types. The open black stars represent the median in morphological types for the R$_c$-band and the thick black line is the fit of the median data. The lime stars represent the median in morphological types for the W1 band and the thick lime line is the fit of the median data. The legends for the three panels are shown in the top panel.}
\label{fig:fig7}
\end{figure}

\subsection{Scale parameters and light distribution of discs and bulges}
\label{sub:discbulge}

From Figs. \ref{fig:fig4} to \ref{fig:fig7} we display the correlations between the photometric parameters obtained for the R$_c$-band photometry and we compare those trends with the ones obtained using the MIR 3.4 $\mu$m of WISE photometry presented in \citet{Korsaga+2018}. The resulting parameters corresponding to those figures are also listed in Table \ref{tab:photometry}.

Surprisingly, the left panel of Fig. \ref{fig:fig4}, the top panel of Fig. \ref{fig:fig6}, the middle and bottom panels of Fig. \ref{fig:fig7}, show several late-type Sdm/Im galaxies displaying a faint bulge structure.

The y=x line of Fig. \ref{fig:fig4}, left panel clearly shows, as expected, that for most of the cases, bulge central surface brightnesses are brighter than disc central surface brightnesses. The slope of this correlation is almost identical in the MIR, this means that both bands lead to roughly the same bulge-to-disc decompositions. The zero point is about one magnitude lower in the MIR than in the R$_c$-band and closer to the y=x line, which means that the stellar populations of the bulge and of the disc are closer in the MIR than in the R$_c$-band. However, one has to be careful since this shift of one magnitude could be partially due to beam smearing effects (seeing): the spatial resolution being higher in the R$_c$-band (ranging from $\sim$ 1.5 to 3 arcsec, depending on the seeing) than in the MIR WISE W1-band (limited at 6 arcsec by the pixel sampling).  As expected, the right panel of Fig. \ref{fig:fig4} shows a correlation between disc central surface brightness and morphological types. The slope is again similar in both bands. 

The left and middle panels of Fig. \ref{fig:fig5} show that disc scale lengths in the R$_c$-band are less correlated to the morphological types than the ones in the MIR-band.  The scatter of those relations being very high, the slope difference in the trend is maybe not significant. As expected, for a given morphological type (from t = 0 to $\sim$7) disc scale lengths are larger/smaller for bright/faint galaxies as it is underlined by the median fit lines. The right panel of Fig. \ref{fig:fig5} compares the disc scale lengths in the MIR- and in the R$_c$-bands, which follow well the y=x line  and as expected faint galaxies have smaller disc scale lengths than bright ones, which seems to support the fact that the difference of slope observed in the left and middle panels is not significant.

Despite the large dispersion, the top panel of Fig. \ref{fig:fig6} shows some correlation between the disc scale length and the bulge effective radius; galaxies with small/large disc scale lengths ($\rm h$) tend to display small/large bulge effective radius ($\rm r_{e,b}$).  This correlation is the same as the one observed in the MIR. 

The middle panel of Fig. \ref{fig:fig6} displays a large scatter between discs scale lengths and central surface brightnesses.  As expected, early type galaxies ($\rm Sa-Sab$) tend to have larger disc scale lengths and late type galaxies ($\rm Sdm-Im$) fainter central surface brightness but we do not see any correlation between central surface brightness and disc scale length or the morphological type from $\rm Sb$ to $\rm Sd$ galaxies. The correlation found in the top panel and the absence of correlation observed in the middle panel mean that the characteristic sizes of the disc and of the bulge are identically correlated in R$_c$- and in MIR-bands while their central surface brightnesses may differ. As expected, the bottom panel of Fig. \ref{fig:fig6} shows a clear correlation between isophotal radii $\rm R_{25}$ in the B-band and discs scale lengths $\rm h_R$ in the R$_c$-band. For spiral galaxies, the 25 mag arcsec$^{-2}$ isophote is, on averaged, reached at 3.2 h$_B$, where h$_B$ is the disc scale length in the B-band \citep{Freeman+1970}. Bottom-right panel of Fig \ref{fig:histogram} provides an average value of 3.2$\pm 1.2 $ h  for this ratio. This value is smaller than the value of 3.5$\pm 1.2 $ h  found in the MIR \citep{Korsaga+2018} but the scatters are rather large. 

As expected, the top and middle panels of Fig. \ref{fig:fig7} show that both discs and bulges luminosities are clearly anti-correlated with morphological types and that early morphological type galaxies show more luminous discs and bulges than later type galaxies. These plots demonstrate that the slope of the correlation is almost the same in the MIR than in the R$_c$-band, which means that the $\rm M/L $ in both bands will display the same trend with respect to the morphological type. The bottom panel of Fig. \ref{fig:fig7} quantifies the bulge--to--disc ratio versus the morphological type, which decreases, as expected. We nevertheless note a large scatter in this relation but here again, consistently with the two previous plots, the slope of the anti-correlation is the same for both bands (MIR- and R$_c$-bands) from morphological type t=0 (S0) to t=7 (Sd). One should note that most bulge-to-disc ratios of t=8 (Sdm) and t=9 (Im) galaxies are quite large and do not fit the general trend observed for earlier types. This is not due to their bulge luminosity that fits the general trend (Fig. \ref{fig:fig7}, middle panel) but to the faint disc luminosity (Fig. \ref{fig:fig7}, top panel). 
From the whole sample of 100 galaxies, only 6 of them are pure discs, i.e. their profiles do not exhibit any bulge, bar(s), ring(s), lense(s) or spiral arm(s) components, thus have not been decomposed. Among the 100 galaxies, only 17 do not exhibit a bulge component. Surprisingly, among the 83 galaxies with a bulge, 13 of them are Sdm or Im galaxies. In addition, the mean $\rm L_b/L_d$ (0.32) of those 13 galaxies is greater that the mean $\rm L_b/L_d$ (0.09) of the 45 Sc and Sd galaxies. This is explained by the fact that, on average, in our sample, the difference of the luminosity between Sdm/Im and Sc/Sd galaxies is larger for the disc than for the bulge.
Indeed, the mean disc luminosity decreases from $\simeq$ 199$ \times 10^8$ for Sc/Sd galaxies to $\simeq$ 17$ \times 10^8$ for Sdm/Im. Meanwhile the bulge luminosity drops only from $ \simeq$14$ \times 10^8$ to $ \simeq$ 3$ \times 10^8$.  Thus the luminosity disc ratio between Sc/Sd and Sdm/Im is $ \simeq$ 12 while the bulge ratio is $ \simeq$ 5.

In conclusion of this section \ref{sub:discbulge}, the scale relations and light distributions observed between the different photometric parameters of discs and bulges are not dependant on the photometric band used, from the optical R$_c$ to the MIR 3.4 $\mu$m bands.  

\section{Mass Models }
\label{sect:massmodel}
\subsection{Methods}

We use H$\alpha$ rotation curves in combination with R$_c$-band photometry to construct the mass models of the 100 galaxies. The mass models are constructed using an improved version of the python routines used for \citet{Korsaga+2018}. We performed two main models to describe the mass distribution of the DM halo; the pseudo-isothermal core density profile (ISO) \citep{Begeman+1987} and the $\Lambda$CDM cuspy density profile (NFW)  \citep{Navarro+1996}.
The total circular velocity is given by:
\begin{equation}
V_{\rm cir}(r)= \sqrt{V_{\rm disc}^2 + V_{\rm bulge}^2  + V_{\rm halo}^2} \\
\label{eq4}
\end{equation}
where $V_{\rm cir}$ is the circular velocity, $V_{\rm disc}$ and $V_{\rm bulge}$  are the velocity of the disc and  bulge components respectively and $V_{\rm halo}$ is the DM halo velocity.

For the pseudo-isothermal sphere (ISO) model, the density profile is given by:
\begin{equation}
\rho_{\rm iso}(r) = \frac{\rho_0}{\left[1+\left(\frac{r}{r_0}\right)^2\right]}
\label{eq5} 
\end{equation}
and the corresponding velocity by:
\begin{equation}
V_{\rm iso}^2(r)= 4\upi G\rho_0 r_0^2 \left[1-\frac{r_0}{r} \arctan\left(\frac{r}{r_0}\right)\right]
\label{viso}
\end{equation}
which is an increasing function of $r$ , asymptotically reaching $V_{\rm max}=V(r=\infty)=\sqrt{4\pi G\rho_0 r_0^2}$\\ where $\rho_0$ is the central density of the DM halo and $r_0$ its scaling radius.

The Navarro-Frenk-White (NFW) model derived from $\Lambda$CDM numerical simulations generates a cuspy halo where the density profile is given by: 
\begin{equation}
\rho_{\rm NFW}(r) = \frac{\rho_i}{(\frac{r}{r_s})(1+\frac{r}{r_s})^2}
\label{eq8}
\end{equation}
 where $\rho_i$ is the density of the universe at the time of collapse and $r_s$ is a scale radius.
 
 The velocity is given by:
\begin{equation}
V_{\rm NFW}^2(r)= V_{200}^2\left[\frac{\ln(1+cx)-cx/(1+cx)}{x[\ln(1+c)-c/(1+c)]}\right]
\label{vnfw}
\end{equation}
where $V_{200}$ is the velocity at the virial radius, $c$ is the concentration parameter of the halo and $x=r/r_s$. The scale radius $r_s$ can be inferred from $V_{200}$ and $c$:
\begin{equation}
r_s = \frac{V_{200}}{10 c\times \text{H}_0}
\label{rs_v200_c}
\end{equation}
where H$_0$ is the Hubble constant.
 
\begin{figure*}
\hspace*{-0.00cm} \includegraphics[width=0.35\textwidth]{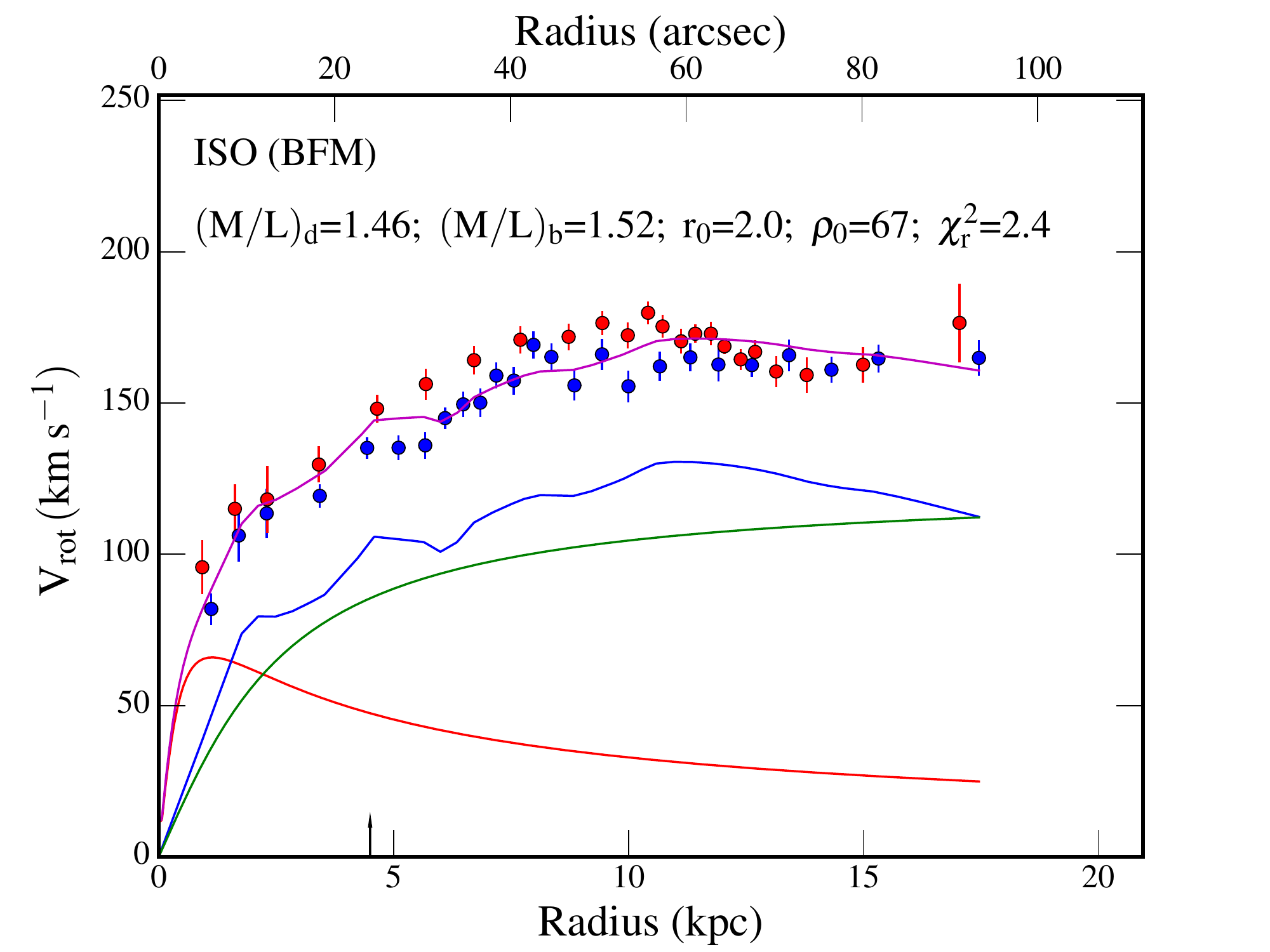}
\hspace*{-0.75cm} \includegraphics[width=0.35\textwidth]{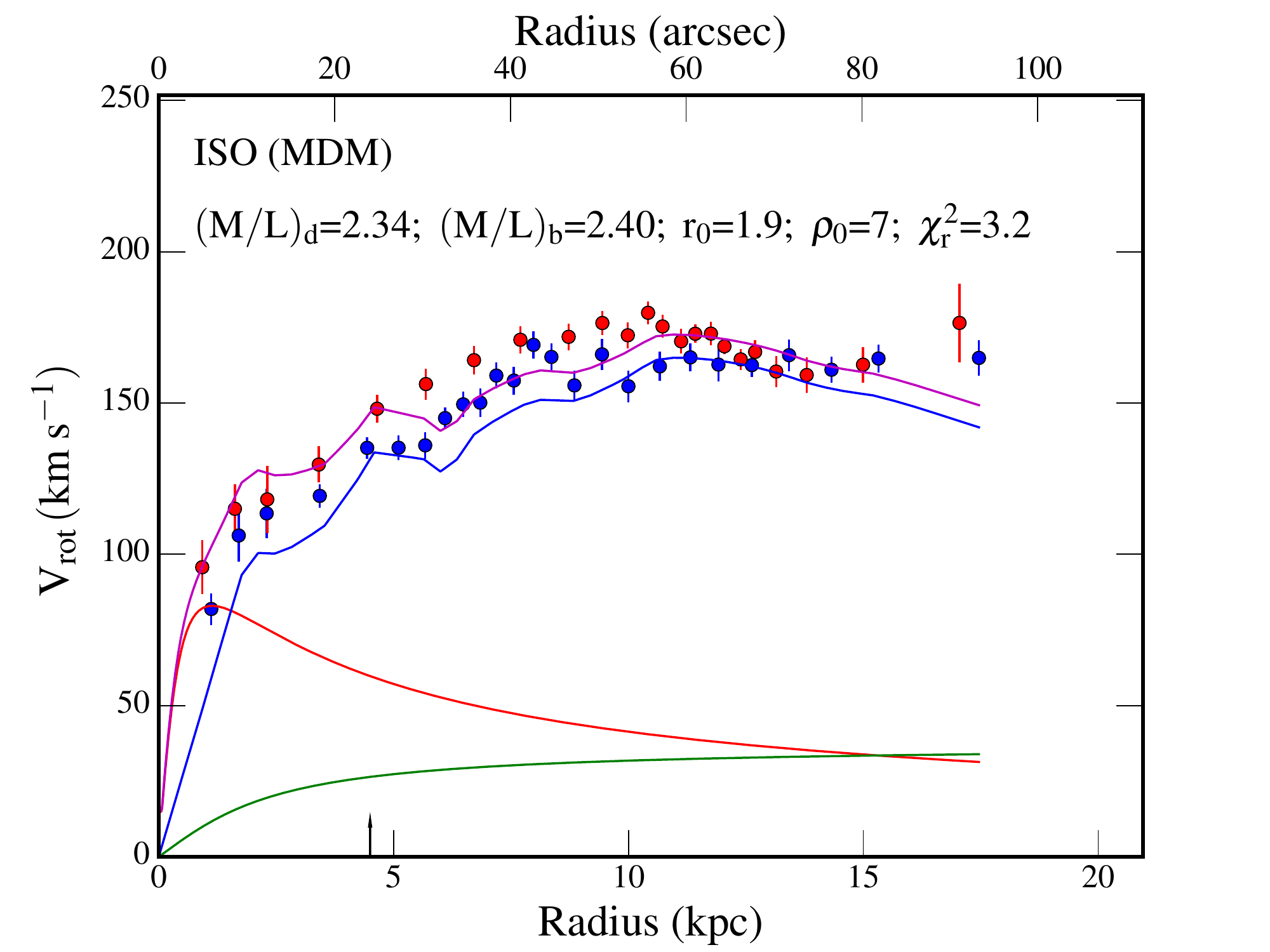}
\hspace*{-0.75cm} \includegraphics[width=0.35\textwidth]{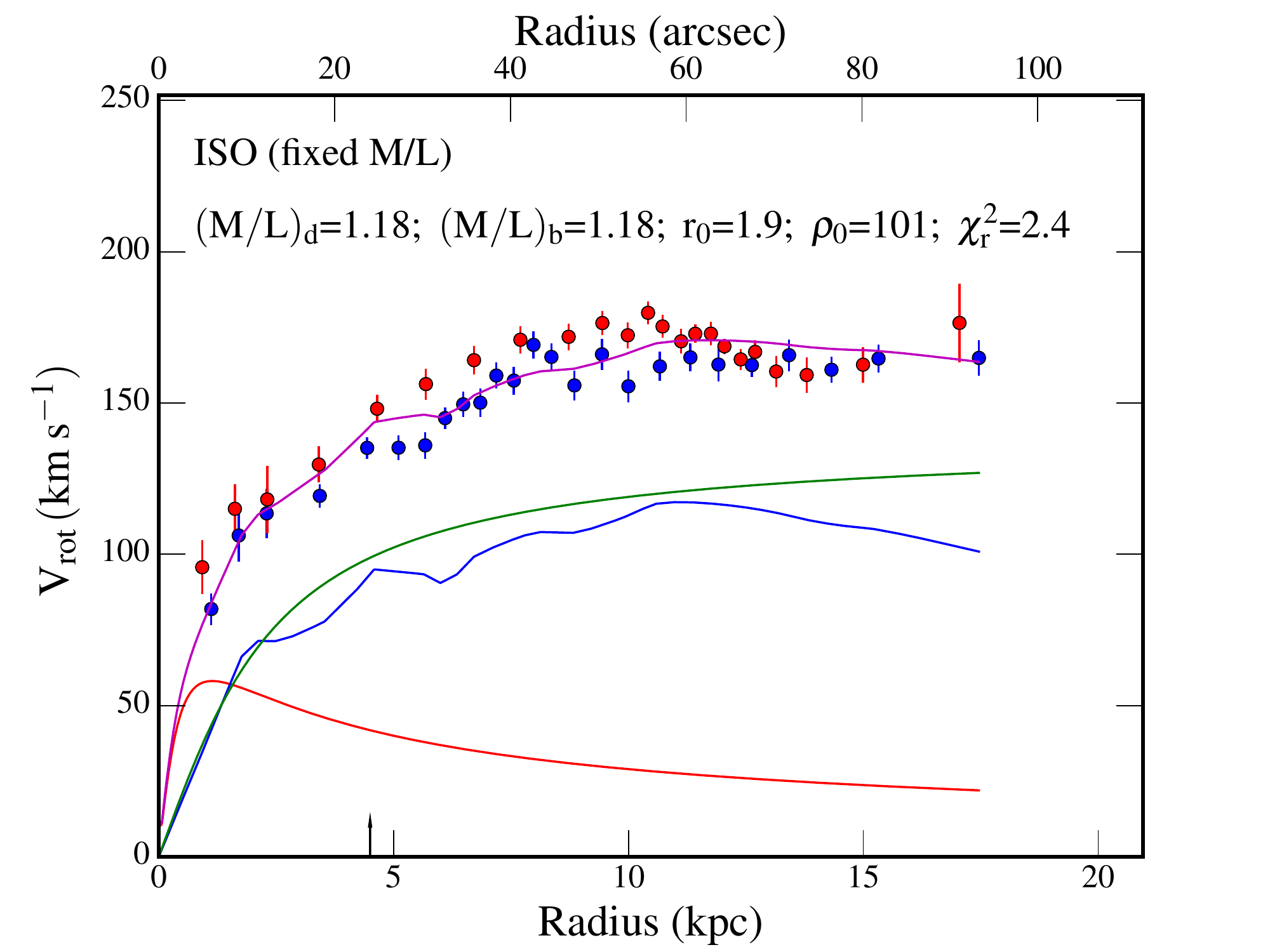}\\
\hspace*{-0.00cm} \includegraphics[width=0.35\textwidth]{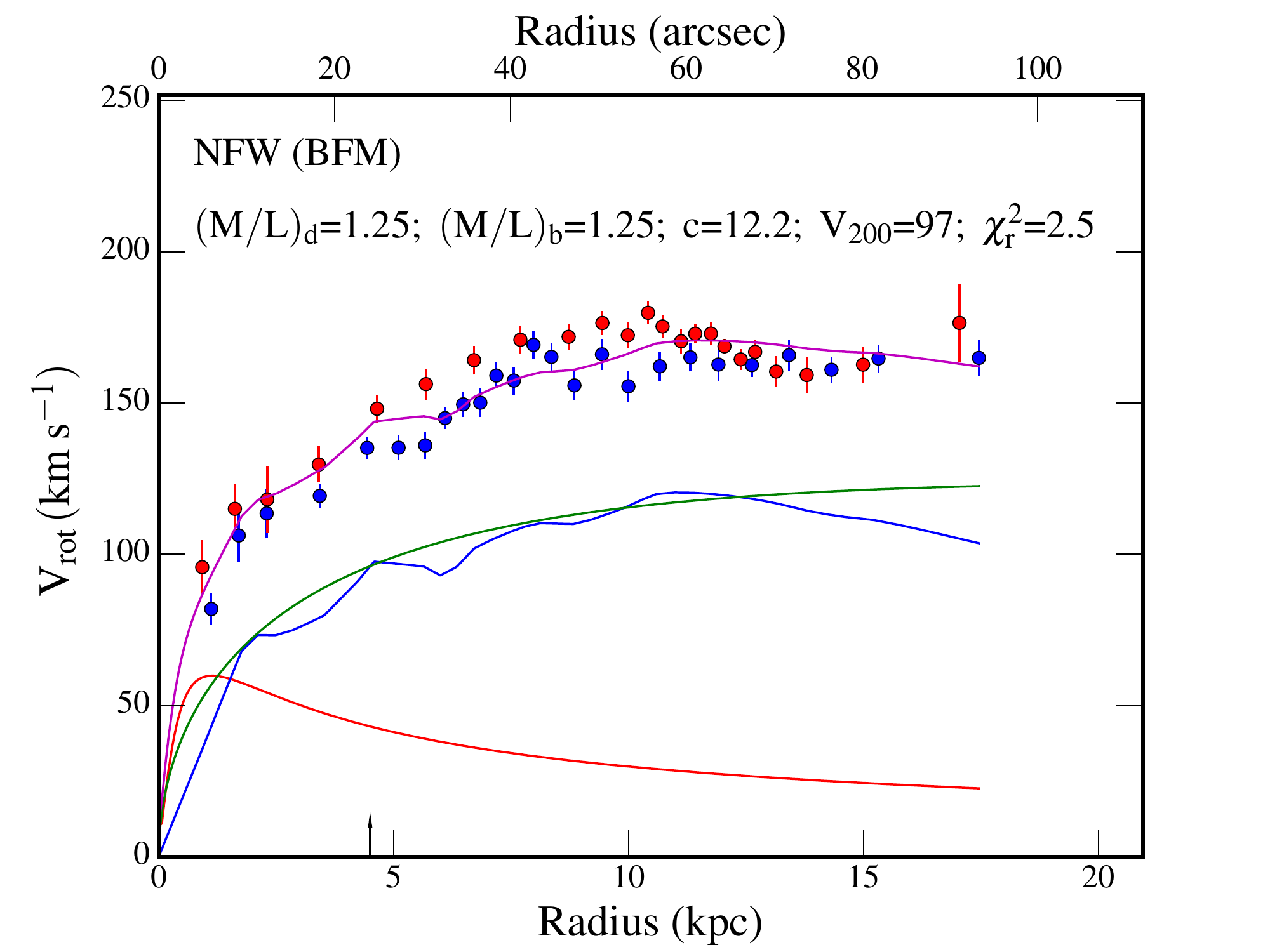}
\hspace*{-0.25cm} \vspace{-1.25cm} \includegraphics[width=0.31\textwidth]{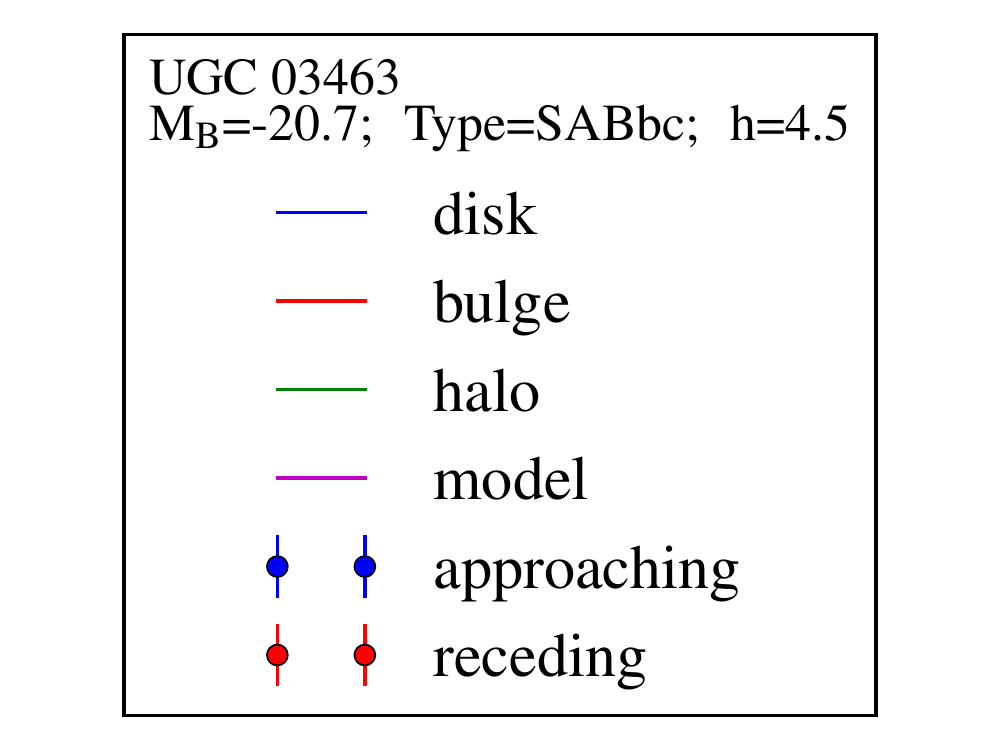} \vspace{1.25cm} \hspace*{-0.5cm}
\hspace*{-0.00cm} \includegraphics[width=0.35\textwidth]{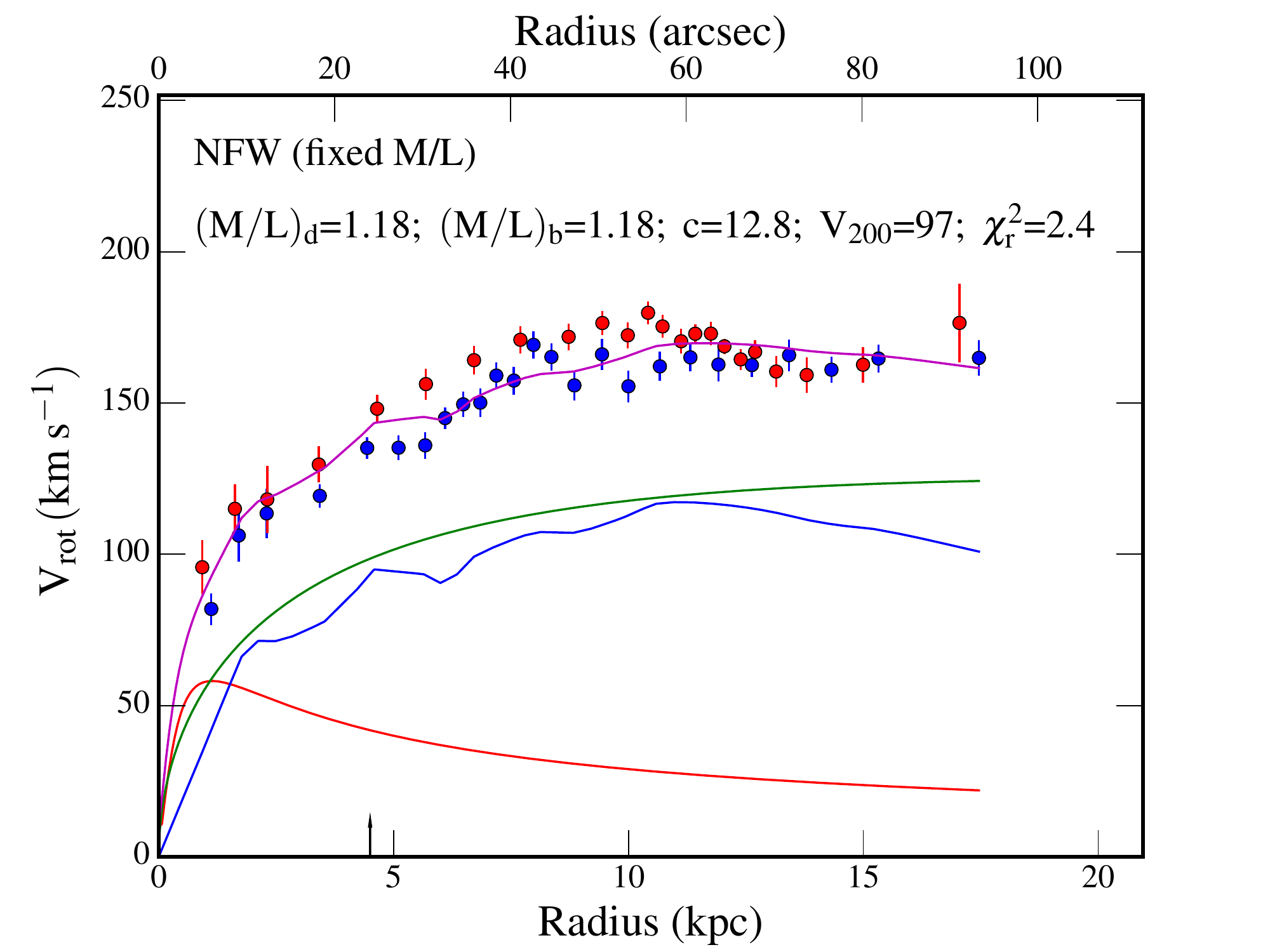}\\
\caption{Example of mass models for the galaxy UGC 3463.  First line: pseudo-isothermal sphere density profiles (ISO). Second line: Navarro, Frenk \& White density profiles (NFW). First column: Best Fit Model (BFM). Second column: Maximum disc Model (MDM) for line 1 (ISO model). Third column: Mass-to-Light ratio M/L fixed using the optical $\rm (B - V)$ color. The name of the galaxy, its B-band absolute magnitude, morphological type and disc scale length have been indicated in the insert located line 2-column 2. For each model, the fitted parameters and the reduced $\chi^2$ have been indicated in each sub-panel.}
\label{fig8:massmodel1}
\end{figure*}

For the different models, we use the same fitting procedures used and discussed in \citet{Korsaga+2018}. 
For the ISO fit, we used three different techniques. The first one is the best fit model (BFM), which lets all the parameters free to be fitted finding the best values corresponding to the minimal $\chi^2$. This leads to three free parameters for galaxies with no bulge (r$_0$, $\rho_0$ and M/L$_{disc}$) or four free parameters for galaxies with a bulge (r$_0$, $\rho_0$, M/L$_{disc}$, M/L$_{bulge}$). Note that M/L$_{bulge}$ is constrained to be equal or larger than M/L$_{disc}$ to account for the older stellar populations in the bulge than in the disc (see section \ref{sub:M/L}). The second technique is the maximum disc fit (MDM) that minimises the halo contribution by maximising the stellar disc (and bulge when present) contribution. We used also three or four free parameters (r$_0$, $\rho_0$, M/L$_{disc}$, M/L$_{bulge}$) depending if the galaxy is has a bulge or not. For the MDM, we allow the $\chi^2$ to increase up to 1.3 times the $\chi^2$ derived from the BFM and the M/L$_{disc}$ is constrained to be higher than the one used for the BFM. The last technique considers a value of M/L (fixed M/L) derived from the $\rm (B-V)$ color as described in section \ref{sub:M/L}. For this technique, we used the same value of M/L for the disc and the bulge due to the fact that the spectrophotometric models do not allow to disentangle them. We are thus left with only the two free parameters of the DM halo (r$_0$, $\rho_0$). To avoid non physical values, a minimal limiting value of r$_0$ = 0.5 kpc was imposed.
For the NFW model, we use two techniques: the BFM and the fixed M/L. The fitting procedures are the same as explained for the ISO model. For all models, we imposed a minimal M/L value of the disc and bulge at 0.1 M$_{\odot}$/L$_{\odot}$. In Fig. \ref{fig8:massmodel1}, we show the different mass models of the galaxy UGC 3463. In this special case, all models give equivalent good fits.

\subsection{Results}
\label{sect:results}

\begin{figure}
             \includegraphics[width=8.5cm]{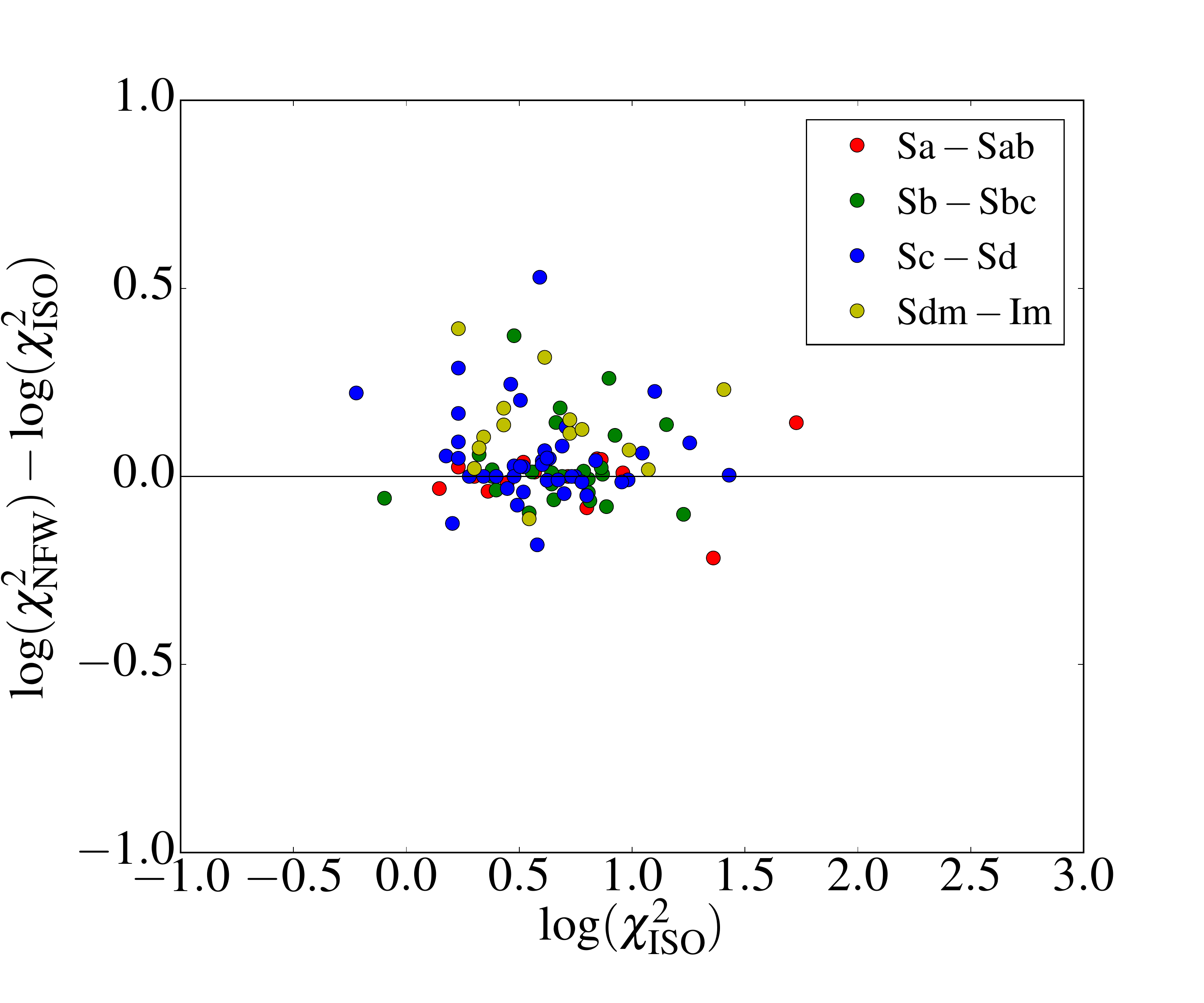}
            \includegraphics[width=8.5cm]{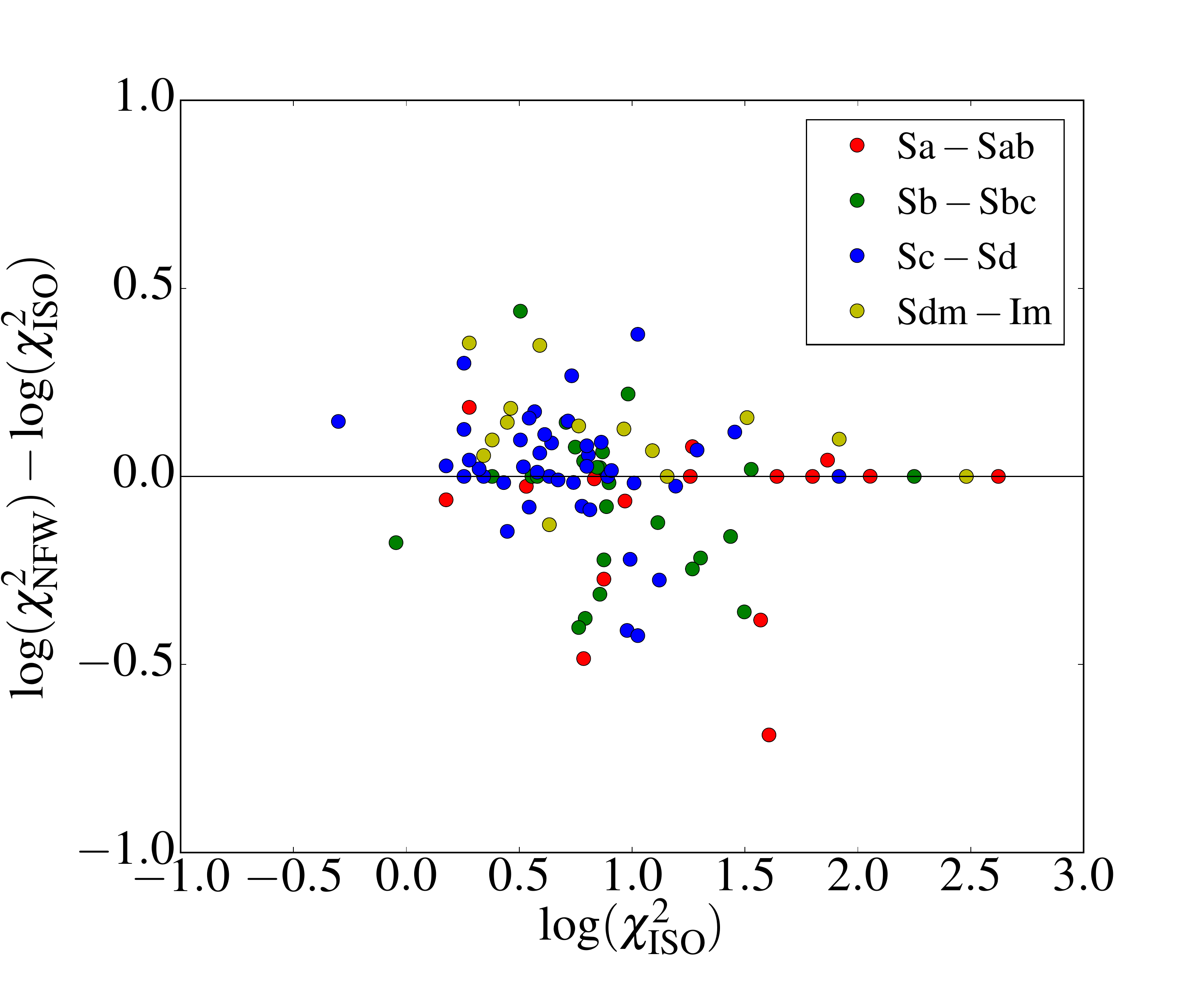}
\caption{Comparison between the reduced $\chi^2$ of NFW and ISO for the best fit model (BFM) (top panel) and for the fixed M/L (bottom panel).}
\label{fig9:nfwiso}
\end{figure}

Most of our observed rotation curves are better described by a central core density profile (ISO) than a cuspy one (NFW). This is clearly represented in Fig. \ref{fig9:nfwiso} for which the smallest $\chi^2$ values correspond to the ISO fits (70 galaxies have higher $\chi^2$ values for NFW compared to ISO for BFM and 66 galaxies for Fixed M/L). The detail of the parameters is shown in Tables \ref{tab:iso} and \ref{tab:nfw}. As mentioned in other studies, the cuspy central density does not describe well most of the faint galaxies \citep{Blok+2003, Gentile+2004}. 

\begin{figure}
	\vspace*{-0.0cm}\includegraphics[width=7.0cm]{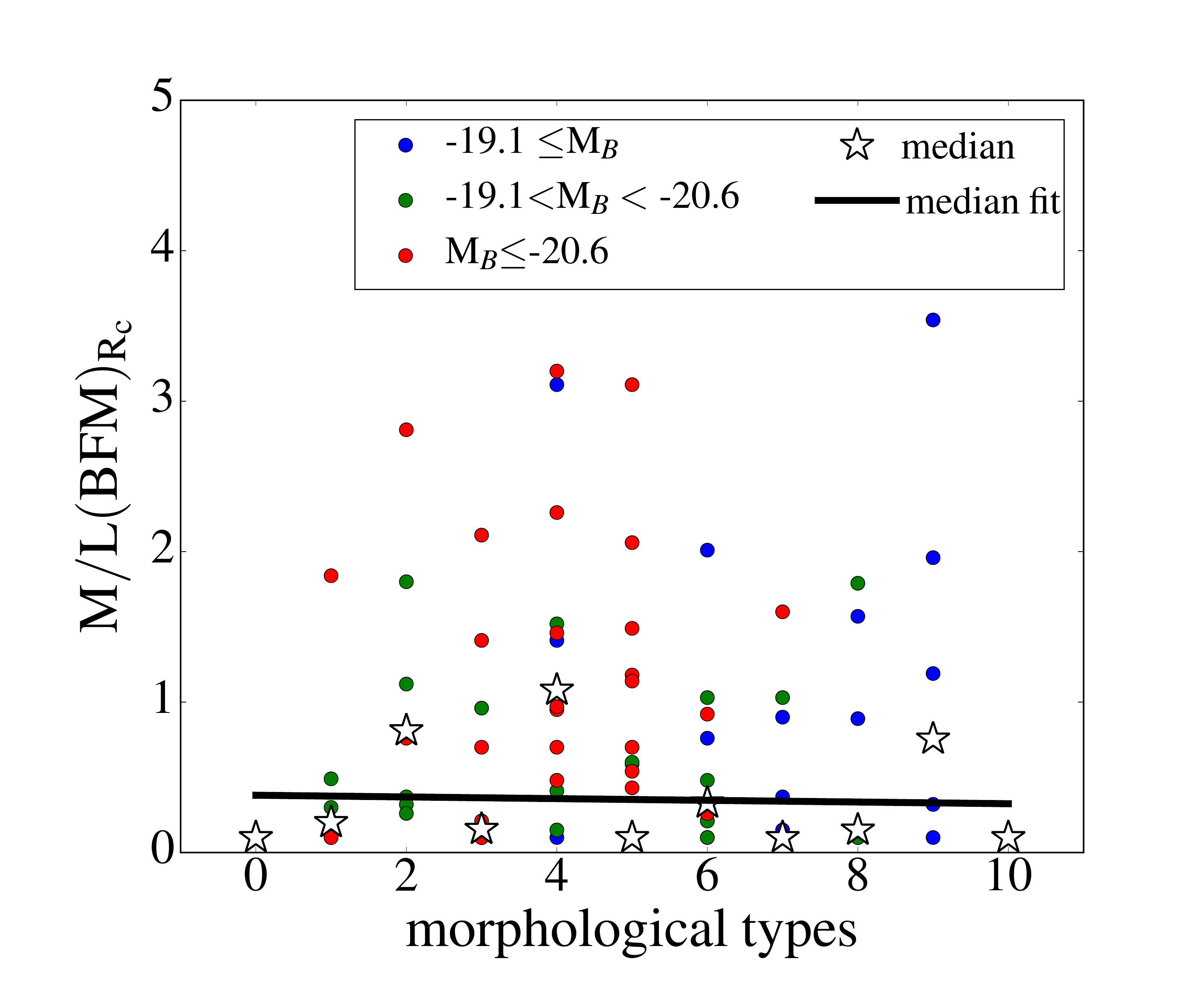}\vspace{-0.65cm}
	\vspace*{-0.0cm}\includegraphics[width=7.0cm]{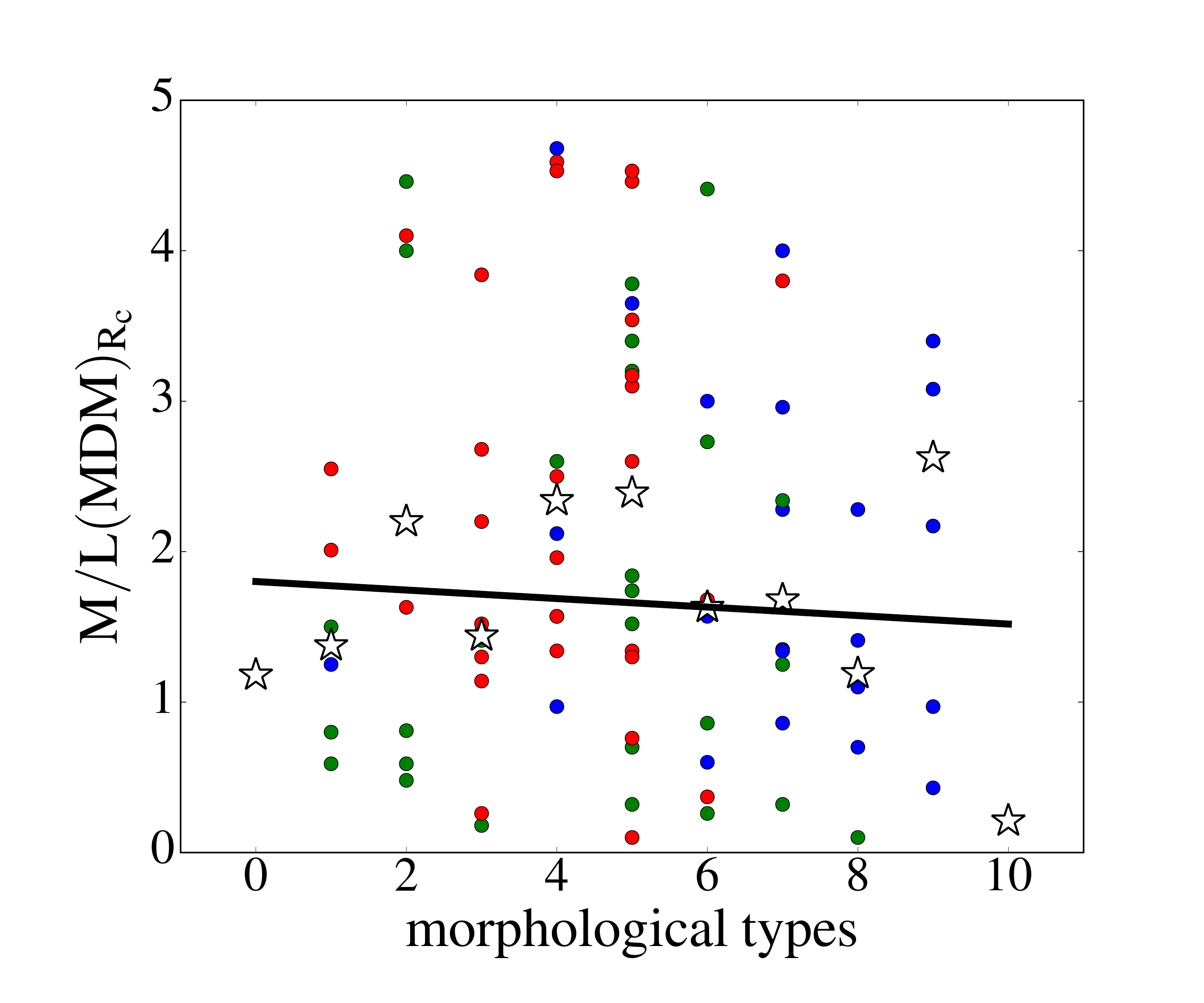}\vspace{-0.65cm}
	\vspace*{-0.0cm}\includegraphics[width=7.0cm]{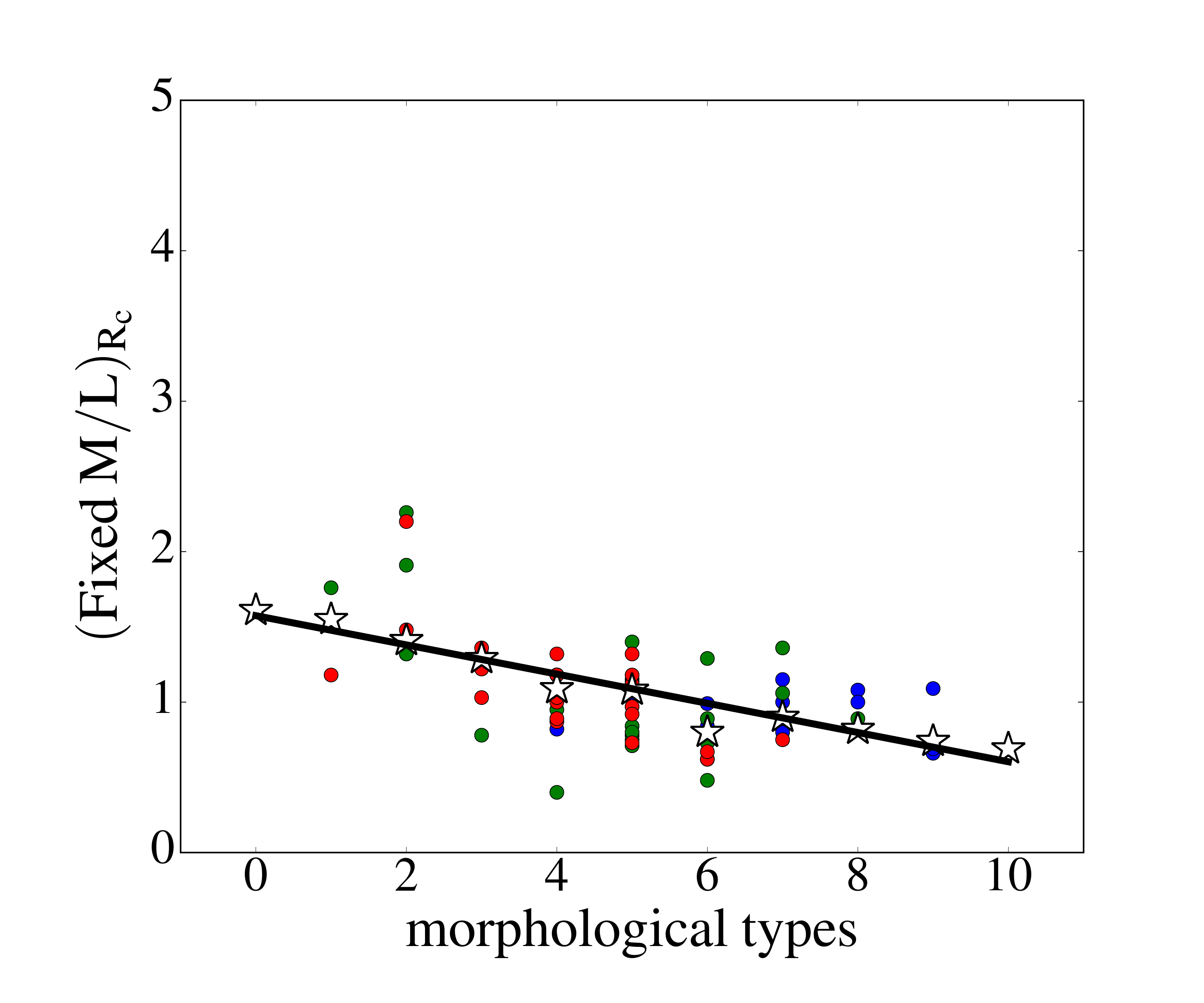}
		\caption{Top to bottom panels: respectively the M/L of ISO (BFM), the M/L of ISO (MDM) and the fixed M/L versus morphological types using the R$_c$-band. These plots represent the disc M/L values. The open black stars represent the median in morphological types. The thick black line is the fit of the median data. The legends for the three panels are shown in the top panel.}
\label{fig10:mltrcw1}
\end{figure}

\begin{figure*}
            \includegraphics[width=5.8cm]{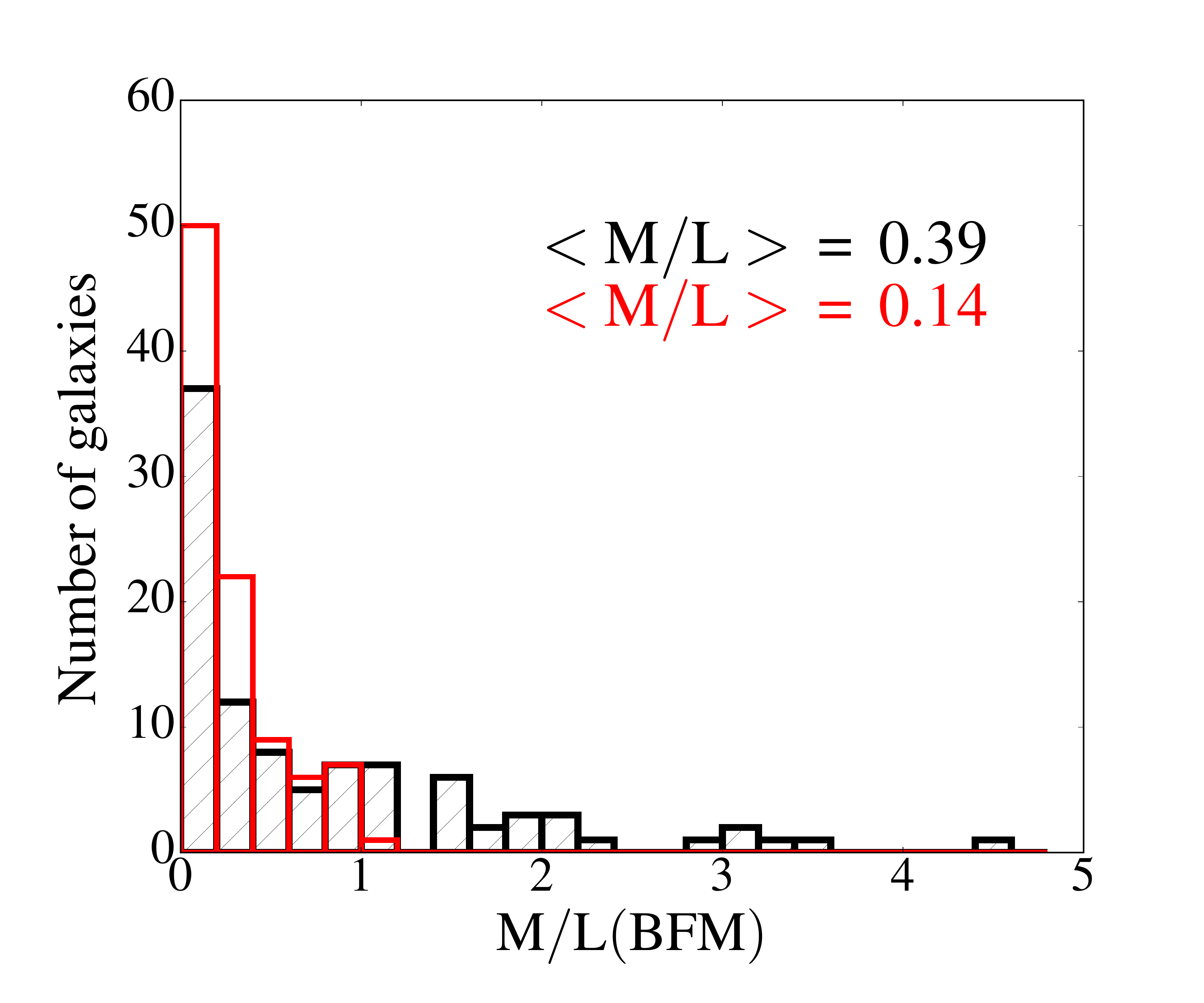}
            \includegraphics[width=5.8cm]{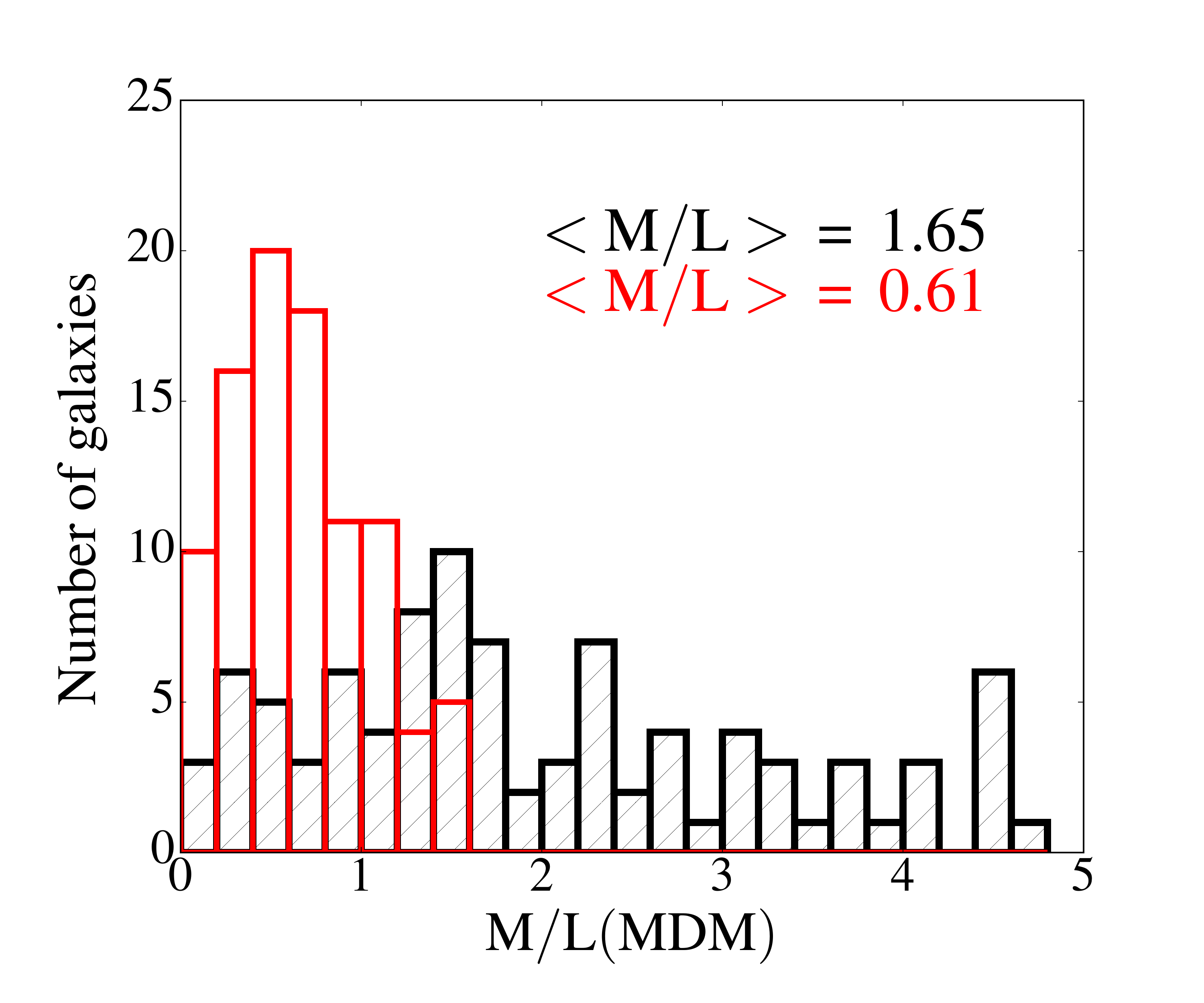}
            \includegraphics[width=5.8cm]{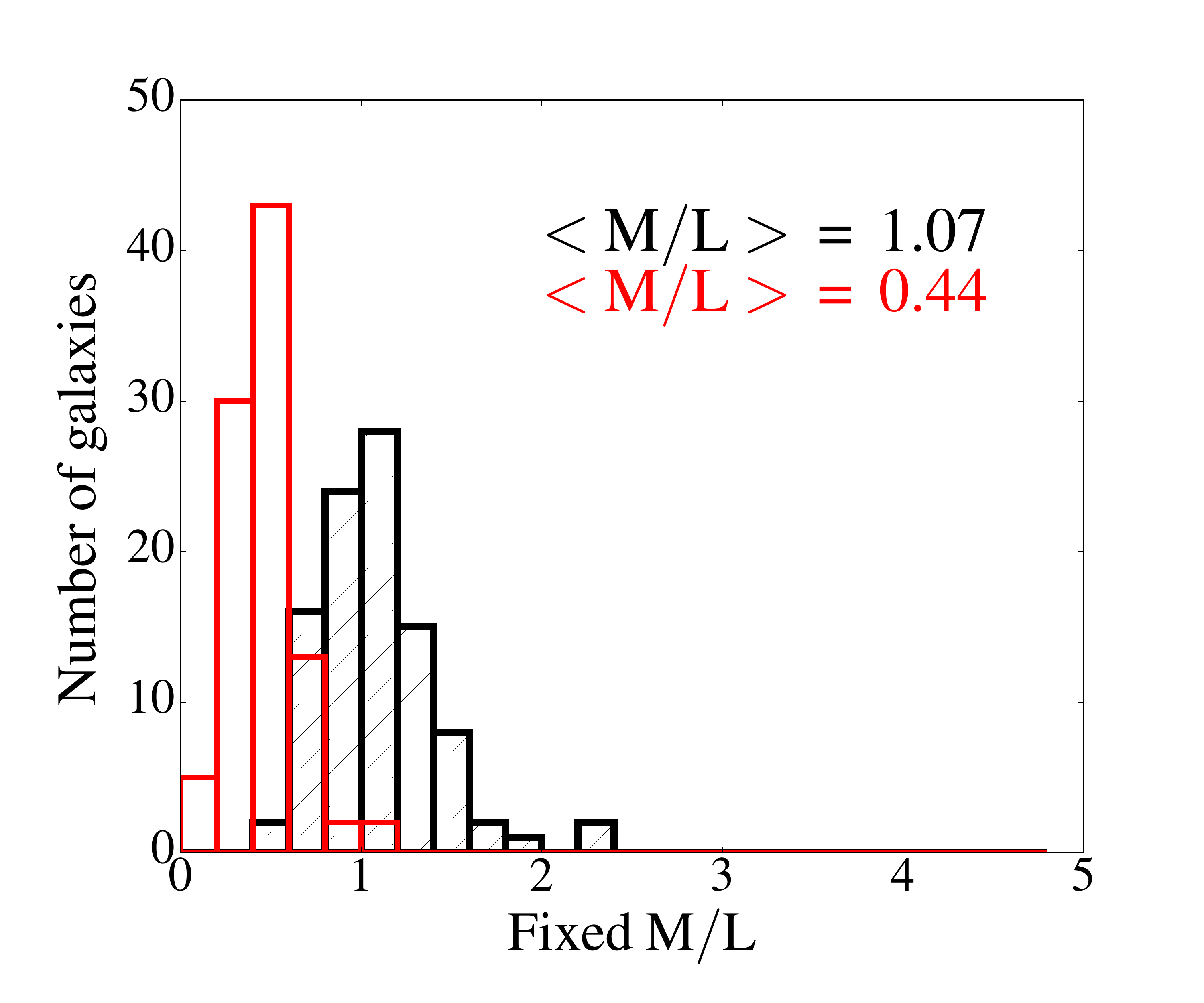}
\caption{Mass-to-light ratio distribution for the isothermal sphere model (ISO) using the R$_c$-band in black and the W$_1$-band in red. The median values are noted in each panel using the same color code. From left to right: best fit model (BFM), maximum disc model (MDM) and value calculated using the optical $\rm (B - V)$ and  MIR $\rm (W_1 - W_2)$ colors respectively. }
\label{fig11:M/L}
\end{figure*}

In Fig. \ref{fig10:mltrcw1}, we show the variation of (M/L) as a function of morphological types for the three ISO models: BFM, MDM and fixed M/L. These plots represent the disc M/L values. As expected, M/L is getting smaller when going from early to late types. 
\citep{faber+1979}. The smallest dispersion is obtained for the fixed (by colour) M/L models. 
The distributions of (M/L)s for the different ISO models are shown in Fig. \ref{fig11:M/L} (black). We find the expected results where the smaller M/Ls are found for the BFM, the larger for the MDM and intermediate values for the fixed M/L.
The M/L median value of the ISO (BFM) is 0.39 M$_{\odot}$/L$_{\odot}$; this value is $\sim 3$ time larger than the one computed for NFW (BFM) which is equal to 0.15 M$_{\odot}$/L$_{\odot}$ (50\% hit the 0,1 M$_{\odot}$/L$_{\odot}$ lower boundary). These low M/L values for NFW profiles are expected \citep{Navarro+1997}; this is due to the intrinsically steeper mass distribution of the cuspy profile which replaces the contribution from the stellar disc in the inner parts \citep[e.g.][]{Blok+2008}.  

Scaling laws between the DM halos parameters (r$_0$ and $\rho_0$) and between the DM halos parameters and the absolute B-magnitude of galaxies are found in \citet{Kormendy+2004} and \citet{Toky+2014}. More precisely, \citet{Kormendy+2004} studied the properties of DM halos in late type and dwarf irregular galaxies and found that high $\rho_0$ tend to have smaller r$_0$, and also less luminous galaxies tend to have smaller r$_0$ and higher $\rho_0$. Similar conclusions had been found by \citet{Toky+2014} who also worked on a sample mostly composed of late type galaxies. In order to check if these scaling laws are still the same whatever the morphological types of galaxies, we plot the DM halos parameters using our sample which covers early type and late type spiral and irregular galaxies. 

\begin{figure}
\begin{center}
\includegraphics[width=9cm]{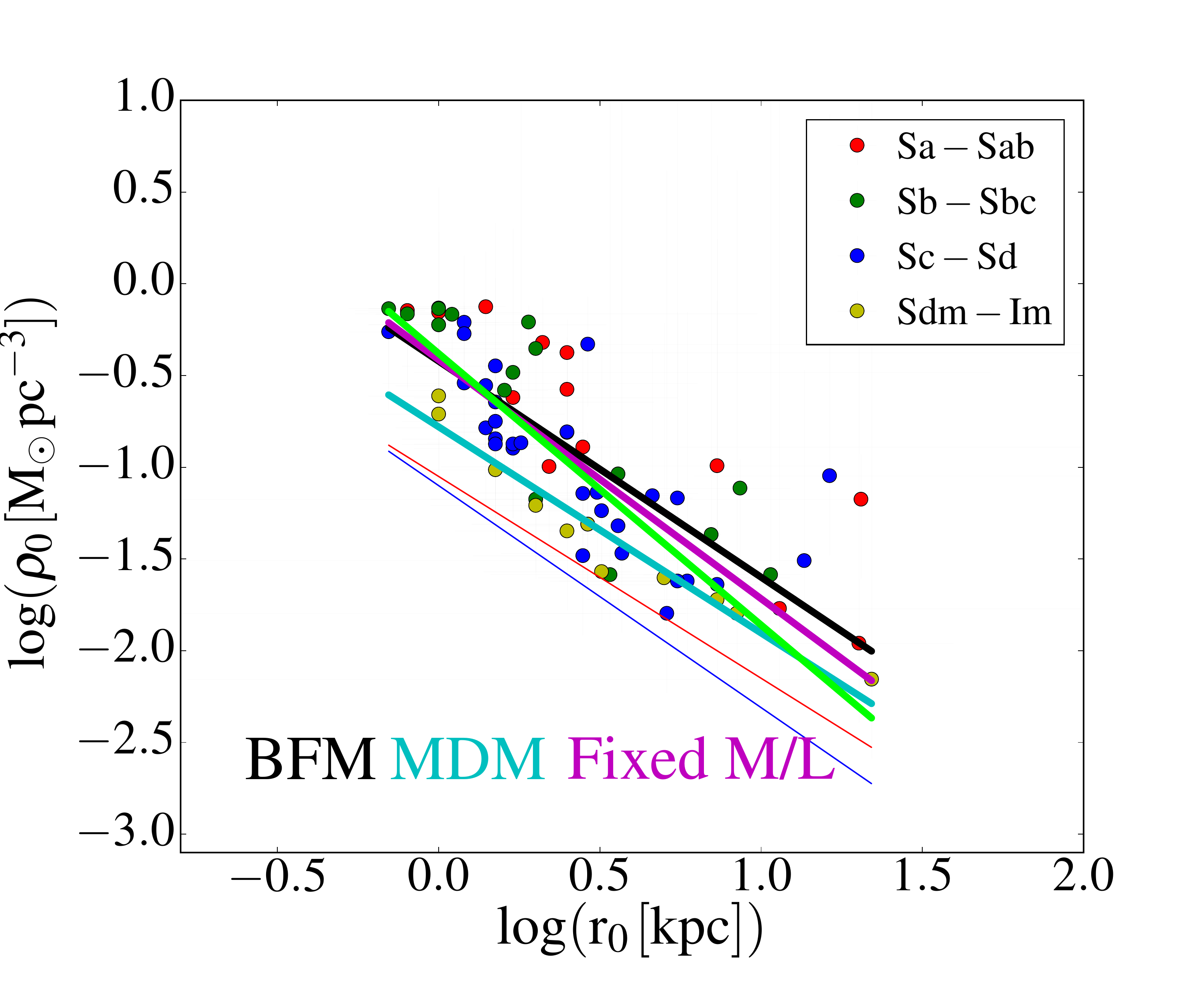}
\caption{Central halo density versus halo core radius for the ISO best fit models (BFM) for the whole sample. The thick black, cyan and magenta lines represent the fit for BFM, MDM and fixed M/L models respectively. The thick lime, thin blue and thin red lines represent respectively the fit of the BFM found using the W1-band, \citet{Kormendy+2004} and \citet{Toky+2014}.}
\label{fig12:all}
\end{center}
\end{figure}

In Fig. \ref{fig12:all}, we plot the relation between the central halo density ($\rho_0$) and the scaling radius (r$_0$) for ISO, which shows a clear anti-correlation meaning that high $\rho_0$ corresponds to smaller r$_0$ for the three techniques (BFM, MDM and fixed M/L). We choose to make the plots using the points of the BFM and overlay the results for the 3 techniques (BFM, MDM and Fixed M/L) to avoid over-crowding. The thick black line is for the BFM, the cyan is for the MDM and the magenta for the fixed M/L. To do the fit of the ISO DM parameters, we excluded galaxies with r$_0$ equal or smaller to 0.5 which corresponds to the minimal limiting value (see Table \ref{tab1:nohalo}). We also excluded galaxies for which a DM halo is not needed to describe the observed rotation curves. In this case, a MDM model describes better the observed rotation curves with only the baryonic matter, without the need of DM. These galaxies (26 galaxies in total) are marked with an asterisk in Table \ref{tab:iso}.
The general relation between $\rho_0$ and r$_0$ is:
\begin{equation}
\begin{array}{l}     
\rm \log\ \rho_0 = (a \pm \delta a)\,   \log\ r_0  + (b \pm \delta b)
\label{eq:iso}
\end{array} 
\end{equation}

where the parameters a, $\delta$a, b and $\delta$b are shown in Table \ref{tab:DM}. 

\begin{figure}
	\vspace*{-0.0cm}\includegraphics[width=8.5cm]{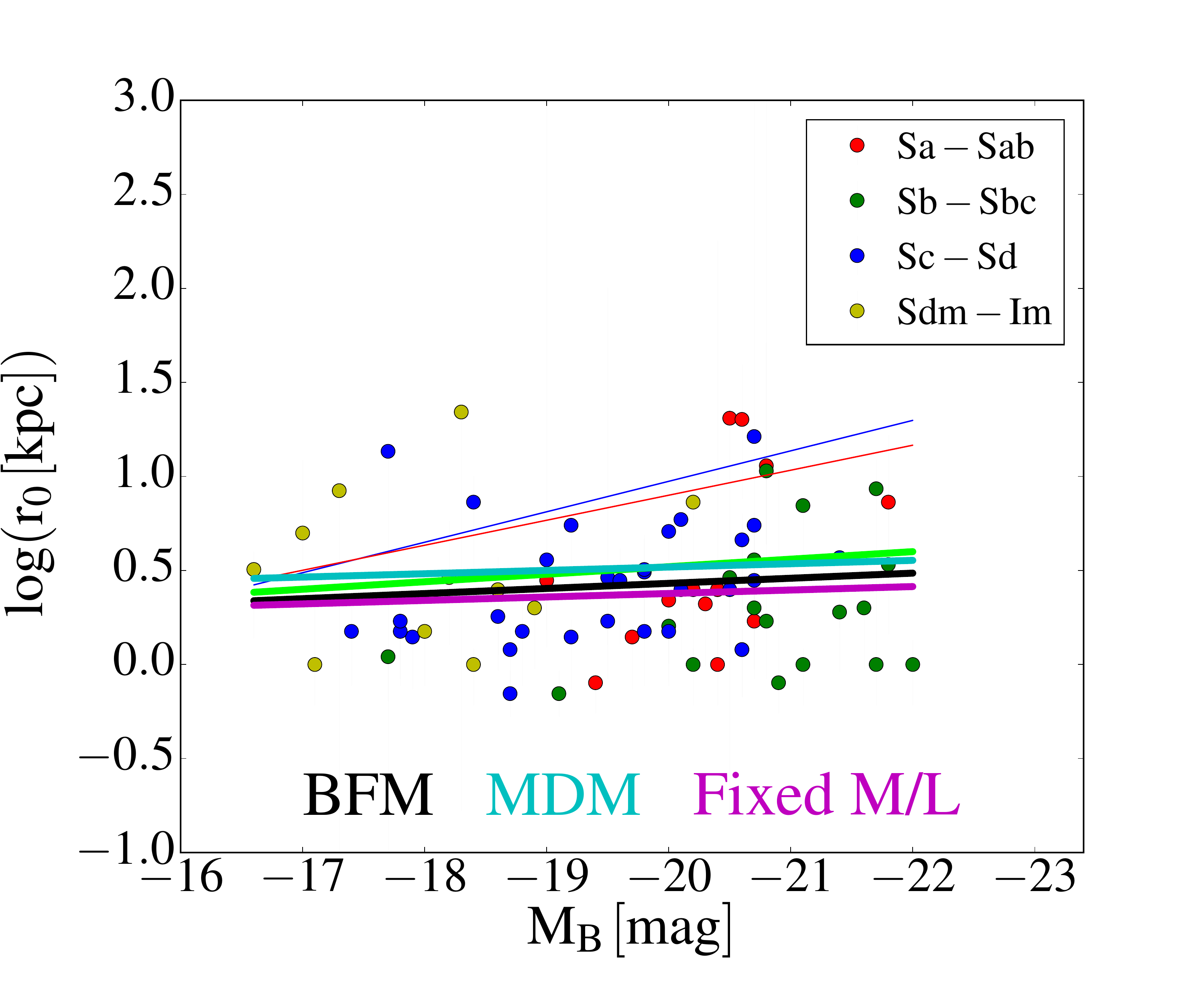}\vspace{-0.76cm}
	\vspace*{-0.0cm}\includegraphics[width=8.5cm]{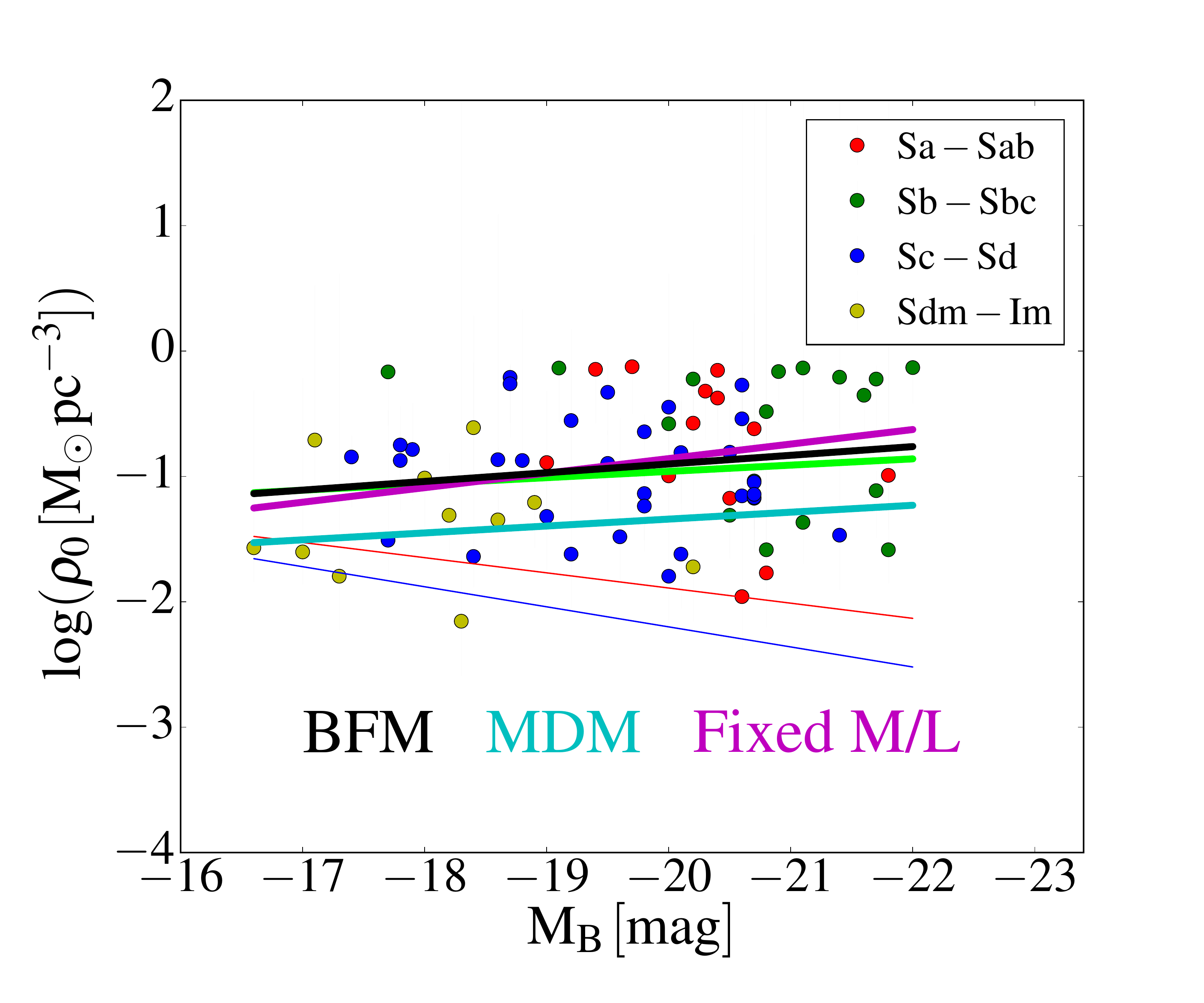}\vspace{-0.76cm}
	\vspace*{-0.0cm}\includegraphics[width=8.5cm]{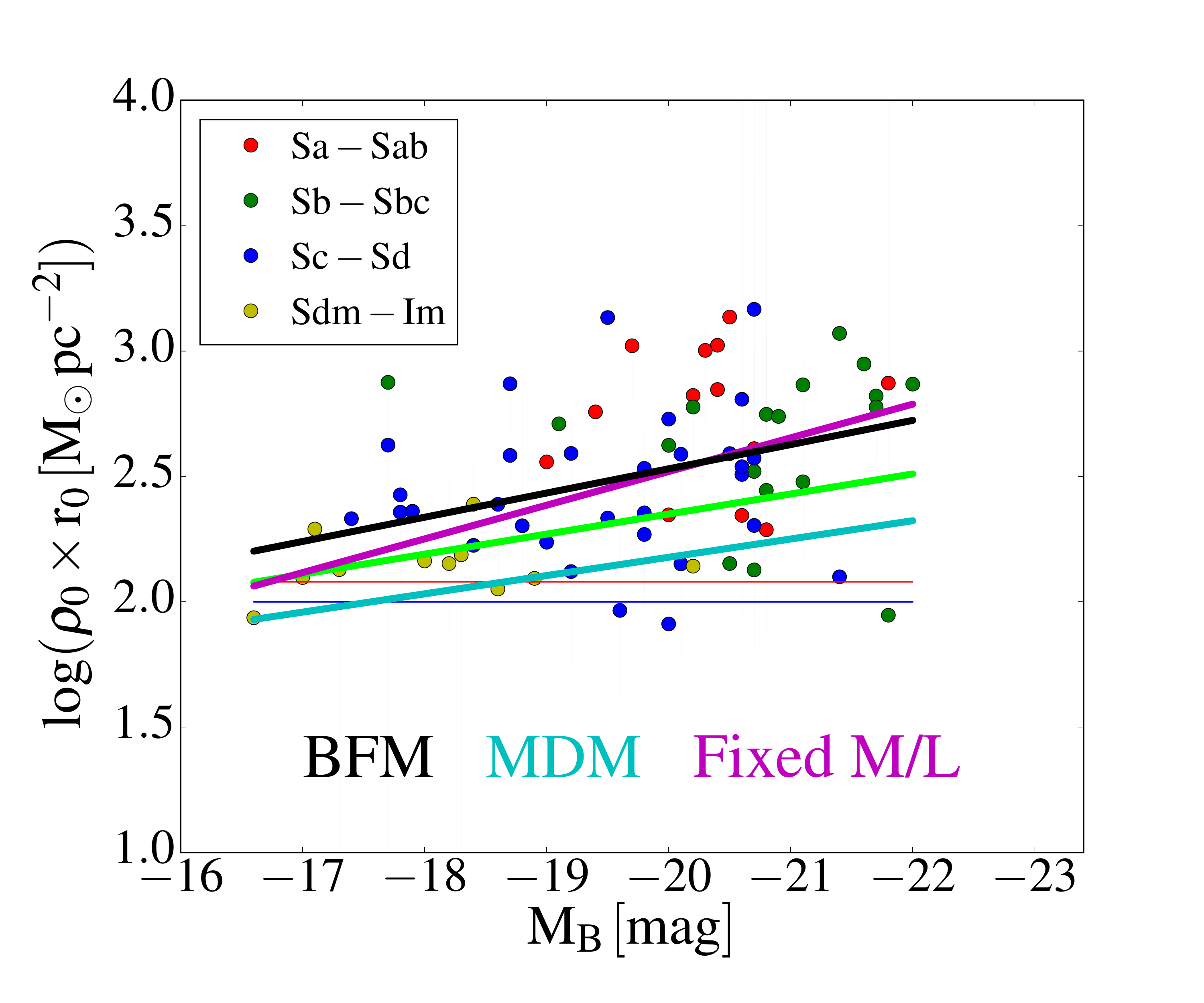}
\caption{Halo scaling radius (top panel), central halo density (middle panel) and  the product of the central halo density with the scaling radius versus the absolute B-band magnitude for the whole sample for ISO (BFM) points. The thick black, cyan and magenta lines are the best fit (BFM), the maximum disc (MDM) and the fixed M/L  models respectively. The thick lime, thin blue and thin red lines represent respectively the fit found using the W1-band, \citet{Kormendy+2004} and \citet{Toky+2014}.}
\label{fig13:allmb}
\end{figure}

Fig. \ref{fig13:allmb} presents the variation of the DM halo parameters r$_0$, $\rho_0$ and $\rho_0\, \times$ r$_0$, as a function of the galaxies absolute magnitude in the B-band. For our total sample, r$_0$ seems independent of the luminosity while brighter galaxies tend to have larger central densities and less luminous galaxies tend to have smaller $\rho_0\, \times$ r$_0$. This seems at odd with previous results by \citet{Kormendy+2004} and \citet{Toky+2014}, which predicted smaller DM scaling radius and higher central density for the weaker dwarf galaxies. The difference between our results and theirs will be explained in section \ref{sub:comp_previous}.

\begin{figure}
	\vspace*{-0.0cm}\includegraphics[width=8.5cm]{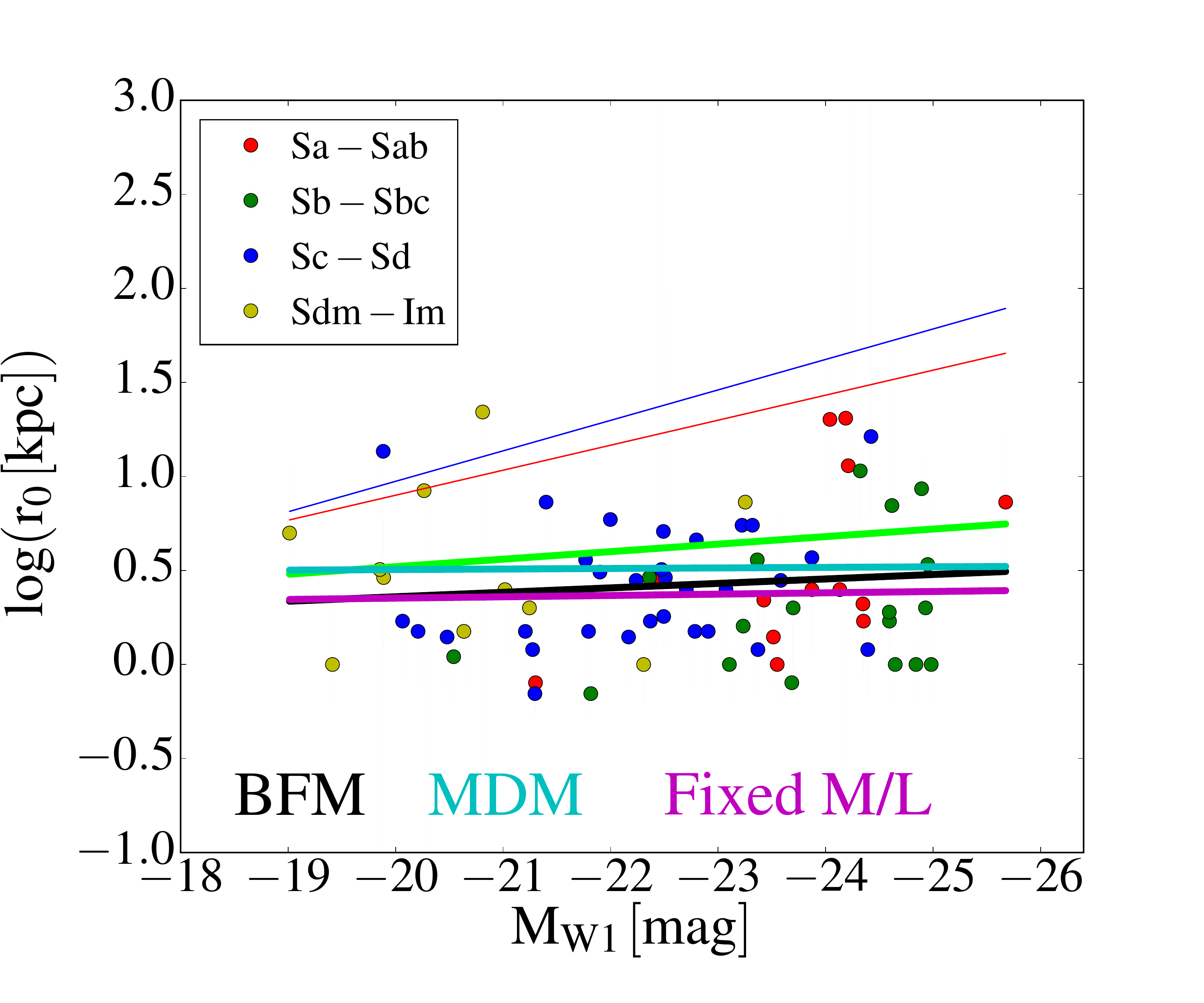}\vspace{-0.76cm}
	\vspace*{-0.0cm}\includegraphics[width=8.5cm]{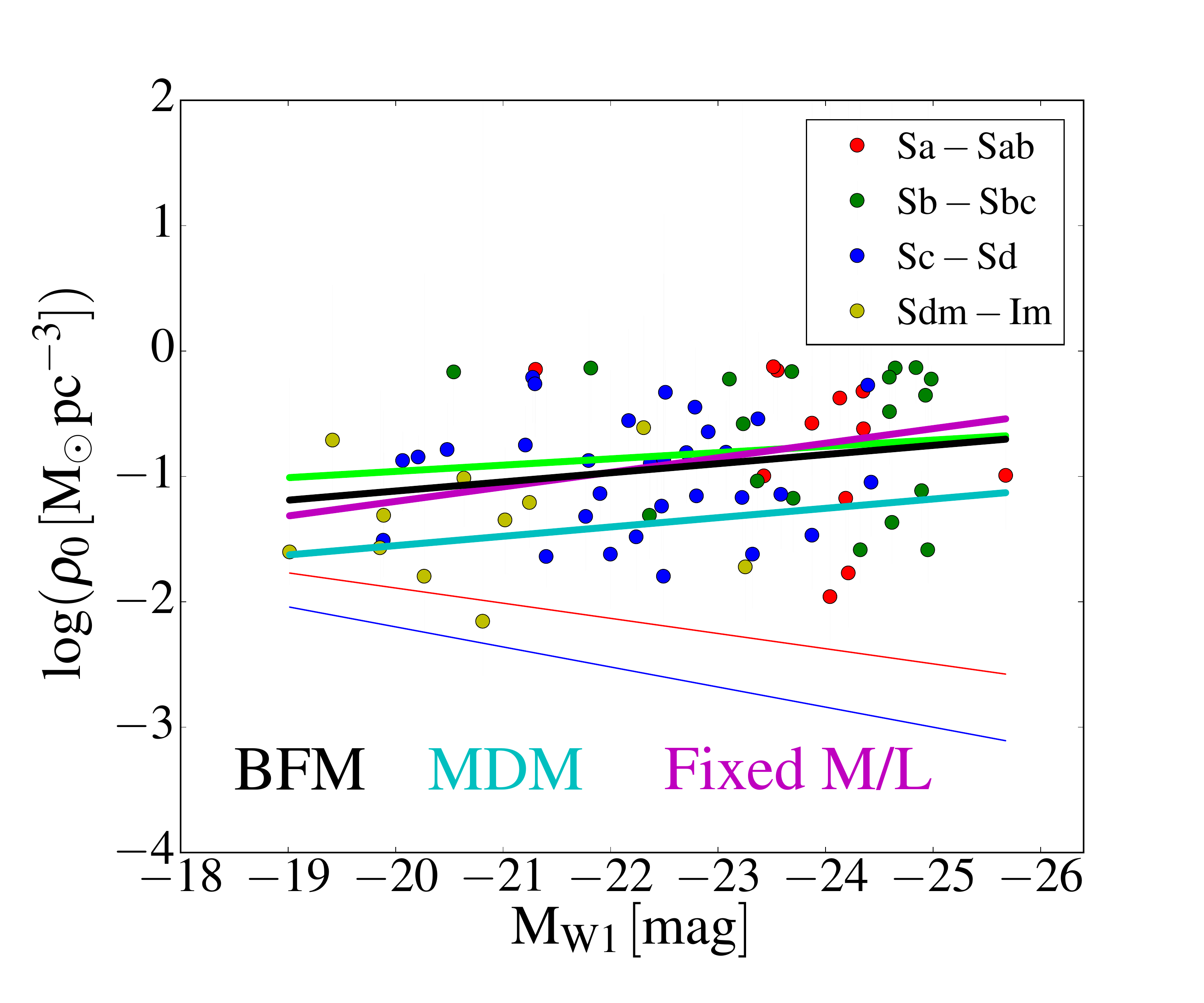}\vspace{-0.76cm}
	\vspace*{-0.0cm}\includegraphics[width=8.5cm]{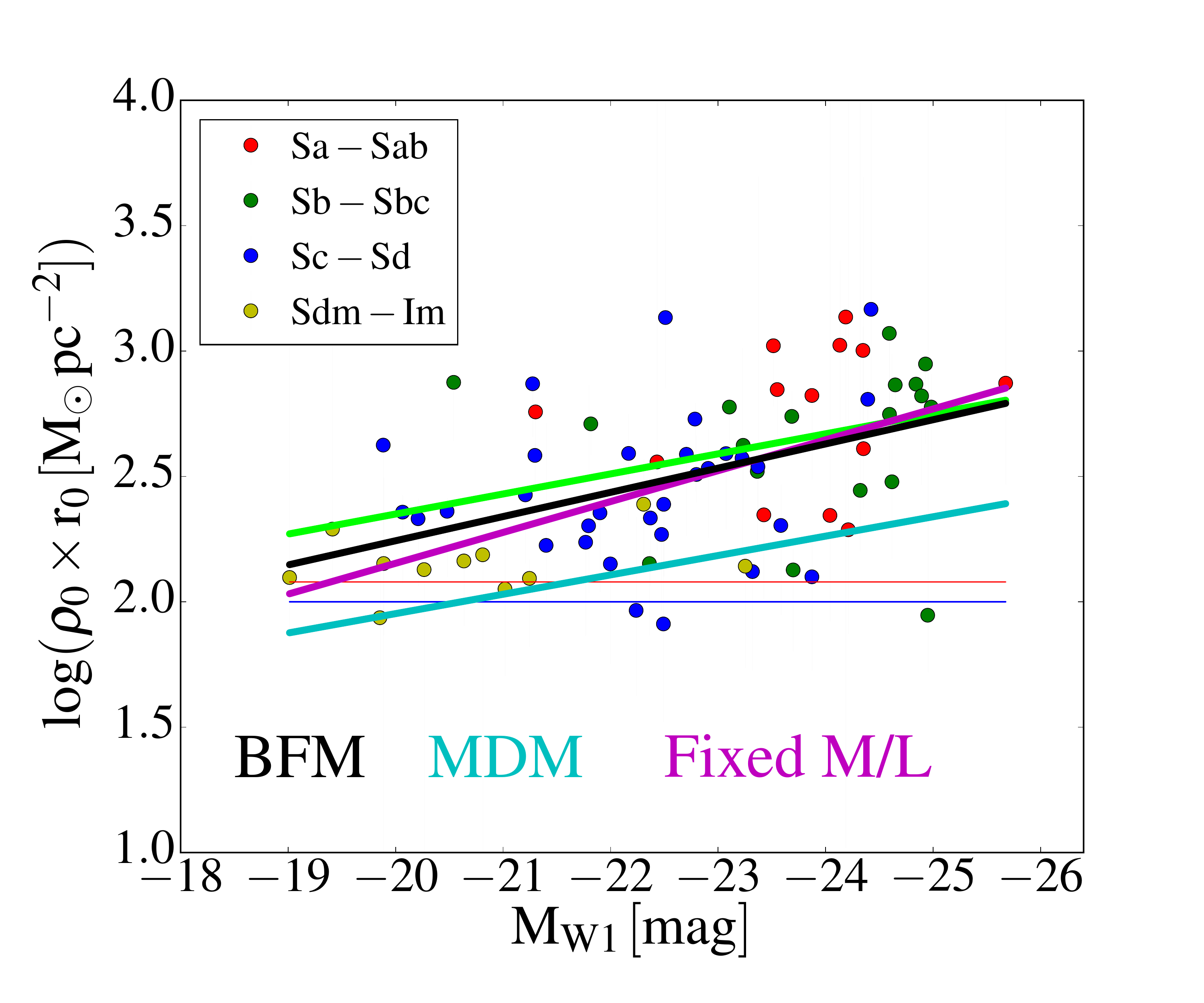}
\caption{Halo scaling radius (top panel), central halo density (middle panel) and  the product of the central halo density with the scaling radius versus the magnitude in W1-band for the whole sample for ISO (BFM) points. The thick black, cyan and magenta lines are the best fit (BFM), the maximum disc (MDM) and the fixed M/L  models respectively. The thick lime, thin blue and thin red lines represent respectively the fit found using the W1-band, \citet{Kormendy+2004} and \citet{Toky+2014}.}
\label{fig:DMw1}
\end{figure}

In order to study the impact of the band used in the correlations found with the DM halo parameters, we used the magnitudes in the W1-band (see Fig. \ref{fig:DMw1}). Interestingly, we found the same trends compared to those found using the absolute B-band magnitude, which means that the different trends with the luminosity do not depend on the band used. In this work, we have used the luminosity in the B-band, because we want to compare our results with the previous authors who used the B-band luminosity.  

\begin{figure}
	\begin{center}
            \vspace*{-0.0cm} \includegraphics[width=8.5cm]{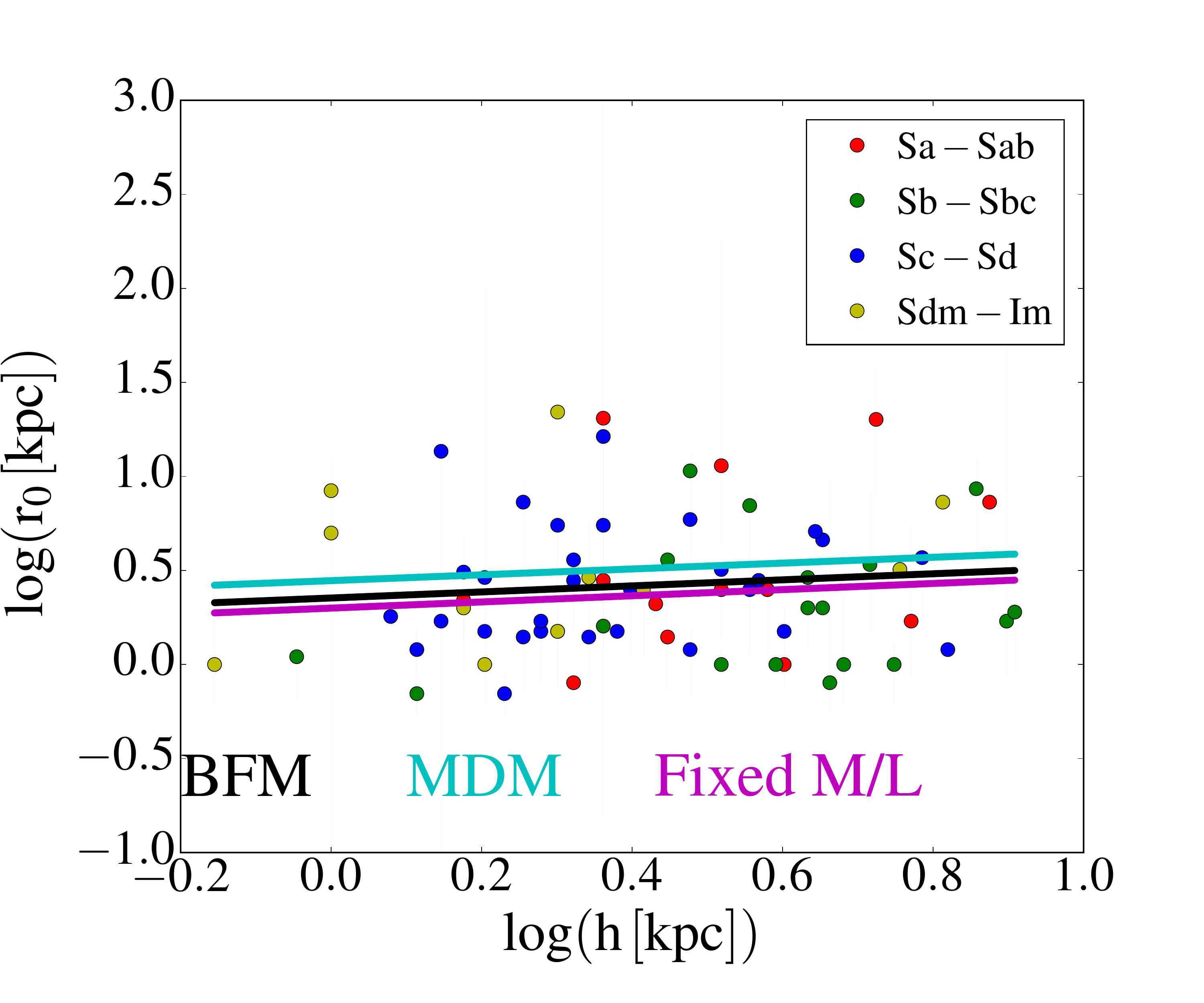}\vspace{-0.76cm}
           \vspace*{-0.0cm}  \includegraphics[width=8.5cm]{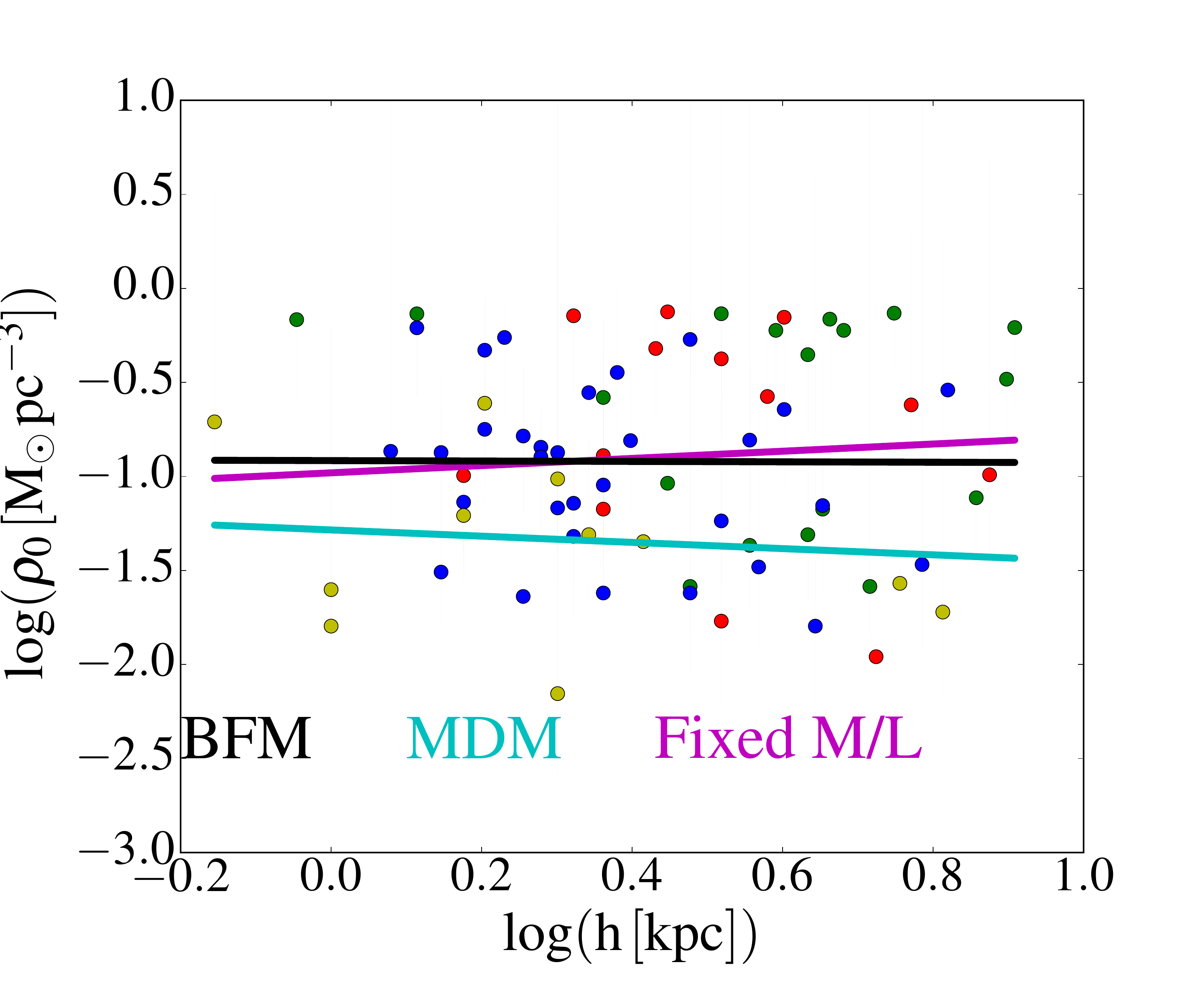}
\caption{Halo parameters as a function of the disc scale length for the whole sample for the ISO (BFM) points. Top and bottom panels show respectively the scale radius and the central halo density as a function of the disc scale length. The thick black, cyan and magenta lines are the best fit (BFM), the maximum disc (MDM) and the fixed M/L  models respectively. The legends for the two panels are shown in the top panel.
}
\label{fig14:ht}
\end{center}
\end{figure}

It is interesting to look also at the relation between the DM halo parameters r$_0$ and $\rho_0$ as a function of the optical size characterized by the scale length h. This is shown in Fig. \ref{fig14:ht}, where both parameters seem quite independent of h for all morphological types.

\begin{figure}
	\includegraphics[width=9cm]{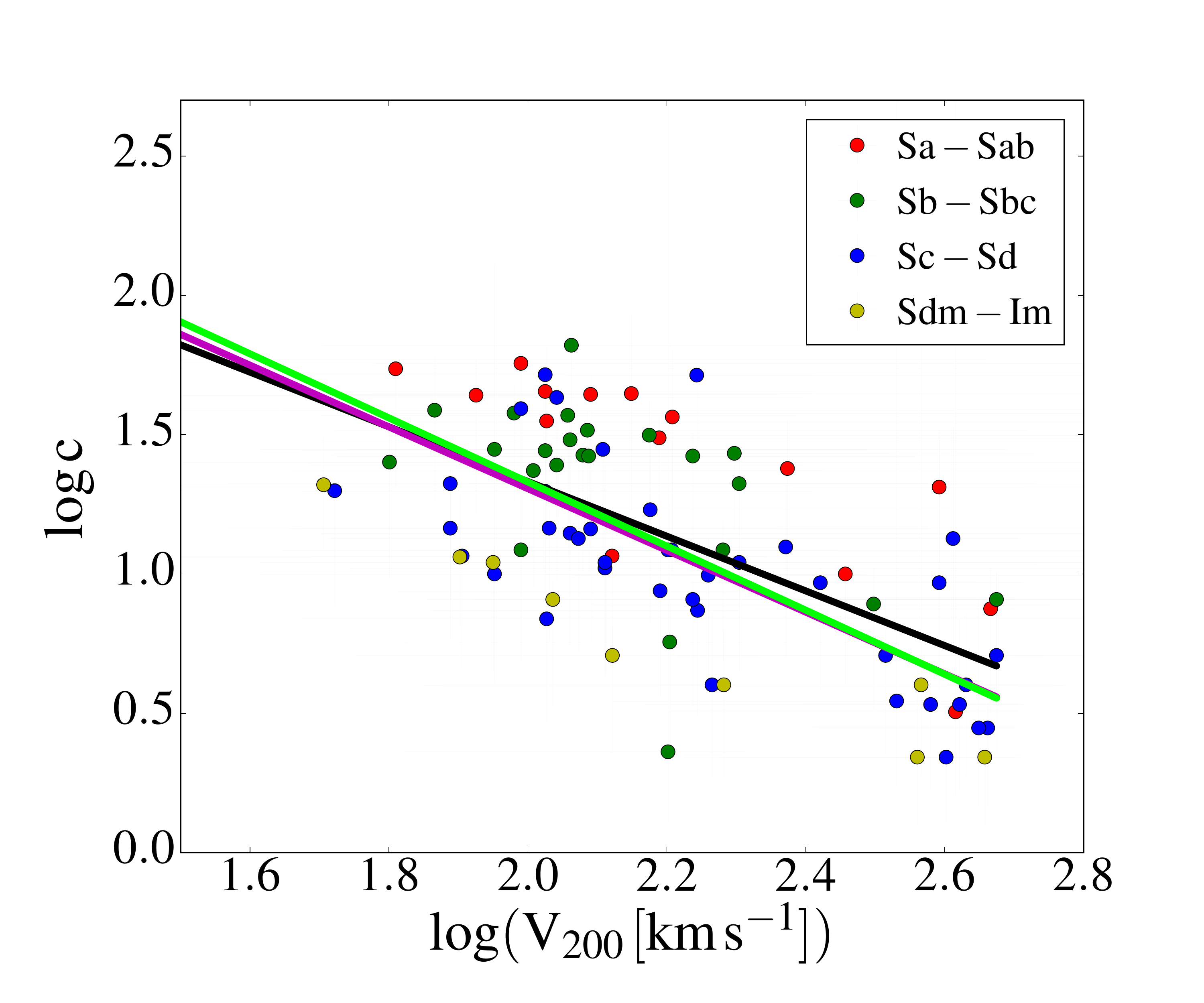}
\caption{Distribution of $\rm \log{c}$ as a function of $\rm \log{(V_{200})}$ for NFW best fit model (BFM). The thick black and magenta lines represent respectively the fit for BFM and fixed M/L. The thick lime line shows the fit for the BFM found using the W1-band.}
\label{fig15:nfw}
\end{figure}

In Fig. \ref{fig15:nfw}, we plot the halos parameters found by using the NFW (BFM) points, that is, the concentration $\rm c$ as a function of the velocity at the virial radius (i.e V$_{200}$). We fitted the following linear relation to the BFM and Fixed M/L:
\begin{equation}       
\rm \log\  c = (a \pm \delta a)\,   \log\ V_{200}  +(b \pm \delta b)
\end{equation}
The parameters are provided in Table \ref{tab:DM}.
These fits are represented in thick black and magenta lines respectively. To do the fit, we excluded galaxies (8 galaxies in total) with $\rm c \leq 1$ and V$_{200} \geq$ 500 \kms\ because these values are not physical in the CDM context \citep{Blok+2008}. The different parameters are shown in Table \ref{tab:nfw}. We find that low mass halos are more concentrated than high mass halos. The halos are slightly more concentrated when using the BFM than the fixed M/L. Indeed, the average concentration is c = 13.80 $\pm$ 2.29 and c = 12.02 $\pm$ 1.66 for the BFM and the fixed M/L respectively (see Table \ref{tab:nfwvalue}). We also notice that early type spirals tend to be located above the best fit line and late type galaxies below. A similar trend is found when using the fixed M/L. 

\begin{figure}
\begin{center}
		\vspace*{-0.0cm}\includegraphics[width=8.5cm,height=4.25cm]{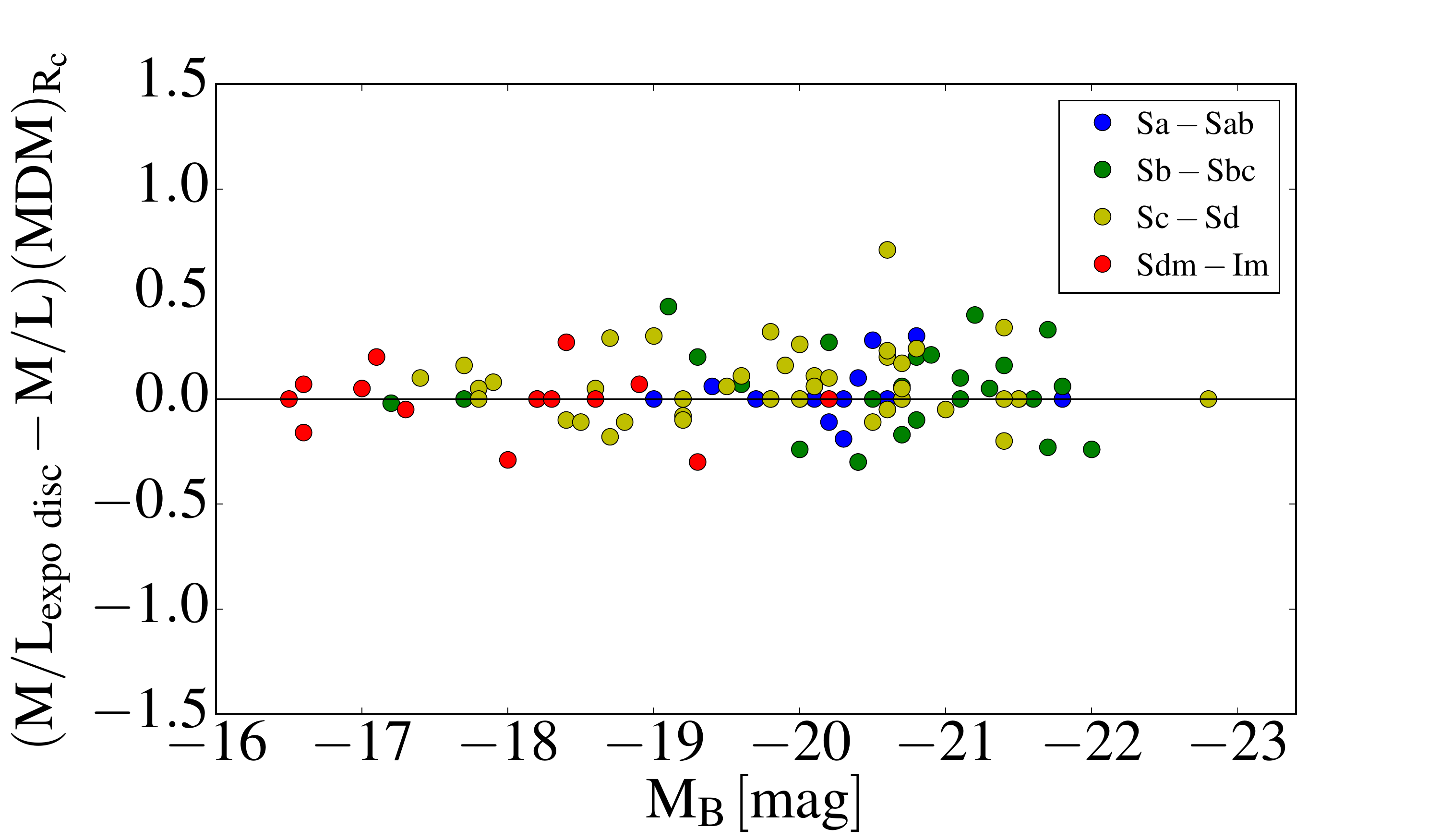}
		\vspace*{-0.0cm}\includegraphics[width=8.5cm,height=4.25cm]{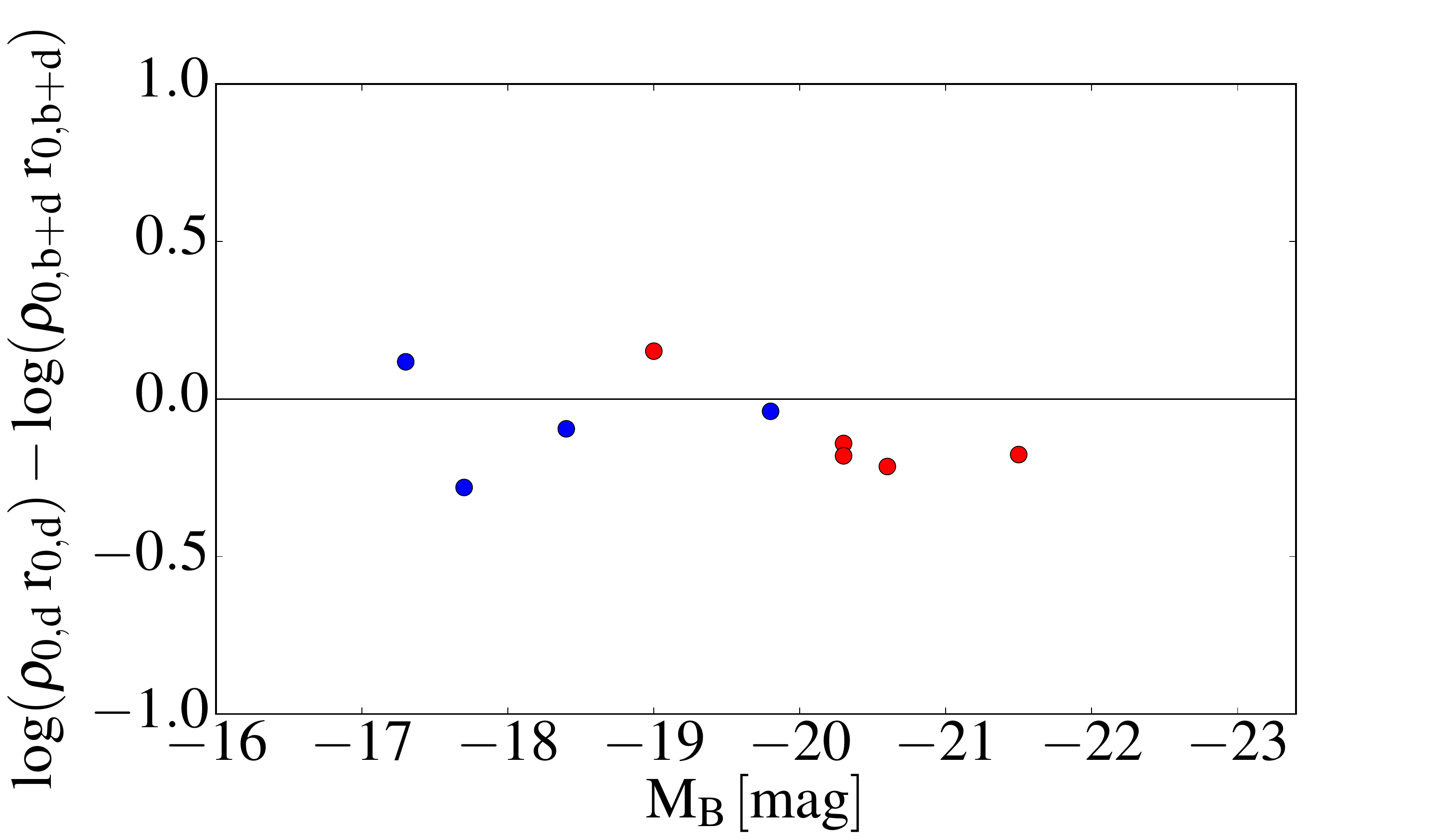}
		\vspace*{-0.0cm}\includegraphics[width=8.5cm,height=4.25cm]{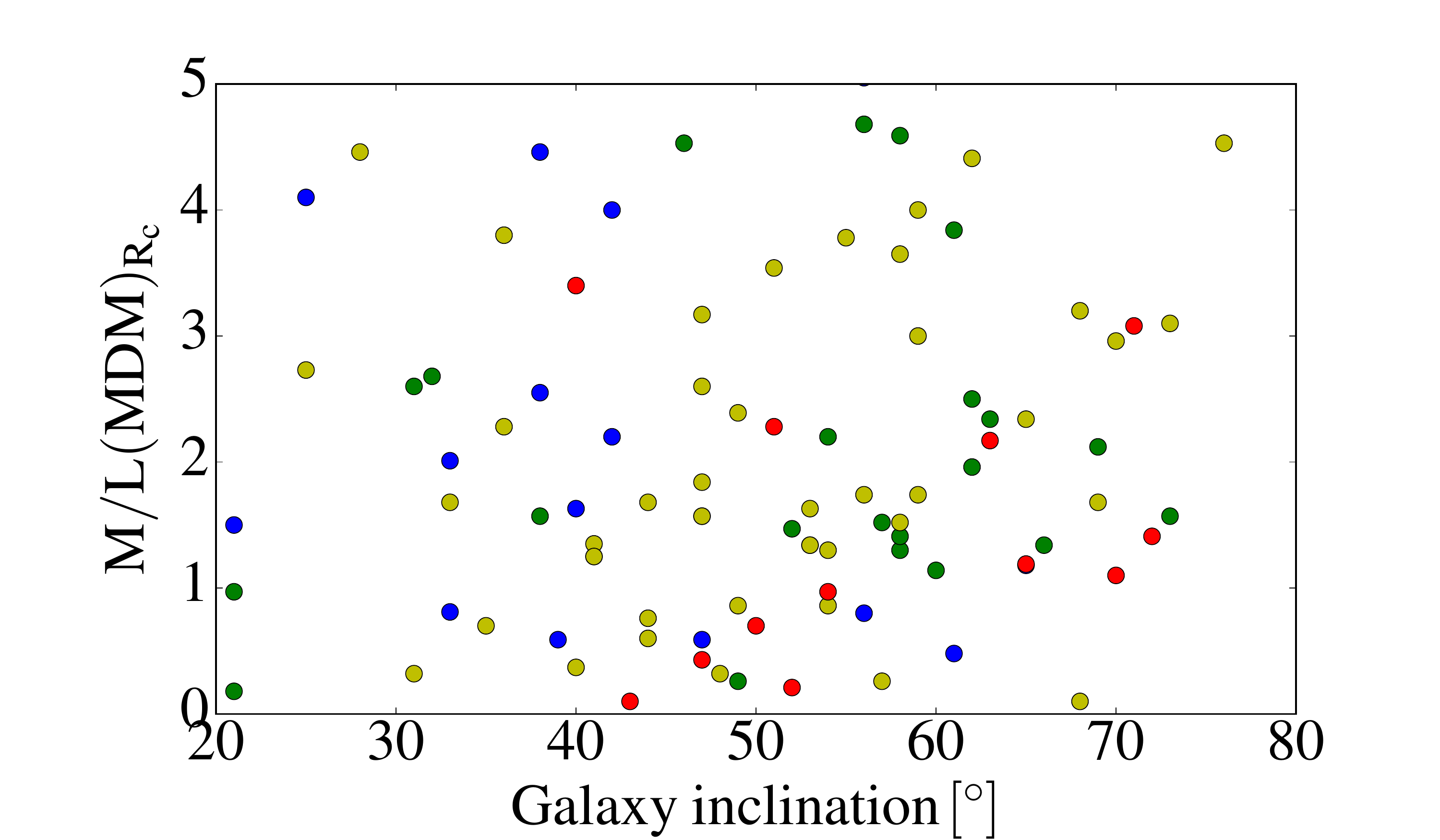}
\caption{The three plots are obtained using the R$_c$-band surface brightness photometry. Top panel: difference between the M/L values obtained using the modelled exponential disc and the actual disc as a function of the absolute M$_B$ magnitude for the ISO (MDM). Middle panel: difference between the log of the product $\rm \rho_{0,d} \times r_{0,d}\ (M_\odot pc^{-2})$  computed without profile decomposition into two distinct components and the log of the  product $\rm \rho_{0,b+d} \times r_{0,b+d}\ (M_\odot pc^{-2})$ obtained when the profiles are decomposed into a disc and a bulge as a function of M$_B$ for the ISO (BFM). Bottom panel: M/L of disc versus the galaxy inclination for the ISO (MDM). The colour code represents the morphological type and is the same than for the top panel.}
\label{fig:mlinc}
\end{center}
\end{figure}

To search of possible bias in the analysis, we describe in this paragraph the three tests we did.
In order to understand if the mass distribution of baryonic and DM halos depends or not on the method used to determine the disc component, we made two tests for which we constructed new mass models: (1) using pure exponential discs and (2) including the bulge light distribution into the disc light distribution. For the first test (1), we decomposed each surface brightness profile, if requested, into a disc and a bulge component, as we did previously but we fit a theoretical exponential disc (see equation \ref{equation1}) instead of the actual light distribution accounting for wiggles due to non-circular motions (spiral arms, bars, …).  This test allowed to check that the disc and halo's parameters we computed previously are not affected by irregular light distributions within the discs. This is illustrated on the example of the disc M/L for ISO (MDM) by the top panel of Fig. \ref{fig:mlinc} where we plot the difference between the M/L values obtained using the modelled exponential disc and the actual disc.
The median difference is close to zero and the scatter is small and does not depend much on the absolute magnitude, meaning that M/L is not strongly affected by non-circular motions.
We found for the ISO models the same trends between the scaling radius and the central halo density, and for the scaling radius or the central halo density as a function of the luminosity, whichever disc is used.
For the NFW models, the same trend is found between the halo concentration and the velocity at the virial radius whichever disc is used but the halo is more concentrated when using the exponential disc. 
In the case of the second test (2),  we selected the nine galaxies of the sample having the strongest bulge, using the criterium $\rm L_B/L_D$ > 0.2.  For those galaxies we do not decompose the surface brightness distribution into two distinct components, a disc and a bulge, even if it is obviously requested by the light distribution. Instead, we use the raw surface brightness profile and consider that all the the stellar contribution lie within the disc component. This test, canceling the bulge component, reduces the number of degrees of freedom of the model and overcomes a possible dependence on the method used to decompose the light profile. We find very similar trends for the DM parameters to what was found when considering the bulge and disc decomposition.  It is well-known that for a given galaxy mass the rotation curve is expected to peak at a velocity $\sim$1.3 times larger if the whole mass is in a flat disc rather than in a spherical bulge component. This means that the M/Ls of the bulges are typically $1.3^2$ $\sim$1.7 times larger in that case. Meanwhile the bulge mass distribution is slightly shifted toward larger radii in the spherical case with respect to the disc case. Consequently, this affects very weakly the halo parameters because these two effects cancel each other and therefore the central slopes and core radii of the halos do not change significantly. This is illustrated by the middle panel of Fig. \ref{fig:mlinc}, in the case of ISO (BFM), by the difference between the log of the product $\rm \rho_{0,d} \times r_{0,d} \times \ (M_\odot pc^{-2})$  when it is computed without profile decomposition into two distinct components and the log of the  product $\rm \rho_{0,b+d} \times \times r_{0,b+d}\ (M_\odot pc^{-2})$ obtained when the profiles are decomposed into a disc and a bulge.  This second test implies that when one does not have sufficient spatial resolution to disentangle the bulge from the disc component as is the case for high redshift galaxies, we can nevertheless probe the luminous-to-dark matter distribution by fitting the surface brightness profile with, for instance, a single exponential disc. Finally, we made a third test: (3) we checked that the different correlations are not artefacts resulting from incorrect galaxy inclinations as it could be the case if the inclinations were not correctly estimated in the mass models or even in the rotation curves. Indeed, the M/L disc ratios and the halo parameters depend on the maximum rotation velocities, which depend on the sinus of the inclination and, to transform the projected surface brightness distribution into baryonic circular velocity distribution in the plane of the galaxy, one has also to account for the inclination of the galaxy. This is illustrated for the ISO (MDM) on the bottom panel of Fig. \ref{fig:mlinc} that clearly shows that the M/L ratios are randomly distributed and thus are not biased by a possible incorrect inclination.

\begin{table}
\begin{center}
\begin{tabular}{c | c |c| c| c | c c }
\hline 

& &a	&$\delta$ a&	 b &$\delta$ b\\
\hline
\multirow{5}{*}{\parbox{1.2cm}{R$_c$-band: $\rho_0$ vs r$_0$ (ISO)}} & BFM	  &-1.18	&0.10		&-0.42	&0.06\\
&MDM	&-1.12	&0.15		&-0.78	&0.09\\
&fixed M/L			&-1.30	&0.12		&-0.41	&0.06\\
&MDM (K$\&$F2004)	&-1.21      & —              &-1.10    &  — \\
&BFM (R$\&$C2014)		&-1.10      & —           & -1.05       &— \\
\hline

\multirow{3}{*}{\parbox{1.2cm}{W1-band: $ \rho_0$ vs $r_0$ (ISO)}} &BFM &-1.48	&0.14		&-0.38	&0.07\\
  &MDM		&-0.81	&0.12		&-0.95	&0.08\\
 & fixed M/L		&-1.07	&0.10		&-0.62	&0.08\\
\hline

\multirow{3}{*}{\parbox{1.5cm}{R$_c$-band: c vs V$_{200}$ (NFW)}} &BFM & -0.98	&0.11		&+3.29	&0.26\\
&fixed M/L		&-1.11	&0.12		&+3.52	&0.26\\
& & & & & \\
\hline

\multirow{3}{*}{\parbox{1.5cm}{W1-band: c vs V$_{200}$ (NFW)}}&BFM	        		&-1.15	&0.14		&+3.63	&0.31\\
&fixed M/L		&-1.28	&0.12		&+3.85	&0.28\\
& & & & & \\
\hline

\end{tabular}
\caption{Relation between the DM halo parameters for ISO and NFW models. K$\&$F2004: \citet{Kormendy+2004} and R$\&$C2014: \citet{Toky+2014}. }
\label{tab:DM}
\end{center}
\end{table}

\begin{table}
\begin{center}
\begin{tabular}{c | c |c| c }
\hline
 & BFM & MDM & Fixed M/L \\ 
 & (1) & (2) & (3) \\ 
\hline

ISO (R$_c$)		& 6	&26		&26			\\
NFW	 (R$_c$)& 12 		&	-	&20			\\
\hline
ISO (W1)		& 10 	&32		&24			\\
NFW (W1)	& 16 		&	-	&28			\\
\hline
\end{tabular}
\caption{Number of galaxies for which a DM halo component is not needed for ISO and NFW using the R$_c$ and W1-band photometry. Columns (1), (2) and (3) show respectively the BFM, the MDM and the Fixed M/L.}
\label{tab1:nohalo}
\end{center}
\end{table}

\begin{table}
\begin{center}
\begin{tabular}{c | c |c }
\hline
c & BFM & Fixed M/L \\ 
 & (1) & (2) \\ 
\hline

NFW (R$_c$)&  13.80 $\pm$ 2.29		&12.02 $\pm$ 1.66				\\
\hline
NFW (W1)   & 11.22 $\pm$ 1.78 		&8.32 $\pm$ 1.58				\\
\hline
\end{tabular}
\caption{Values of the average concentration c using the R$_c$ and W1-band photometry. Columns (1) and (2) represent the values of BFM and Fixed M/L respectively.}
\label{tab:nfwvalue}
\end{center}
\end{table}

\section{Discussion}
\label{sect:rcw1}
\subsection{Comparison between R$_c$ and W1}
In this section, the results obtained using optical R$_c$ photometry are compared to the results using MIR W1 photometry. 
Because some improvements were done to the mass model package since \citet{Korsaga+2018}, we reran the models using the W1 band photometry before doing the comparison with the R$_c$ band photometry.
In this way, it should be possible to evaluate if early mass modelling results, mainly based on optical photometry, are still valid or if they are superseded by more recent results using MIR photometry.

\subsubsection{Photometry}

If we look at the photometry, we found in Figs. \ref{fig:fig4} right panel and \ref{fig:fig7} the same trends for the photometric parameters as a function of morphological types. Probably the most interesting one is in Fig. \ref{fig:fig4} (right) where we see clearly, in both bands, that the central surface brightnesses of the discs are far from being constant but show a clear trend where early type discs are brighter and late-type discs fainter. \color{black}
The distribution of the baryonic matter is characterized by the adopted M/L ratios. Fig. \ref{fig11:M/L} shows that the median values of the M/L ratios in the R$_c$-band (0.39, 1.65 and 1.07 M$_{\odot}$/L$_{\odot}$ respectively for BFM, MDM and fixed M/L) are, as expected, higher than the values found in the MIR (0.14, 0.61 and 0.44). The distributions in both bands are however quite similar. The MDM maximises the contribution from the discs; the M/Ls obtained using the MDM method are indeed $\sim$4 times larger than the BFM model ones.  

\begin{figure*}
	\hspace*{-0.85cm} \includegraphics[width=5.87cm]{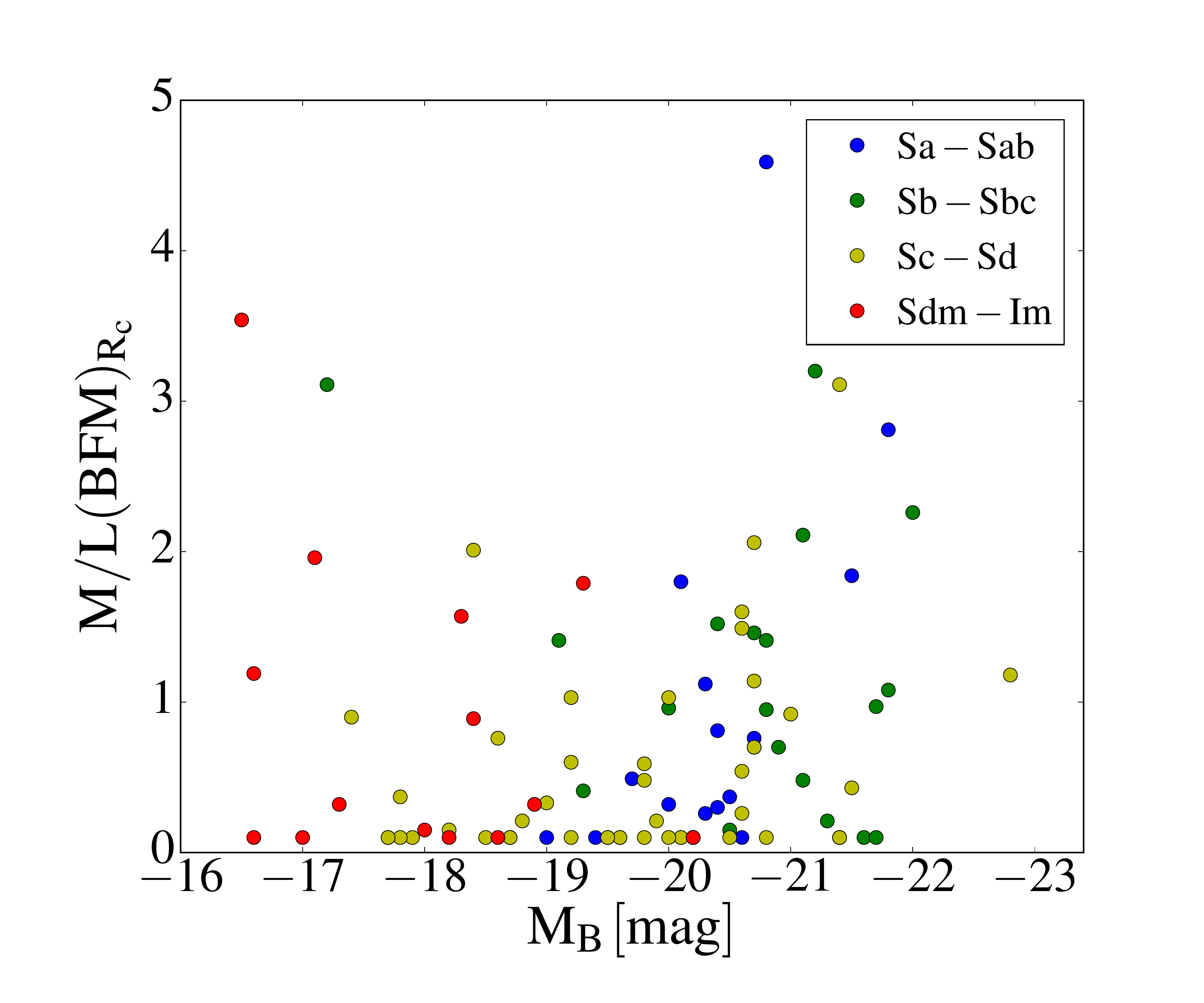}\vspace{-0.17cm}
	\hspace*{-0.35cm} \includegraphics[width=5.87cm]{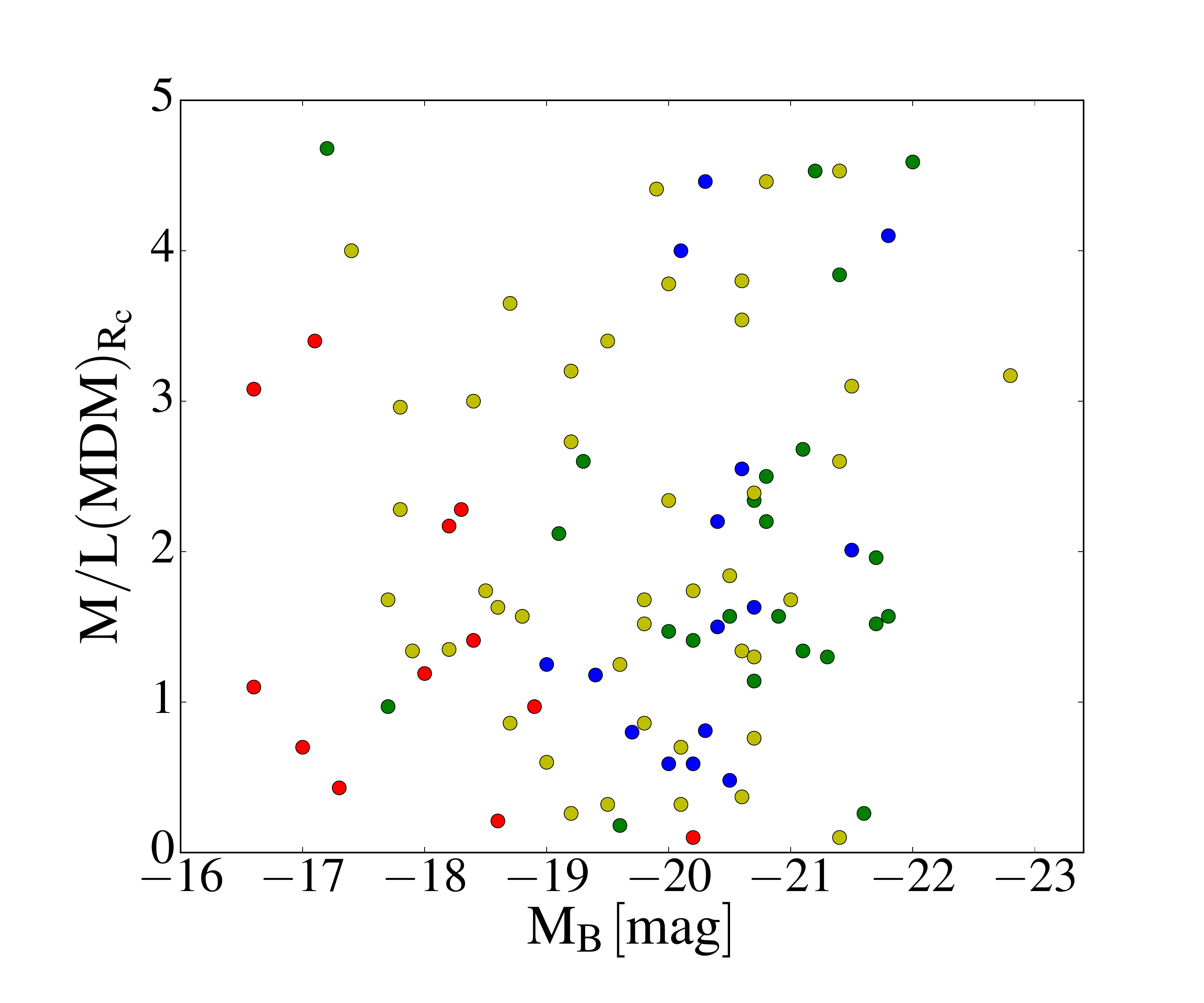}\vspace{-0.17cm}
	\hspace*{-0.35cm} \includegraphics[width=5.87cm]{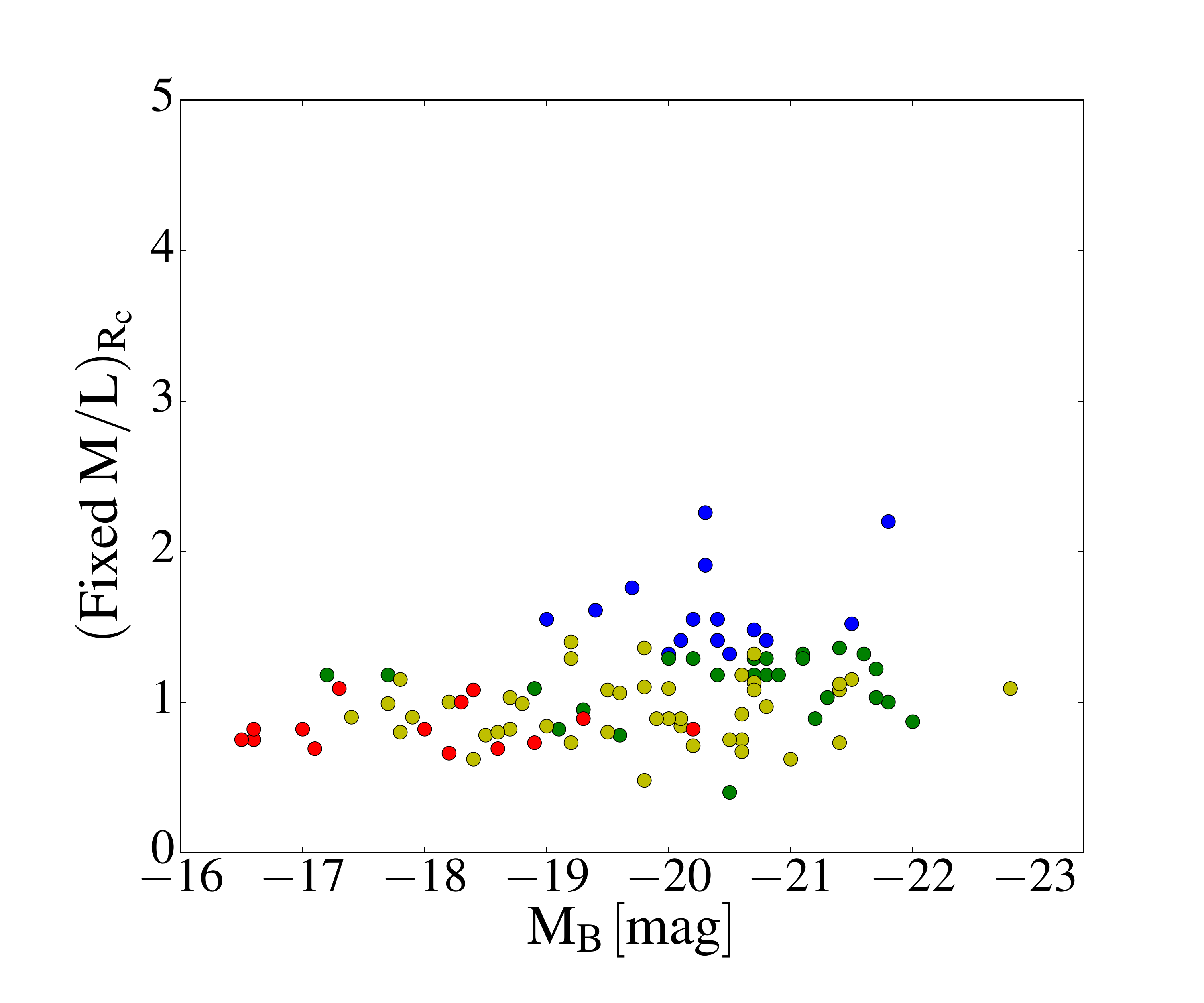}\vspace{-0.17cm}
	\hspace*{-0.85cm} \includegraphics[width=5.87cm]{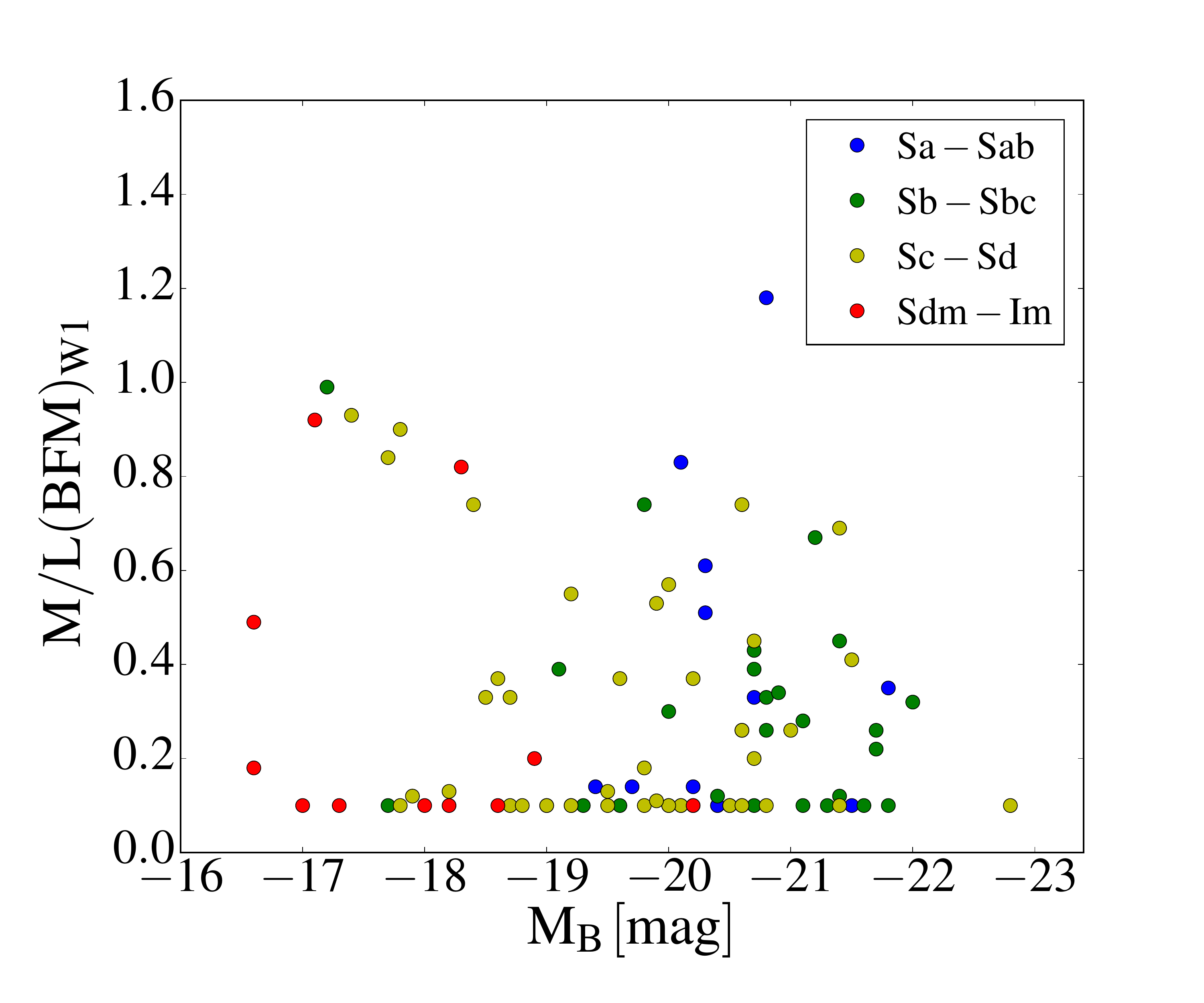}
	\hspace*{-0.35cm} \includegraphics[width=5.87cm]{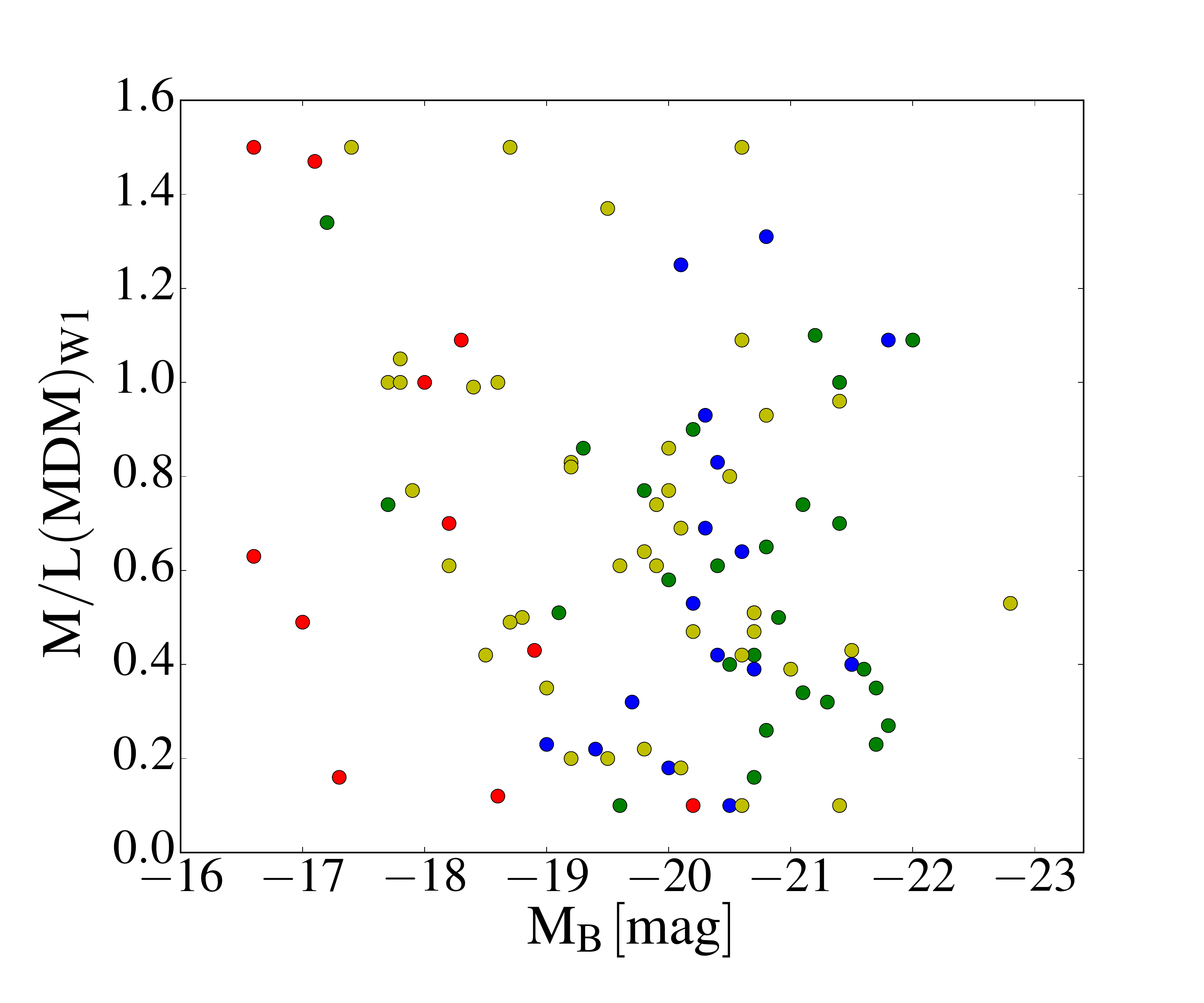}
	\hspace*{-0.35cm} \includegraphics[width=5.87cm]{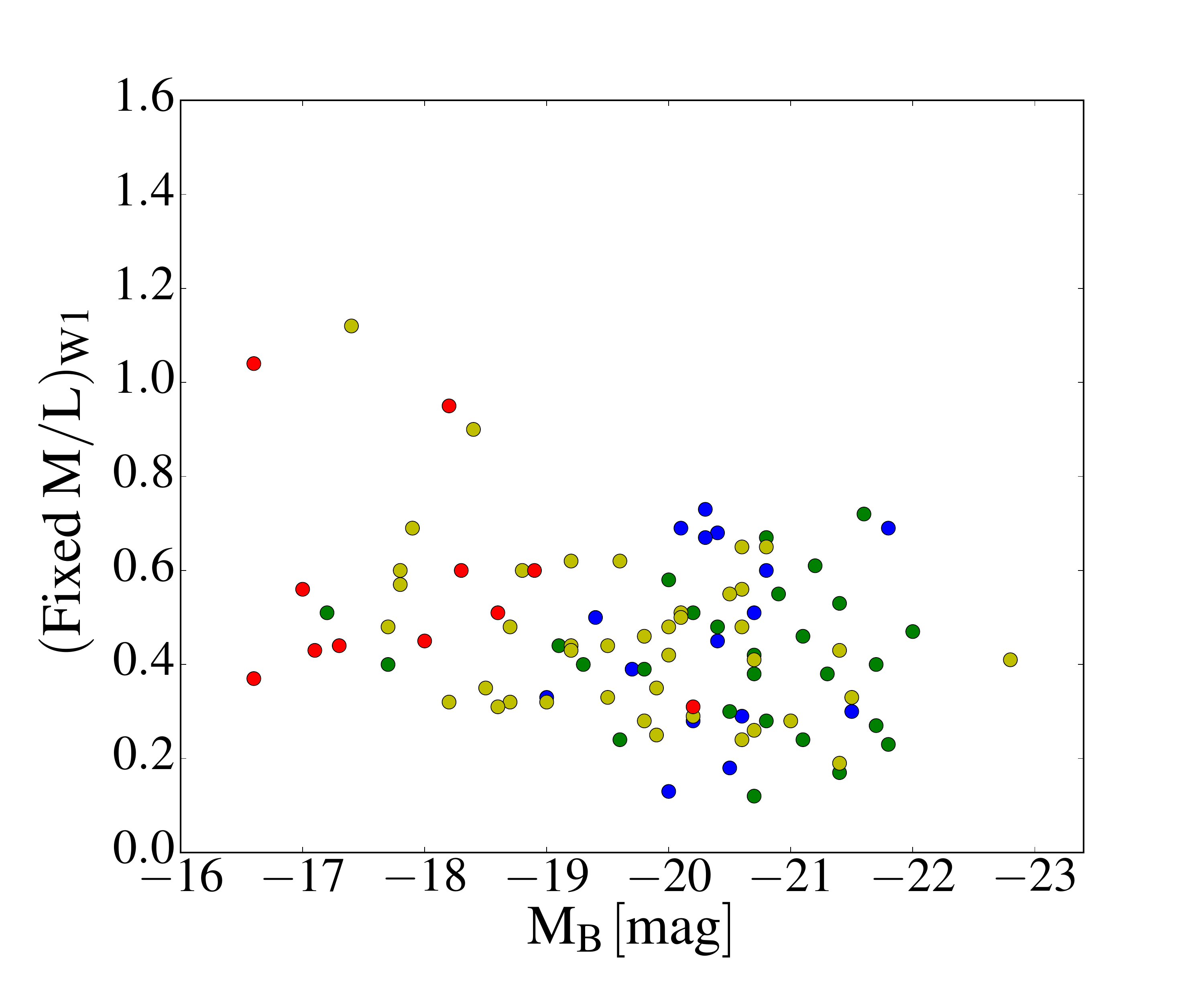}
	\caption{M/L ratio versus absolute magnitude M$_B$ using the R$_c$-band (top) and the W1-band (bottom). From left to right: M/L of ISO (BFM), M/L of ISO (MDM) and fixed M/L. The legends are shown in the left panels.}
\label{fig16:mlmbrcw1}
\end{figure*}

The dispersion in the M/L values are smaller in the MIR than in the optical for BFM and MDM and fixed M/L. This can be seen in Table \ref{tab:sigma} (see columns 3 and 6) and Fig. \ref{fig16:mlmbrcw1}, where the M/L values range from 0 to 5 in the R$_c$-band and from 0 to 1.6 in the W1-band. Nevertheless, in order to compare those dispersions with respect to the median M/L value for each band, we have to consider the normalized dispersions.  Table \ref{tab:sigma} (see columns 4 and 7) shows that the normalized dispersion is also higher in the optical than in the MIR for BFM and MDM while it is smaller in the optical than in the MIR when M/L is fixed. This is probably due to the relation used to calculate the M/L as a function of color. In the W1 band, they used the relation described by \citet{Cluver+2014} while in the R$_c$ band, we used the relation shown in equation \ref{eq7}. Using the fixed M/L, we found that 19 galaxies have non physical M/L values and do not provide good fit (the model is higher than the observed rotation) for W1 while we found only 9 such galaxies in the R$_c$-band.

\begin{table}
\begin{center}
\begin{tabular}{c | c c c | c c c}
\hline
\hline
M/L & \multicolumn{3}{c |}{R$_c$}&\multicolumn{3}{c}{W1} \\ 
 M$_{\odot}$/L$_{\odot}$   & Med &$\sigma$&$\sigma_N$	& Med&$\sigma$ &$\sigma_N$ \\ 
(1) & (2) & (3) & (4) & (5) & (6) & (7) \\ 
\hline

BFM		& 0.39		&0.99		&2.54			& 0.14 	&0.26		&1.86\\
MDM	& 1.65 		&1.29		&0.78			& 0.61	&0.37		&0.61\\
Fixed M/L	&1.07		& 0.33		&0.31			& 0.44      &0.18		&0.41\\
\hline

\hline
\end{tabular}
\caption{Values of the M/L for the BFM, MDM and fixed M/L.  Columns (2) to (4) show respectively the median value of M/L, the dispersion and the normalized dispersion for the R$_c$- band. Column (5) to (7) show the same description using the W1- band.}
\label{tab:sigma}
\end{center}
\end{table}

\subsubsection{Mass models}

The anti-correlation found in Fig. \ref{fig12:all} between $\rho_0$ and r$_0$ is comparable to that found using the 3.4 $\mu$m surface brightness (which is represented by a thick lime line and corresponds to the fit of the BFM. This relation is slightly different from the one published in \citet{Korsaga+2018} because the sample contains here 100 galaxies instead of 121.
The values of the different parameters for ISO (BFM, MDM and fixed M/L) are shown in Table \ref{tab:DM}. These parameters are very similar to those found in the R$_c$-band.
The different weak trends found in the top and middle panels of Fig. \ref{fig13:allmb} for ISO, where no clear correlation is found between the scaling radius and the luminosity of galaxies while less luminous galaxies tend to have smaller central halo densities, are in agreement with \citet{Korsaga+2018}. However, previous authors found that less luminous dwarf galaxies tend to have smaller scaling radii and higher central densities \citep[e.g,][]{Carignan+1988}. 

\citet{Korsaga+2018} explored the reasons of the discrepancy with the literature by splitting the sample either into (i) very high and acceptable quality rotation curves, (ii) barred (SB), moderately barred (SAB) and no-barred galaxies (SA), or into (iii) bulge-poor and bulge-rich galaxies. They concluded that the difference is due to the presence of bulge-rich galaxies for which the bulge has an influence on the derived halo contribution.

\begin{figure}
\begin{center}
	\vspace*{-0.0cm}\includegraphics[width=8.5cm]{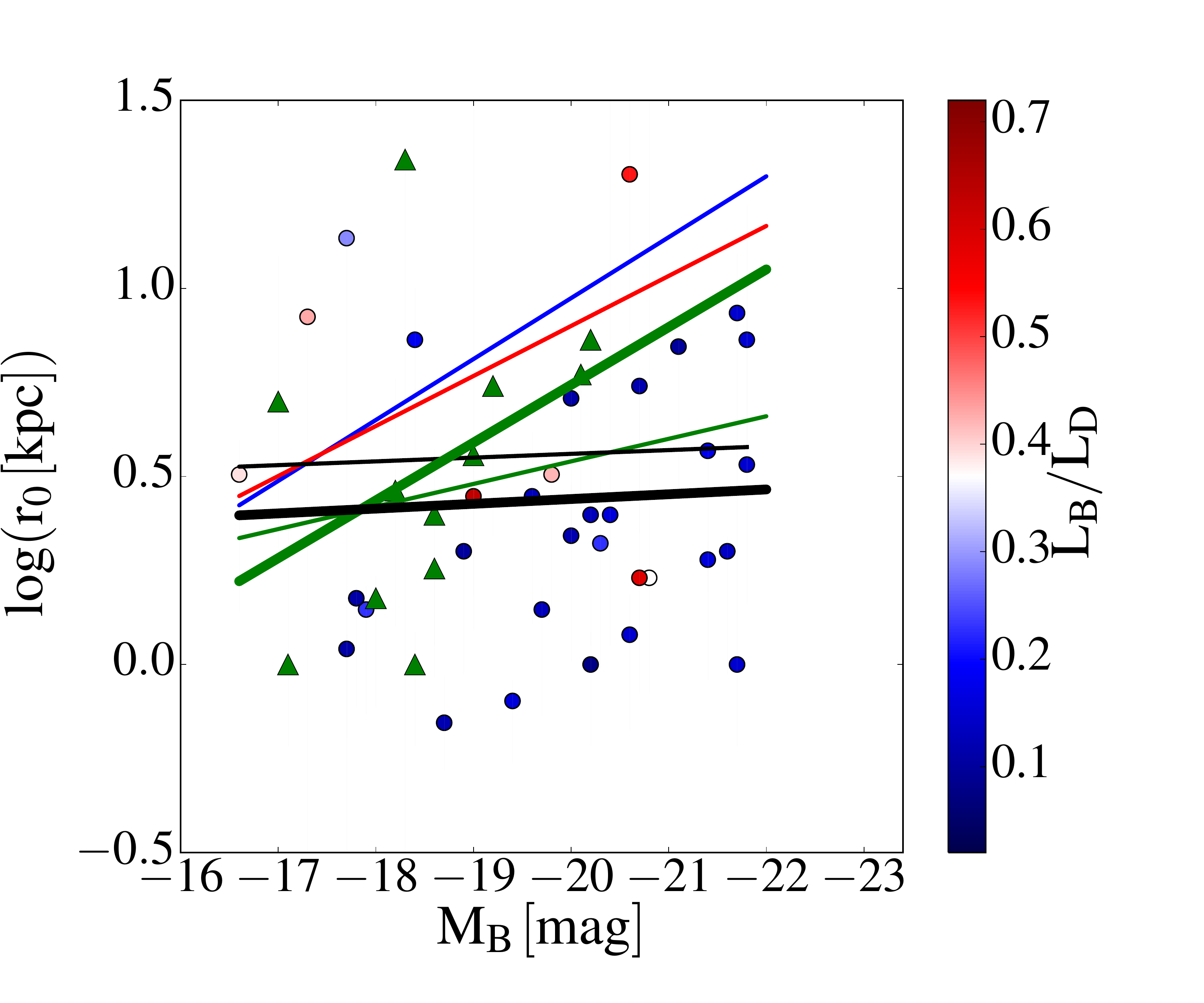}\vspace{-0.75cm}
	\vspace*{-0.0cm}\includegraphics[width=8.5cm]{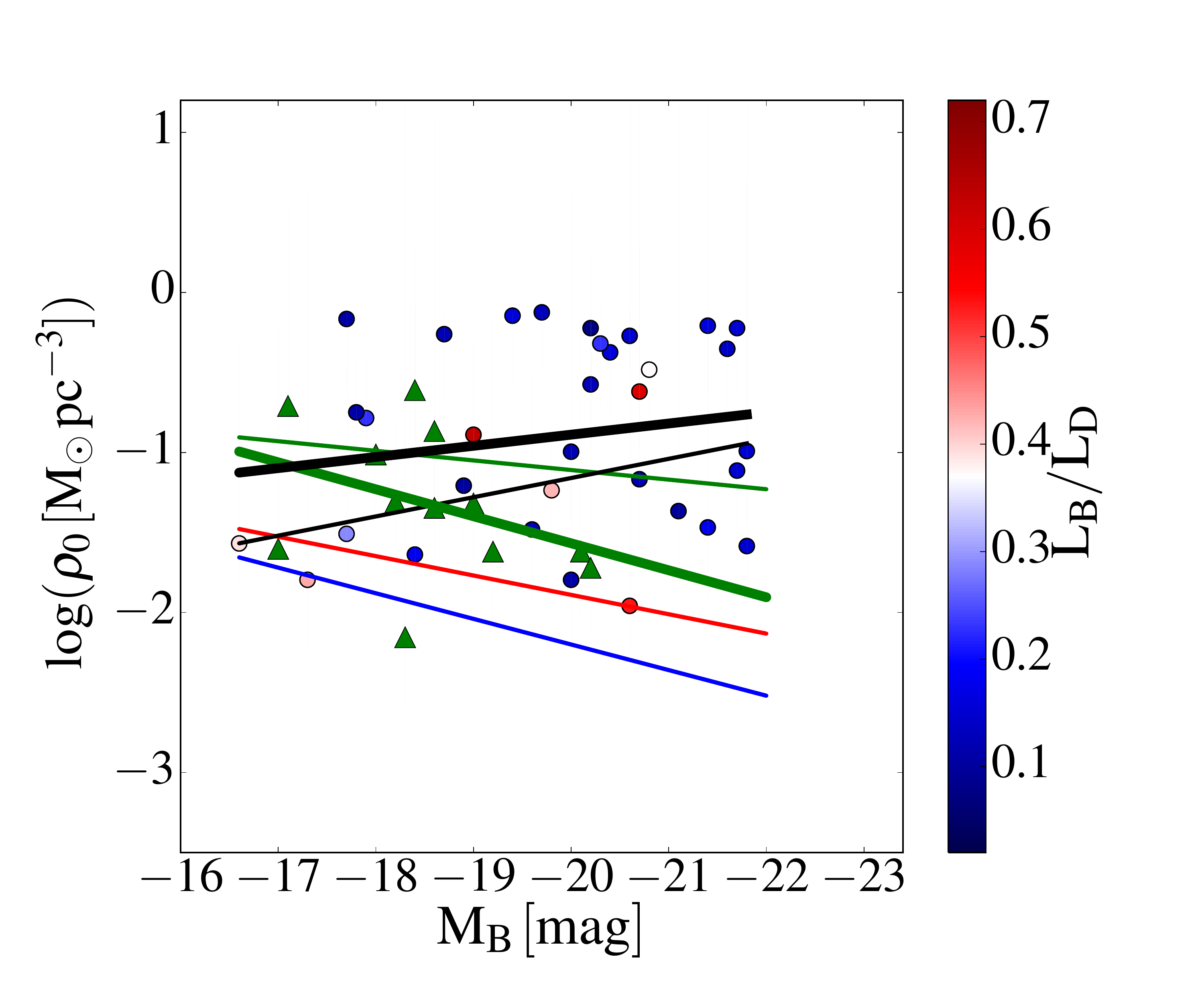}\vspace{-0.75cm}
	\vspace*{-0.0cm}\includegraphics[width=8.5cm]{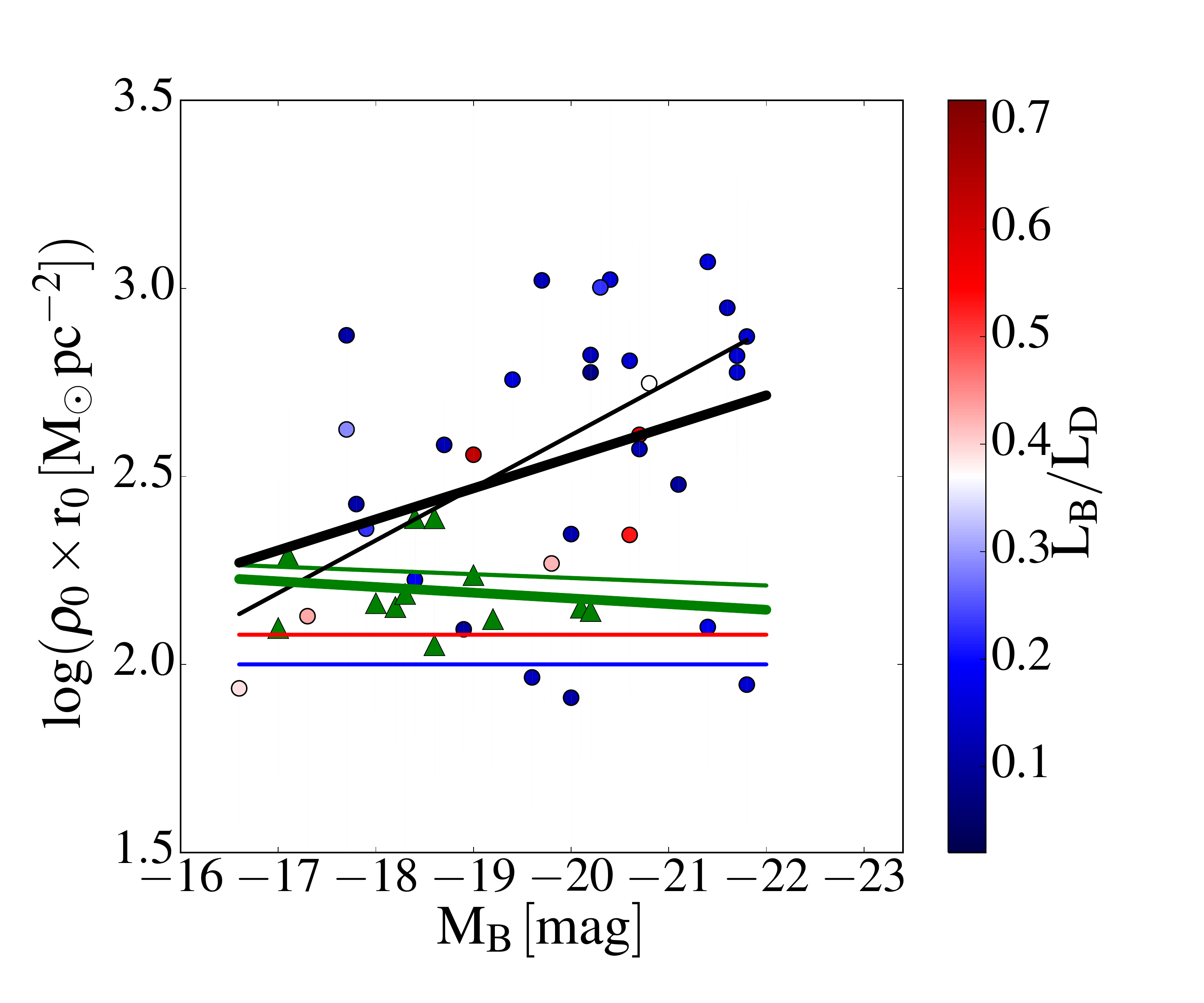}
\caption{Halo parameters (scaling radius at the top, central density in the middle and $\rho_0\, \times$ r$_0$ at the bottom) for ISO (BFM) versus absolute magnitude for the bulge-poor (triangles) and bulge-rich (dots) sub-samples. The colors of the dots represent the importance of the bulge (see color bar on the right). The thick green line represents the fit for the bulge-poor galaxies, which corresponds to $\rm L_{bulge}/L_{total} < 0.02$ and the thick black line for the bulge-rich galaxies which corresponds to $\rm L_{bulge}/L_{total} > 0.07$. The thin blue  and red lines represent respectively the fit found by \citet{Kormendy+2004} and \citet{Toky+2014}, mainly for late-type galaxies. The thin green line represents the fit for the bulge-poor and the thin black line for the bulge-rich galaxies found using the W1-band.}  
\label{dbr_rc_rho_BFM_mag}
\end{center}
\end{figure}

In the bottom panel of Fig. \ref{fig13:allmb}, less luminous galaxies tend to have smaller $\rho_0\, \times$ r$_0$, which is consistent with the results found in the W1-band (thick lime line). 
In order to understand if the various trends we found between the halo scale radius and the central density as a function of M$_B$ are also due to bulges, we defined two sub-samples like \citet{Korsaga+2018} did with the W1-band photometry: a bulge-poor sample composed of galaxies with $\rm L_{bulge}/L_{total} < 0.02$ (mostly composed of late-type spiral galaxies) and a bulge-rich sample with $\rm L_{bulge}/L_{total} > 0.07$ (composed of early type spirals). The DM halo parameters of these sub-samples are plotted as a function of M$_B$ in Fig. \ref{dbr_rc_rho_BFM_mag}. The top panel of this figure shows the scaling radius as a function of M$_B$; a clear correlation appears for bulge-poor galaxies (thick green line) while we notice a very weak correlation for bulge-rich galaxies (thick black line). The middle panel of Fig. \ref{dbr_rc_rho_BFM_mag} represents the central density of the halo as a function of the luminosity M$_B$ where an anti correlation is found for bulge-poor galaxies (thick green line) and a correlation for bulge-rich galaxies (thick black line). The bottom panel  shows $\rho_0\, \times$ r$_0$ as a function of M$_B$; we found no correlation for bulge-poor galaxies (thick green line) while a clear correlation is seen for bulge-rich galaxies (thick black line). All the trends we found in this work for bulge-poor and bulge-rich galaxies are similar to those found in W1-band (represented by thin green and black lines respectively for bulge-poor and bulge-rich galaxies). Those results seem to point out that the results of the mass models are quite independent of the photometric band used.

For the NFW model, the average concentration values found are c = 13.80 $\pm$ 2.29 and c = 12.02 $\pm$ 1.66 for the BFM and the fixed M/L respectively. These values are higher than what found using the 3.4 $\mu$m luminosity profile (c = 11.22 $\pm$ 1.78 and 8.32 $\pm$ 1.58 for BFM and fixed M/L respectively). As seen in Fig. \ref{fig15:nfw}, an anti-correlation is found between c and V$_{200}$ showing that high concentrations trend to have smaller mass halos. Similar results are seen in the W1-band. 

In Fig. \ref{fig:isorcw1}, the top panel shows the comparison between r$_0$ for the two bands R$_c$ and W1 using ISO (BFM); we notice that the scaling radius are smaller in the R$_c$-band than in the W1-band. The bottom panel represents the comparison between $\rho_0$ for R$_c$ and W1 bands, $\rho_0$ are higher in the R$_c$-band. 
When looking at the NFW models (Fig. \ref{fig:nfwrcw1}), it can be seen that the concentration c are higher and V$_{200}$ smaller in the R$_c$-band compared to the results obtained in the W1 band. This could be due to the fact that those two bands are probing different stellar populations. Most specifically, the difference between the M/L ratios of bulges and discs may be more important for the Rc-band than for the W1-band because the Rc-band is more sensitive to blue stars.
Therefore, if the M/L ratio of the bulge and disc becomes similar, we might have a larger contribution of the bulge in the W1-band which might lead to a lower concentration.
Indeed, the rise of the rotation curve grows as the central concentration c, thus when the contribution of the bulge decreases, the contribution of the DM halo is expected to increase in the inner regions of the galaxies.

Alternatively, this could be the result of the minimal value allowed for the M/L ratio. 
The median of M/L is 3 times larger for the Rc-band than for the W1-band (see Table \ref{tab:sigma}) but the M/L lower limit allowed for both bands are identical (M/L = 0.1).  Thus we expect that more galaxies reach this lower limit for the W1-band than for the Rc-band and that this limit corresponds to a larger contribution of the stellar components for the W1-band.

\begin{figure}
	\begin{center}
             \vspace*{-0.0cm}\includegraphics[width=8.5cm]{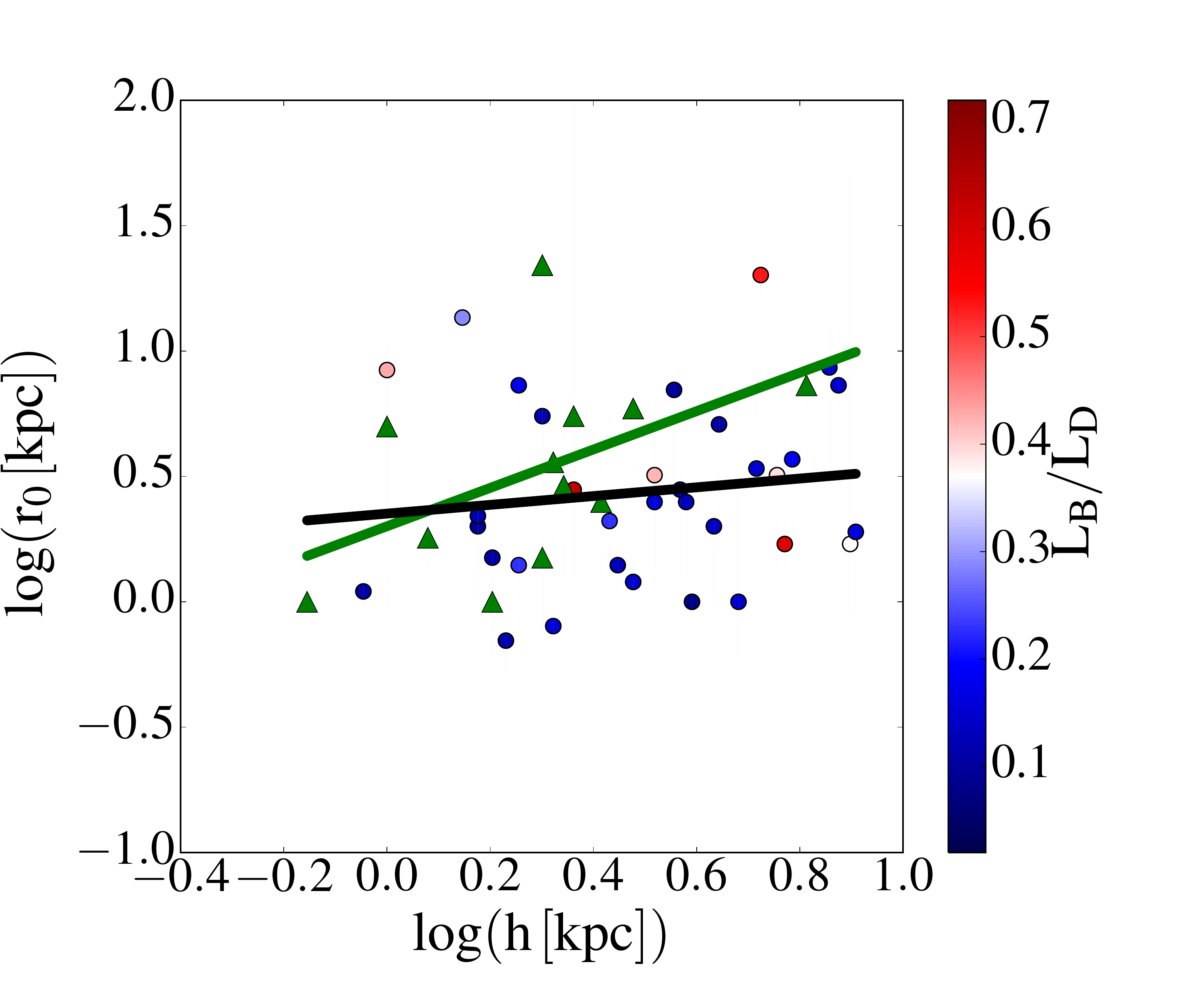}\vspace{-0.75cm}
             \vspace*{-0.0cm}\includegraphics[width=8.5cm]{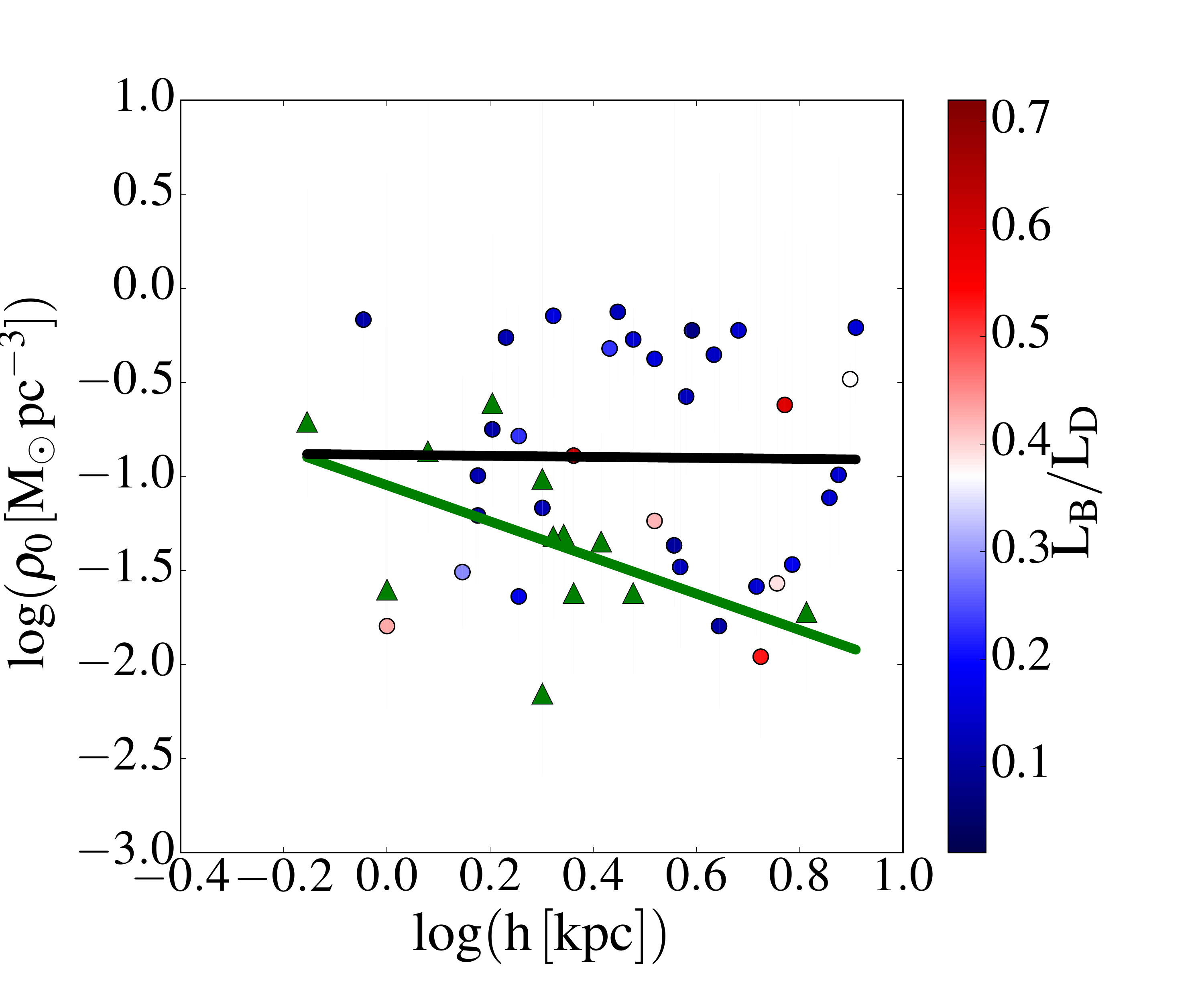}
\caption{The top and bottom panels show respectively the scale radius and the central halo density as a function of the disc scale length for ISO (BFM). The colors and symbols are the same as used in Fig. \ref{dbr_rc_rho_BFM_mag}.
The thick green line represents the fit for the bulge-poor and the thick black line for the bulge-rich galaxies.}
\label{fig:hdbr}
\end{center}
\end{figure}

\begin{figure}
	\begin{center}
             \includegraphics[width=8.5cm]{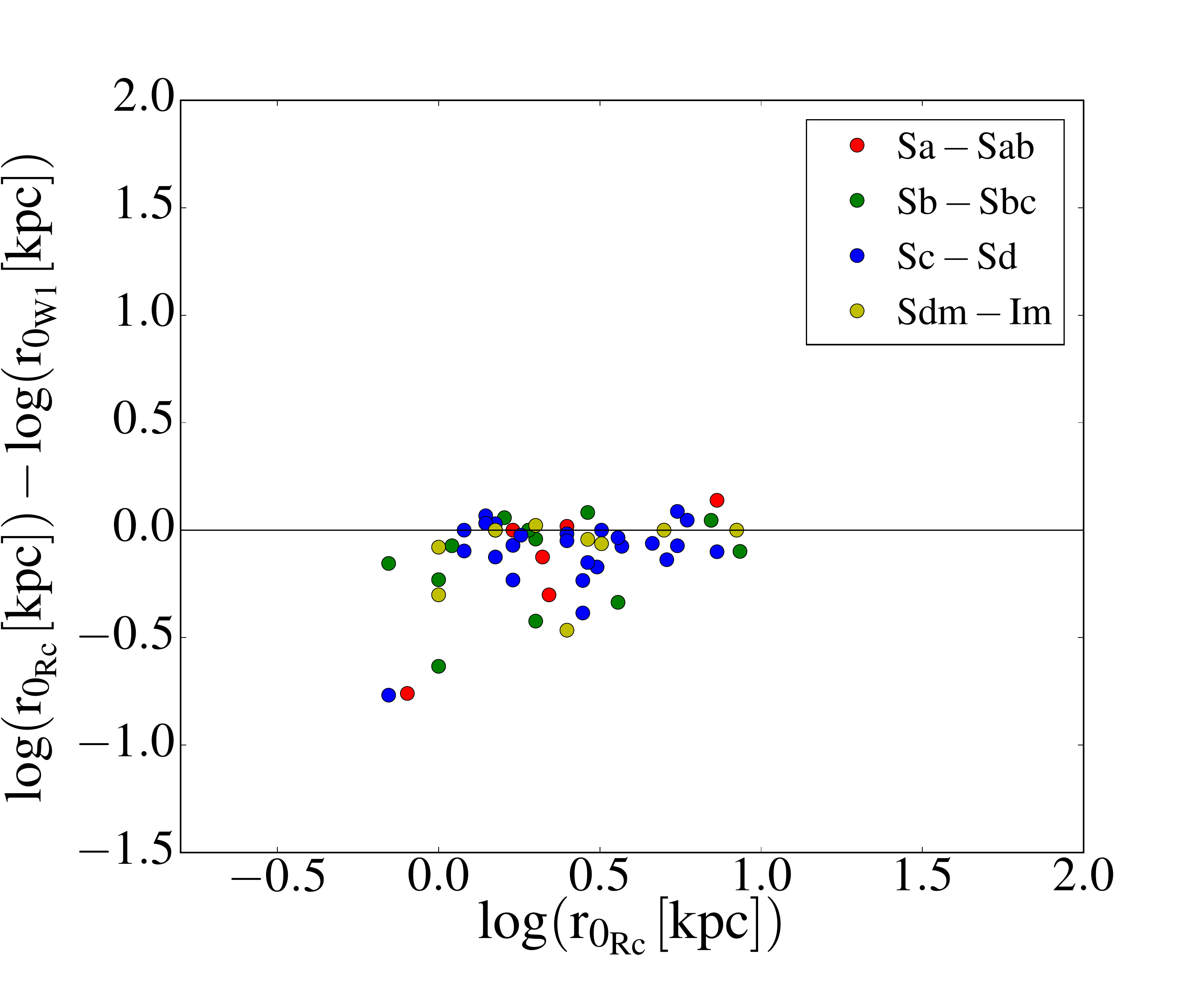}
             \includegraphics[width=8.5cm]{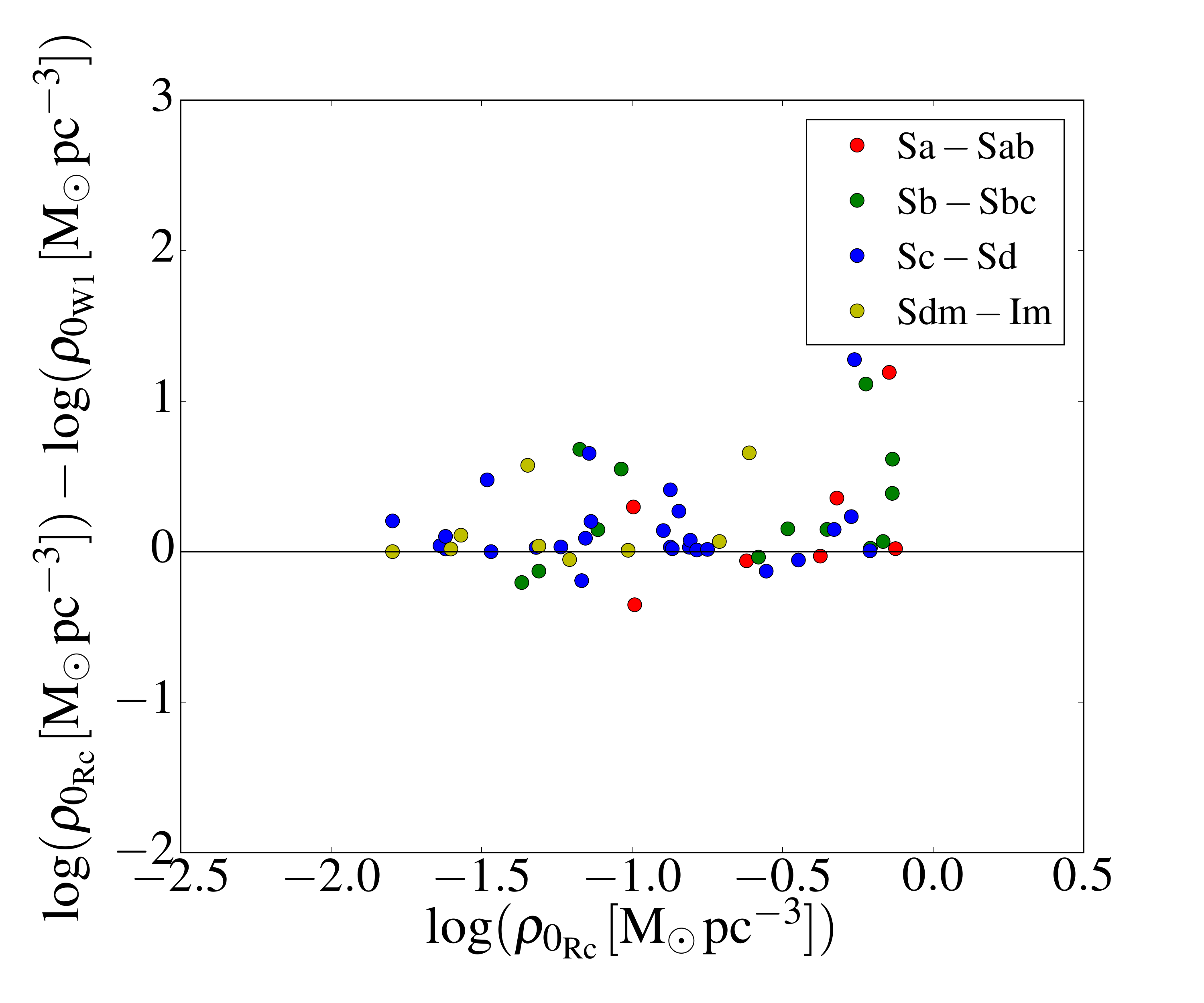}
\caption{Comparison of the results obtained for ISO (BFM) using the optical R$_c$ and MIR W1 bands. Top panel: scaling radius r$_0$. Bottom panel: central density $\rho_0$.}
\label{fig:isorcw1}
\end{center}
\end{figure}

\begin{figure}
	\begin{center}
            \includegraphics[width=8.5cm]{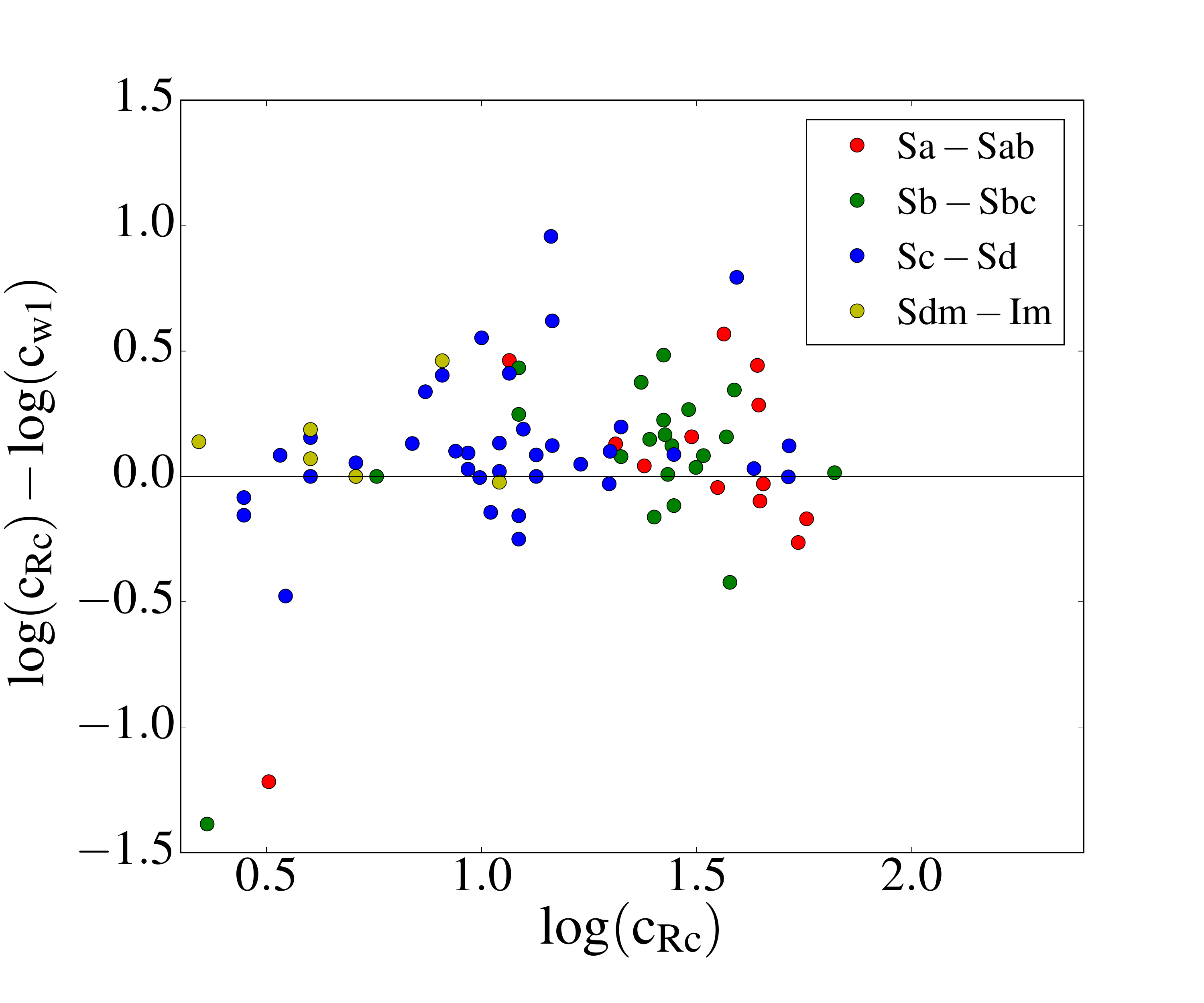}
            \includegraphics[width=8.5cm]{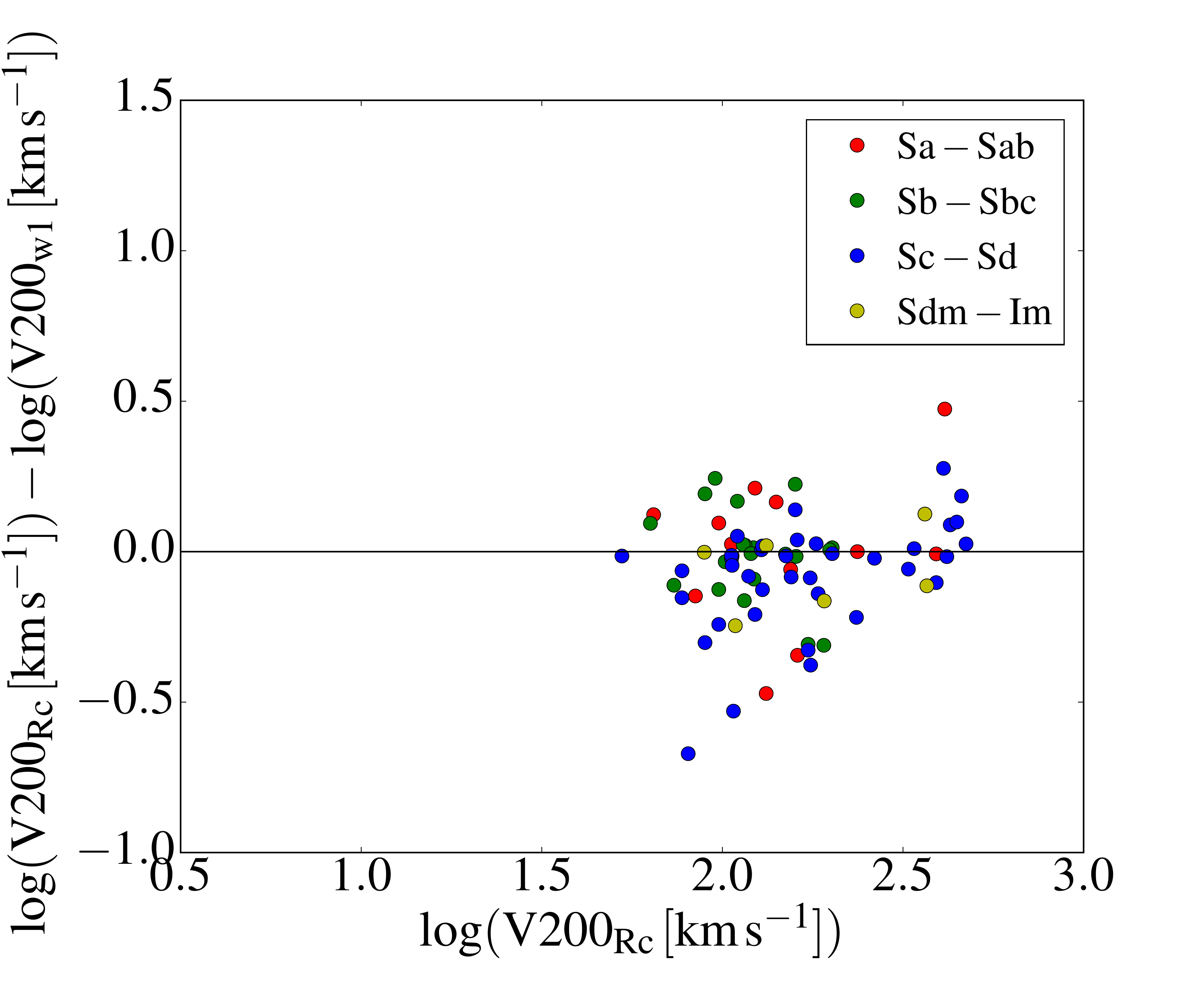}
 \caption{Comparison of the results obtained for NFW (BFM) using the optical R$_c$ and MIR W1 bands. Top panel: concentration c. Bottom panel: velocity at the virial radius V$_{200}$. }
 \label{fig:nfwrcw1}
 \end{center}
\end{figure}

\subsection{Comparison with previous works}
\label{sub:comp_previous}

According to \citet{Swaters+2011}, the M/L ratio is usually high (up to 15 in the R-band) when using the maximum-disc fits of \Hi\, rotation curves for gas-rich irregular galaxies; these high values imply that irregular galaxies are dominated by DM at all radii \citep{Zibetti+2009}. As can be seen in Fig. \ref{fig16:mlmbrcw1}, such large values are surely not seen in this study where the largest values we find are even less than 5. This difference could be partly explained by the fact that the \Hi\ resolution is lower than in the optical, thus the constraints are lower in the optical regions. Moreover, \citet{Lelli+2014}'s study on dwarf galaxies revealed that M/L is $\sim$1.5 in the R$_c$-band, mostly for the blue compact dwarfs (BCDs); this value is close to the one found in this paper when using the ISO (MDM). 
As expected, the median values of the M/L ratio found in this work  (0.39, 1.65 and 1.07 M$_{\odot}$/L$_{\odot}$ respectively for BFM, MDM and fixed M/L) are higher than the previous works which used the MIR photometry band \citep[e.g.][]{McCaugh+2015,Richards+2016,Lelli+2016,Richards+2018,Korsaga+2018}. 

In this work, we find no clear correlation between the scaling radius and the luminosity of the galaxies for the whole sample and less luminous galaxies tend to have smaller central halo densities (see top and middle panels of Fig. \ref{fig13:allmb}). However, our results are not in agreement with previous studies made by \citet{Kormendy+2004} and \citet{Toky+2014} represented by the thin blue and red lines respectively in Fig. \ref{fig13:allmb}. They found that less luminous dwarf galaxies tend to have smaller scaling radii and higher central densities. In the bottom panel of Fig. \ref{fig13:allmb}, we find that less luminous galaxies tend to have smaller $\rho_0\, \times$ r$_0$ which is not in agreement with previous studies \citep[][ thin blue and red lines respectively]{Kormendy+2004, Toky+2014} which suggest that $\rho_0\, \times$ r$_0$ does not depend on the luminosity. 
When comparing the correlations found using the sub-samples in Fig. \ref{dbr_rc_rho_BFM_mag} with previous authors \citep[][ thin blue and red lines respectively]{Kormendy+2004, Toky+2014} who mostly worked on late type galaxies, we find a consistency for bulge-poor galaxies only but not for bulge-rich galaxies. We conclude that the distribution of the DM as a function of the absolute luminosity depends on the morphological type of the galaxy which means that we cannot expect to have the same trend when working with bulge-poor or bulge-rich galaxies.

The average concentration values c = 13.80 $\pm$ 2.29  and c = 12.02 $\pm$ 1.66 for the NFW (BFM) and NFW (fixed M/L) respectively in this study are higher than the standard value found by \citet{Bullock+2001} which is c = 10 but in the range of the values found by \citet{Martinsson+2013} which are between 10 and 20. \citet{Edo+2006} used this standard value to model mass distributions and concluded that NFW fits with $\rm c = 10$ remain comparable to the minimal $\chi^2$ of the isothermal fits. 
The anti-correlation between the concentration and the mass halos for NFW (BFM) and NFW (fixed M/L) found in Fig. \ref{fig15:nfw} is also found by \citet{Wechsler+2002,Barnes+2004,Edo+2006,Martinsson+2013}.

As already mentioned in section \ref{sect:results}, we do not find a clear correlation between r$_0$ and $\rho_0$ as a function of the disc scale length represented in Fig. \ref{fig14:ht}. This is not in agreement with previous studies \citep[e.g.][]{Cote+2000,Donato+2004} who used a sample composed of mostly late type galaxies and found that galaxies with a small disc scale length h tend to have a small r$_0$ and a high $\rho_0$. However, when we use the sub-samples in Fig. \ref{fig:hdbr} (top and bottom panels), we find that bulge-poor galaxies  (thick green line) with a small h tend to have a small r$_0$ and a high $\rho_0$. This correlation is in agreement with the previous authors mentioned. No clear correlation is found for the bulge-rich galaxies (thick black line). Once again, we conclude that the relation between the halo parameters and the optical disc scale length depends on the presence or not of a bulge.

\section{Summary and Conclusions}
\label{sect:conclusion}

This paper presents the study of mass models using H$\alpha$ rotation curves and optical R$_c$-band photometry for a sample of 100 galaxies covering morphological types from Sa to Irr. The high resolution rotation curves are used to understand the stellar mass distribution and the shape of dark halo density distribution. Therefore, we used two main models to describe the DM distribution; the pseudo-isothermal sphere (ISO) and the Navarro-Frenk-White model (NFW) with different fitting procedures (a best fit model (BFM), a maximum disc model (MDM) and a M/L calculated using the colors (fixed M/L)).

(i) Using only optical rotation curves, galaxy mass models request a DM halo for two third of the sample for the ISO  (MDM) models, this fraction rises to 87\% for BFM and to 78\% for the fixed M/L.   For NFW models, a DM halo is requested for 89\%, 78\% of the galaxies for BFM and fixed M/L respectively. \citet{kalnajs+1983} showed more than 30 years ago that many galaxies do not need DM halos to model their RCs when only optical kinematics is available (e.g. M33, NGC 7793). For those galaxies, the need to add a DM halo only arise when HI kinematics going further out is added.

(ii) For the ISO (BFM), the model describes very well the inner and outer part of the rotation curves for both low and high luminosity galaxies. The ISO (MDM) also procures good fits to the observed rotation curves with large values of M/L ratios. The fixed M/L provides acceptable fits but we sometimes found some non-physical fits when using this technique (e.g.: UGC 2045, 2855, 6118). The M/L values found for ISO (MDM) are $\sim$4 times higher and for ISO (fixed M/L) $\sim$3 times higher than the M/L ratio values of the ISO (BFM). The M/L dispersion is higher when using the R$_c$-band compared to the W1 band for all models.

(iii) Isothermal models suggest that smaller scaling radius (r$_0$) tend to have higher central halo density ($\rho_0$) (Fig. \ref{fig12:all}). This correlation does not depend on the morphological types (presence of bulge or not) because the same trend is seen when we compare our results with previous studies which used mostly late type spiral galaxies. A correlation is found between $\rho_0$ and $\rho_0\, \times$ r$_0$ as a function of the luminosity when using the whole sample; less luminous galaxies tend to have smaller $\rho_0$ and smaller $\rho_0\, \times$ r$_0$ while no clear correlation is found between r$_0$ and the luminosity. We need to be careful with these relations in the sense that our work contains galaxies covering all morphological types (from early to late type spirals and irregulars) which means the presence of bulge-rich galaxies. 

(iv) For that, we divided the sample in two sub-samples (bulge-poor and bulge-rich galaxies). 
For bulge-poor galaxies, we found that less luminous galaxies tend to have smaller r$_0$ and higher $\rho_0$, and an independent relation is found between $\rho_0\, \times$ r$_0$ and the luminosity. 
For the bulge-rich galaxies, the correlations found are in agreement with those found using the whole sample. We found no clear correlation between r$_0$ and $\rho_0$ and the optical disc scale length (h). However, a clear correlation is found for bulge-poor galaxies; smaller h corresponds to smaller r$_0$ and higher $\rho_0$. Once again, for bulge-rich galaxies, the correlations found are in agreement with the whole sample. 

In conclusion, we can say that the relations between the halo parameters and the luminosity depend on morphological types, which means that we do not find the same relation when using galaxies with or without a bulge. The same conclusion is also found between the halo parameters and the optical disc scale length. The fact that we found similar correlations when using the whole sample and when using the sub-sample of bulge-rich galaxies is due to the presence of a large number of bulge-rich galaxies in the sample. It is worthwhile to note that these relations are independent of the band used: the same results are being observed when using the B-band and the W1-band luminosities. 

(v) In order to test the strength of the previous results we have modelled the observed disc with an exponential disc. Indeed, one might expect that the disc - DM halo models depend on the local light distribution rather than on the global light distribution.  The correlations discussed in item (ii) remain basically the same.  This means that high-z galaxies, for which it is too challenging to measure the detailed light distribution as a function of the galacto-centric distance, could be modelled by an exponential disc. In other words, the knowledge of the disc scale length should be enough for high-z galaxies.

(vi) In the case of NFW (BFM), the rotation curves can be well explained with this model especially for high luminosity galaxies, but for less luminous galaxies, the fit becomes inconsistent. The fixed M/L ratio provides acceptable fits for most galaxies while the model gives non physical fits for some galaxies. For cuspy profile density, we found that low mass halos are more concentrated than high mass halos. We concluded that our rotation curves are better described by core than cuspy density profiles whatever we use BFM or fixed M/L techniques, as shown in Fig. \ref{fig9:nfwiso}.

(vii) When we compare the M/L ratios and the halo distribution in R$_c$-band with W1-band, we notice that the dispersion of the M/L values is higher for the optical R$_c$-band than the W1 band;  and the halo is more concentrated when using R$_c$-band than the MIR $3.4 \mu$m band for ISO and NFW models.

\section*{Acknowledgements}
We thank the referee for useful comments that helped in improving the paper. Most of the research of MK was done while she was having a PhD Scholarship from the Science faculty of the University of Cape Town. CC’s work is based upon research supported by the South African Research Chairs Initiative (SARChI) of the Department of Science and Technology (DST), the Square Kilometre Array South Africa (SKA SA) and the National Research Foundation (NRF). We acknowledge financial support from “Programme National de Cosmologie et Galaxies” (PNCG) funded by CNRS/INSU-IN2P3-INP (Centre national de la recherche scientifique/Institut national des sciences de l’Univers - Institut national de physique nucleaire et de physique des particules - Institut de physique), CEA (Commissariat à l’Energie atomique et aux Energies alternatives) and CNES (Centre national d’etudes spatiales) in France.





\bibliographystyle{mnras}
\bibliography{exemple_biblio}



\onecolumn
\appendix
\section{Tables}
\label{appendixA}
We list here the global properties and mass models parameters of 8 of the 100 galaxies. The remaining galaxies are available in the online version. 
\input{latex_table_paper}



\section{Surface brightness profile and mass models} 
\label{appendixB}
We present the mass models of 2 of the 100 galaxies using the optical R$_c$-band photometry. The remaining galaxies are available in the online version. The surface brightness profiles decomposition for the 27 galaxies from SDSS are shown in this work while we do not present the remaining 73 galaxies from OHP because they have already been published in \citet{Barbosa+2015}. Note that 4 of the 27 SDSS galaxies have not been decomposed, for these galaxies, we just show the observed surface brightness profiles.
\FloatBarrier
\begin{figure*}
\hspace*{-0.00cm} \includegraphics[width=0.35\textwidth]{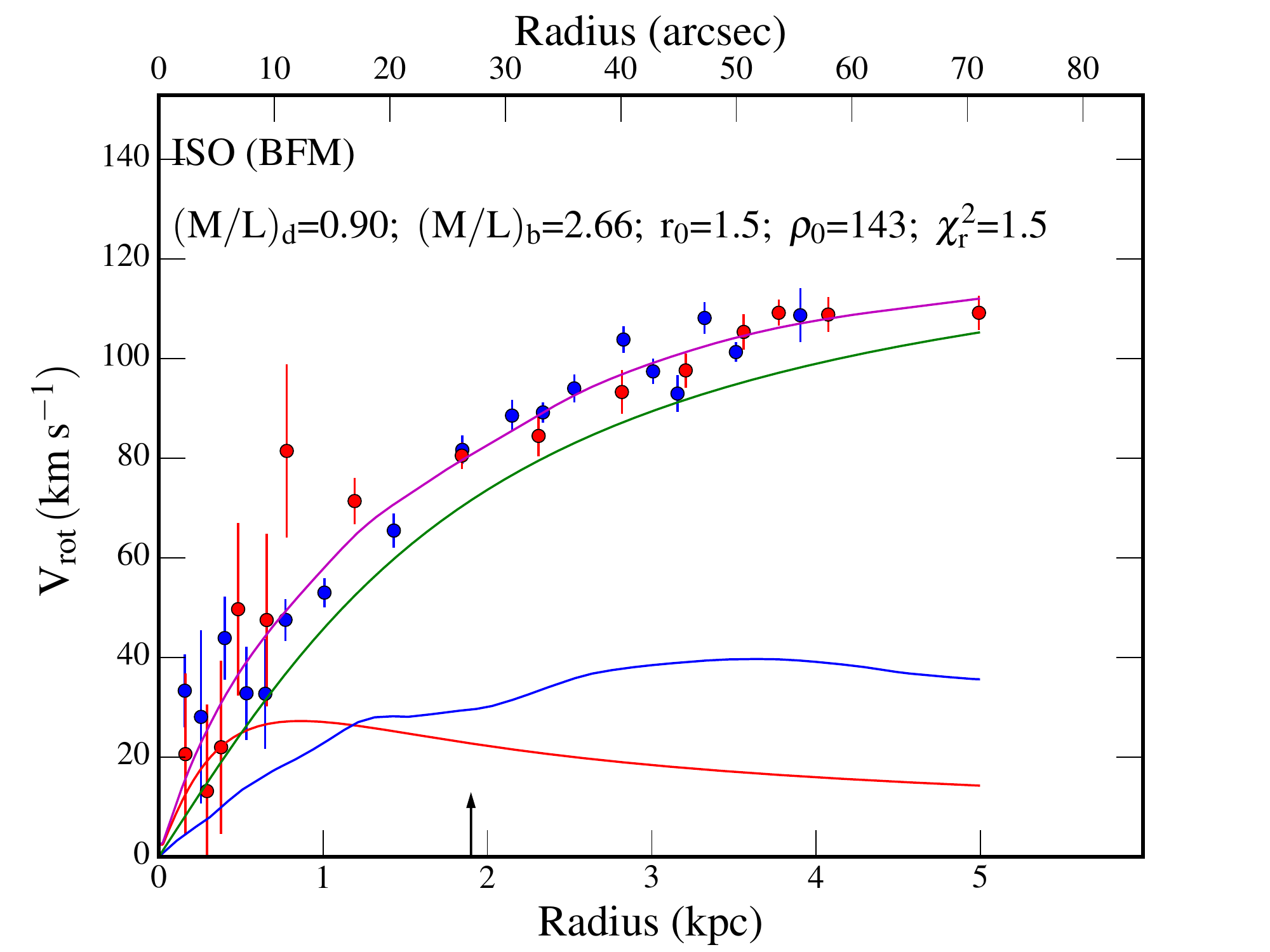}
\hspace*{-0.75cm} \includegraphics[width=0.35\textwidth]{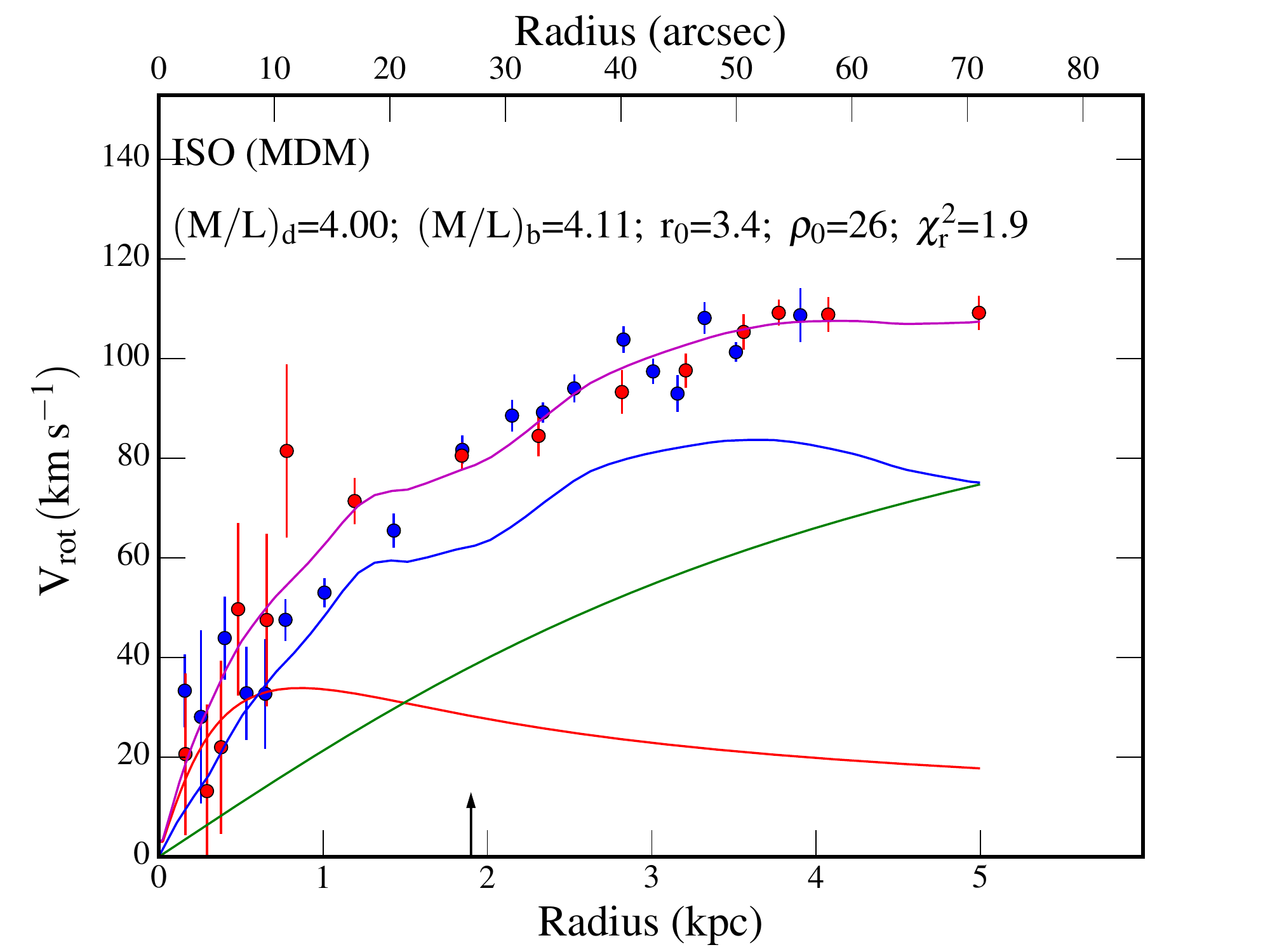}
\hspace*{-0.75cm} \includegraphics[width=0.35\textwidth]{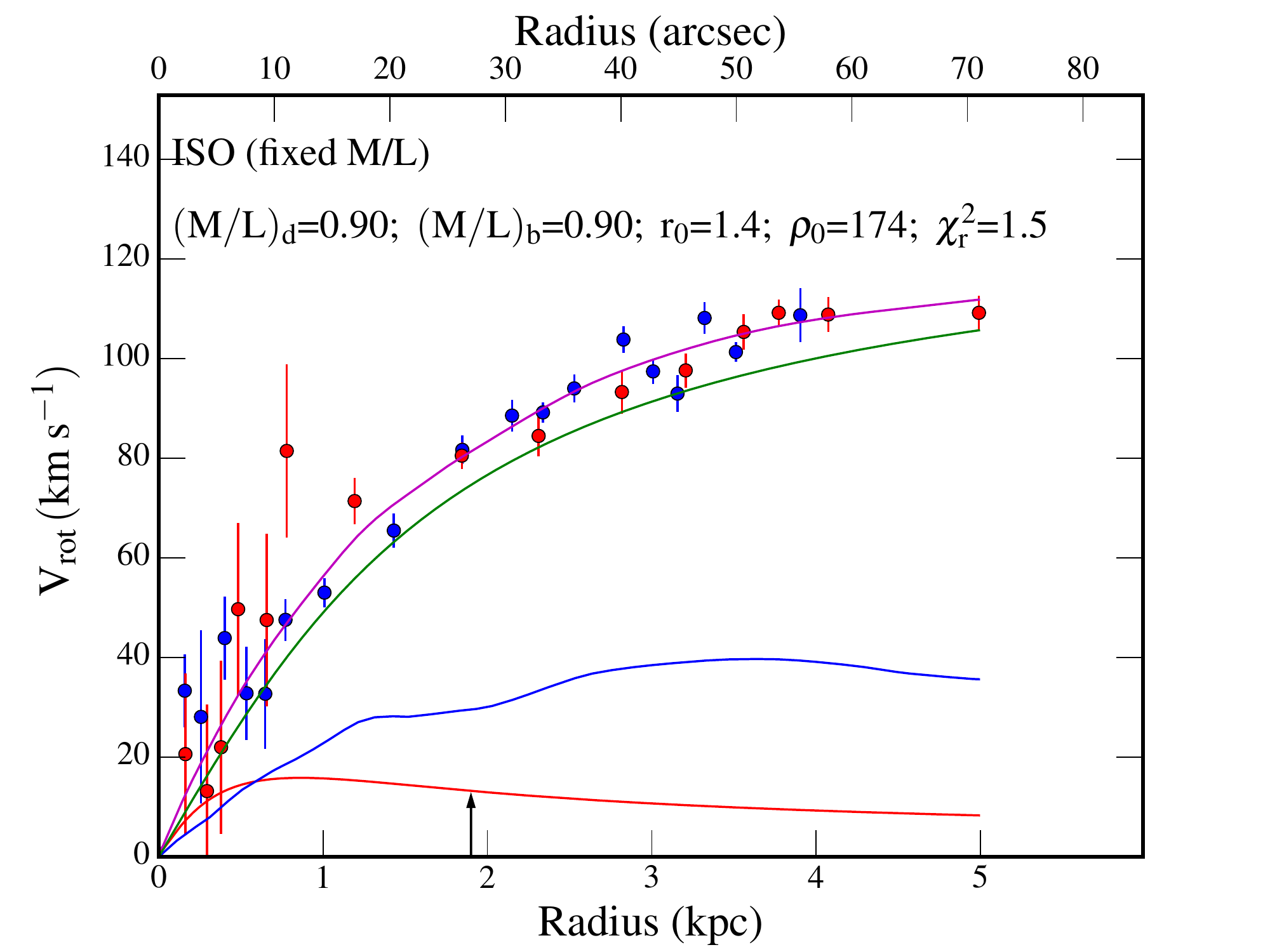}\\
\hspace*{-0.00cm} \includegraphics[width=0.35\textwidth]{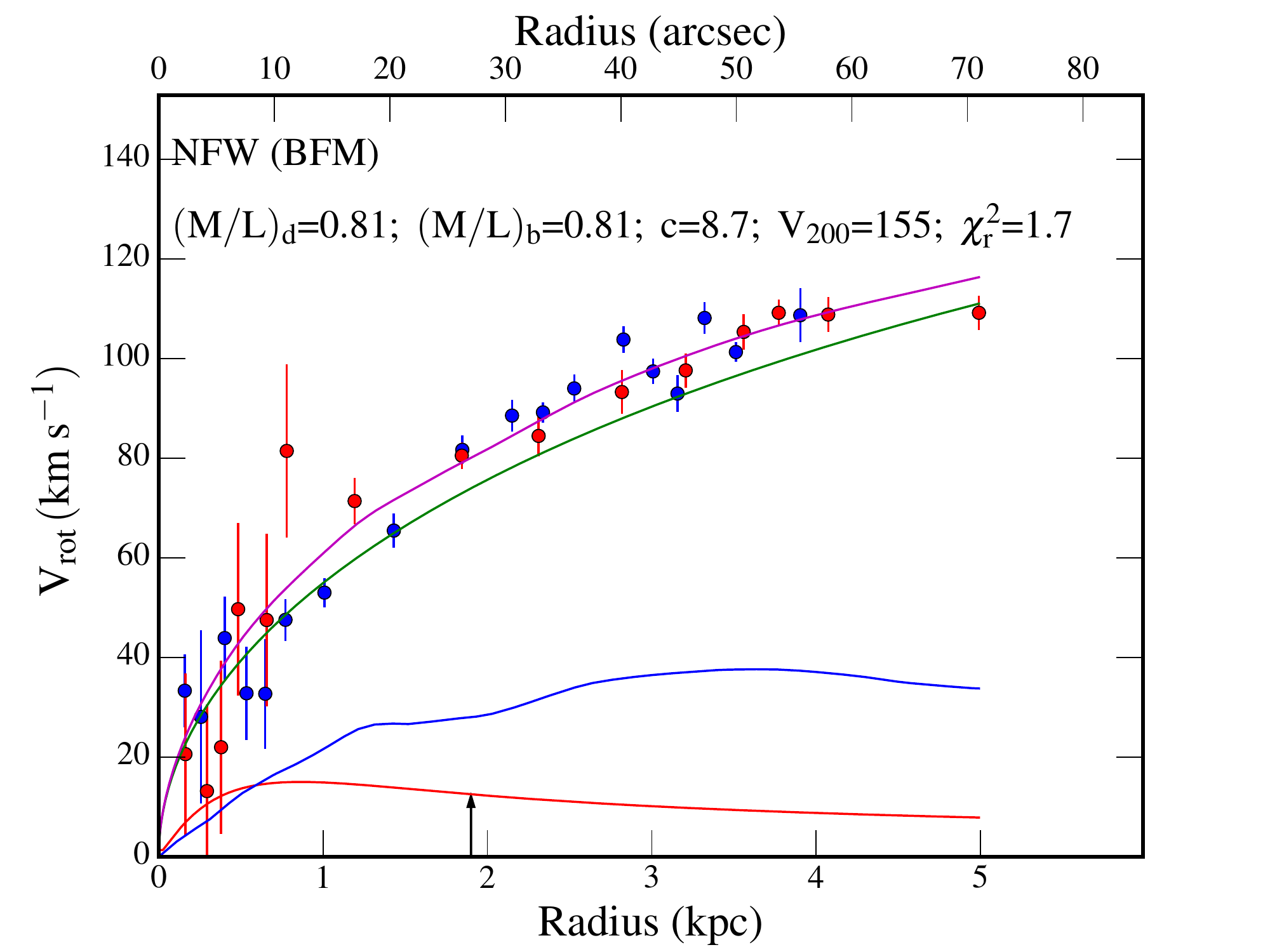}
\hspace*{-0.25cm} \vspace{-1.25cm} \includegraphics[width=0.31\textwidth]{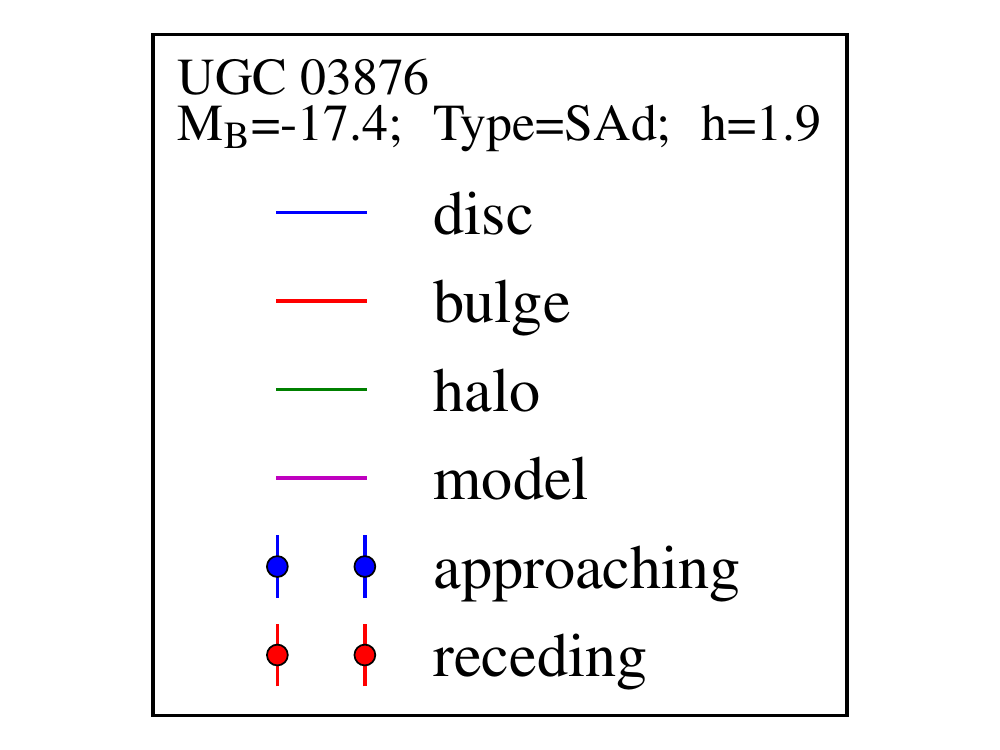} \vspace{1.25cm} \hspace*{-0.5cm}
\hspace*{-0.00cm} \includegraphics[width=0.35\textwidth]{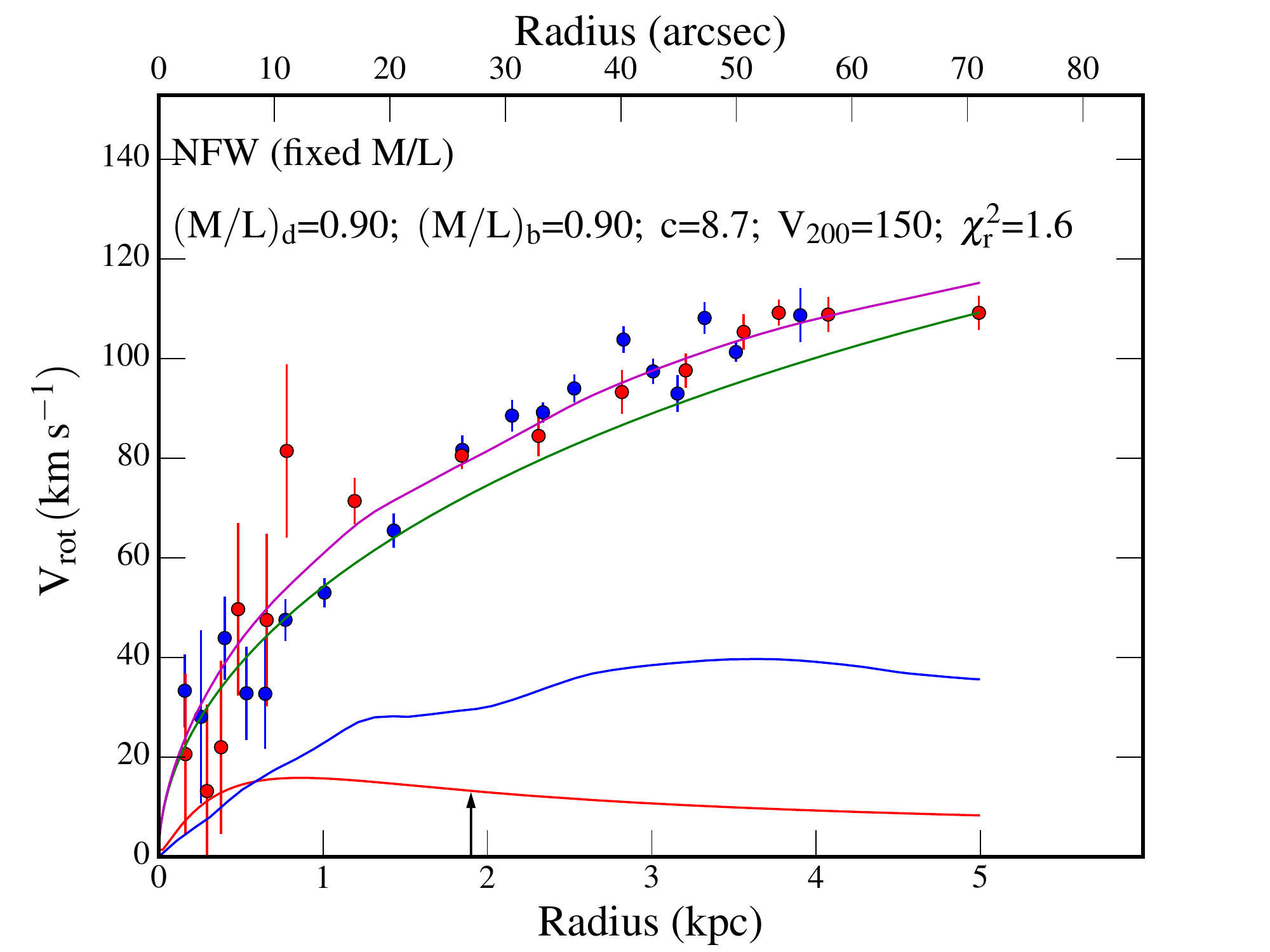}\\
\caption{Example of mass models for the galaxy UGC 3876. First line: pseudo-isothermal sphere density profiles (ISO). Second line: Navarro, Frenk \& White density profiles (NFW). First column: Best Fit Model (BFM). Second column: Maximum disc Model (MDM) for line 1 (ISO model). Third column: Mass-to-Light ratio M/L fixed using the optical $\rm (B - V)$ color. The name of the galaxy, its B-band absolute magnitude, morphological type and disc scale length have been indicated in the insert located line 2-column 2. For each model, the fitted parameters and the reduced $\chi^2$ have been indicated in each sub-panel.}
\label{massmodel1ap}
\end{figure*}

\begin{figure*}
\hspace*{-0.00cm} \includegraphics[width=0.35\textwidth]{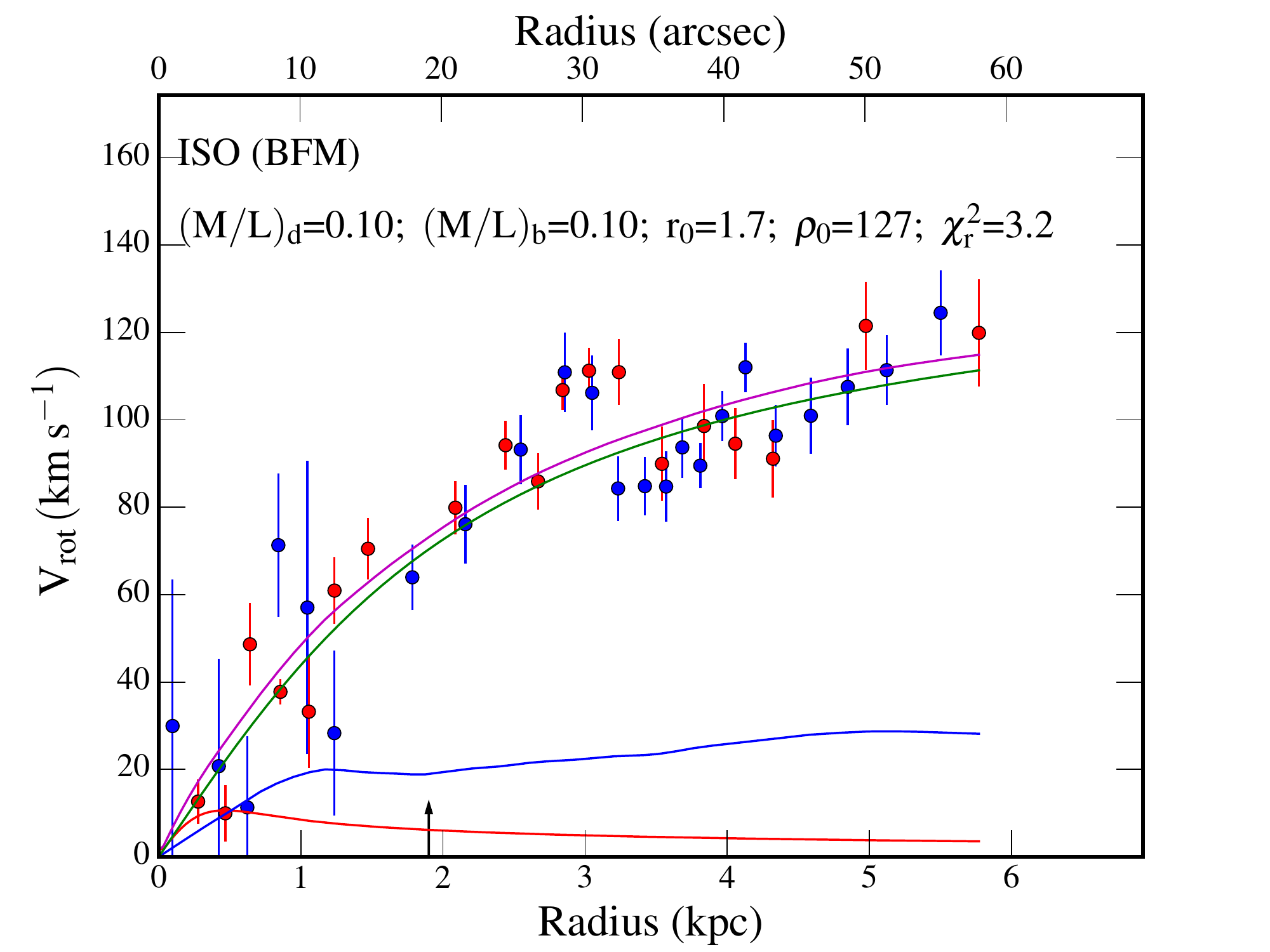}
\hspace*{-0.75cm} \includegraphics[width=0.35\textwidth]{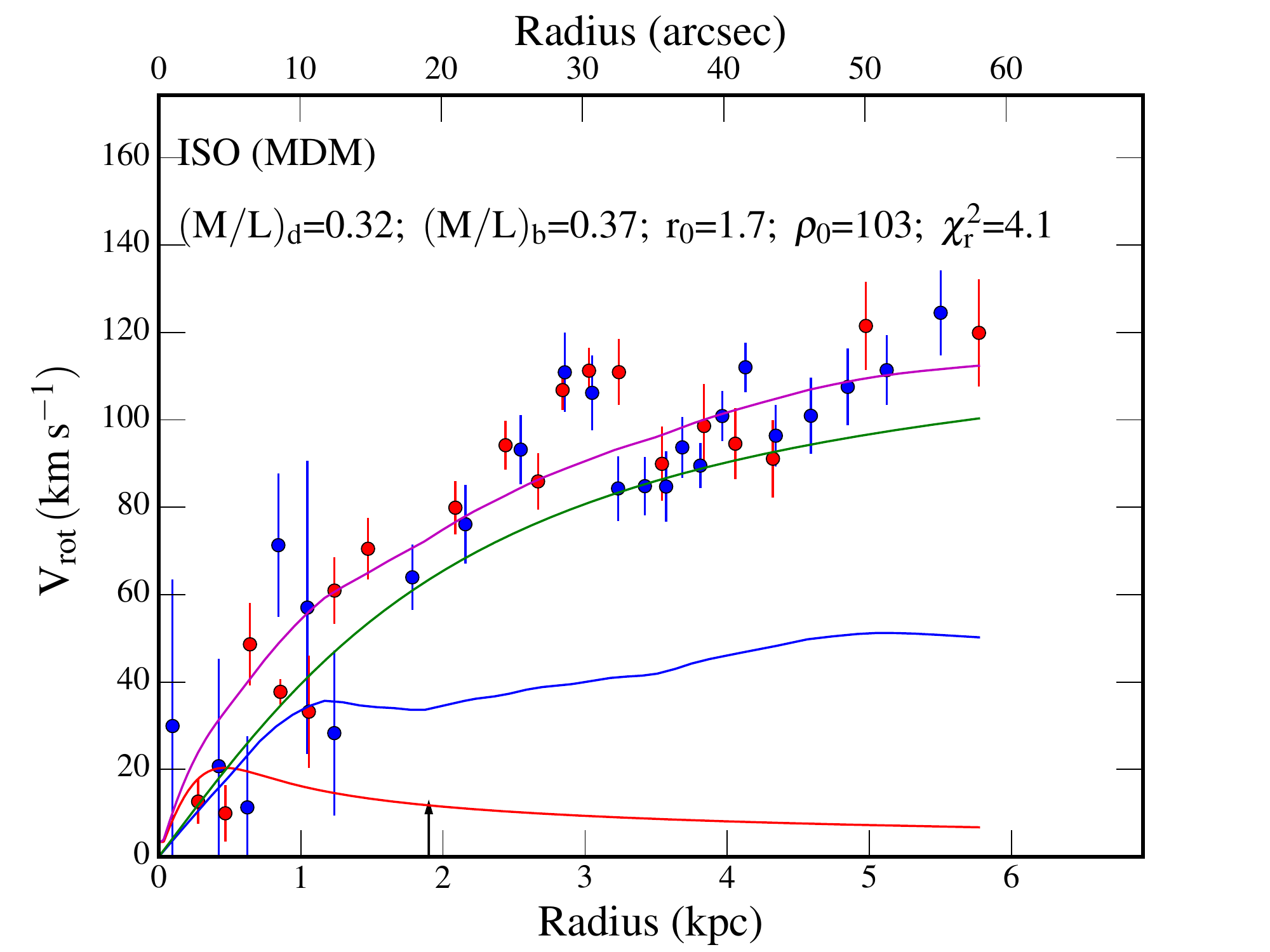}
\hspace*{-0.75cm} \includegraphics[width=0.35\textwidth]{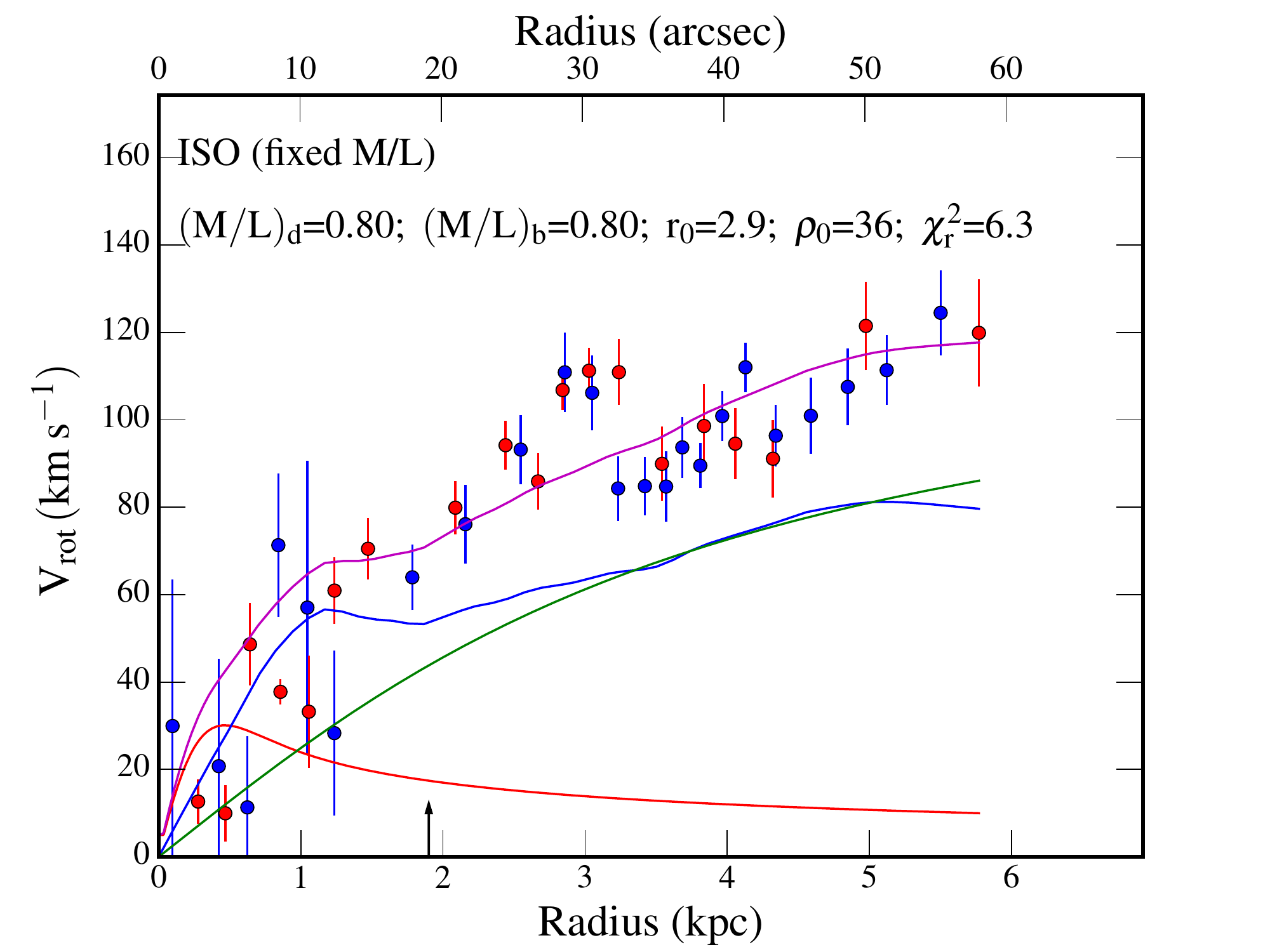}\\
\hspace*{-0.00cm} \includegraphics[width=0.35\textwidth]{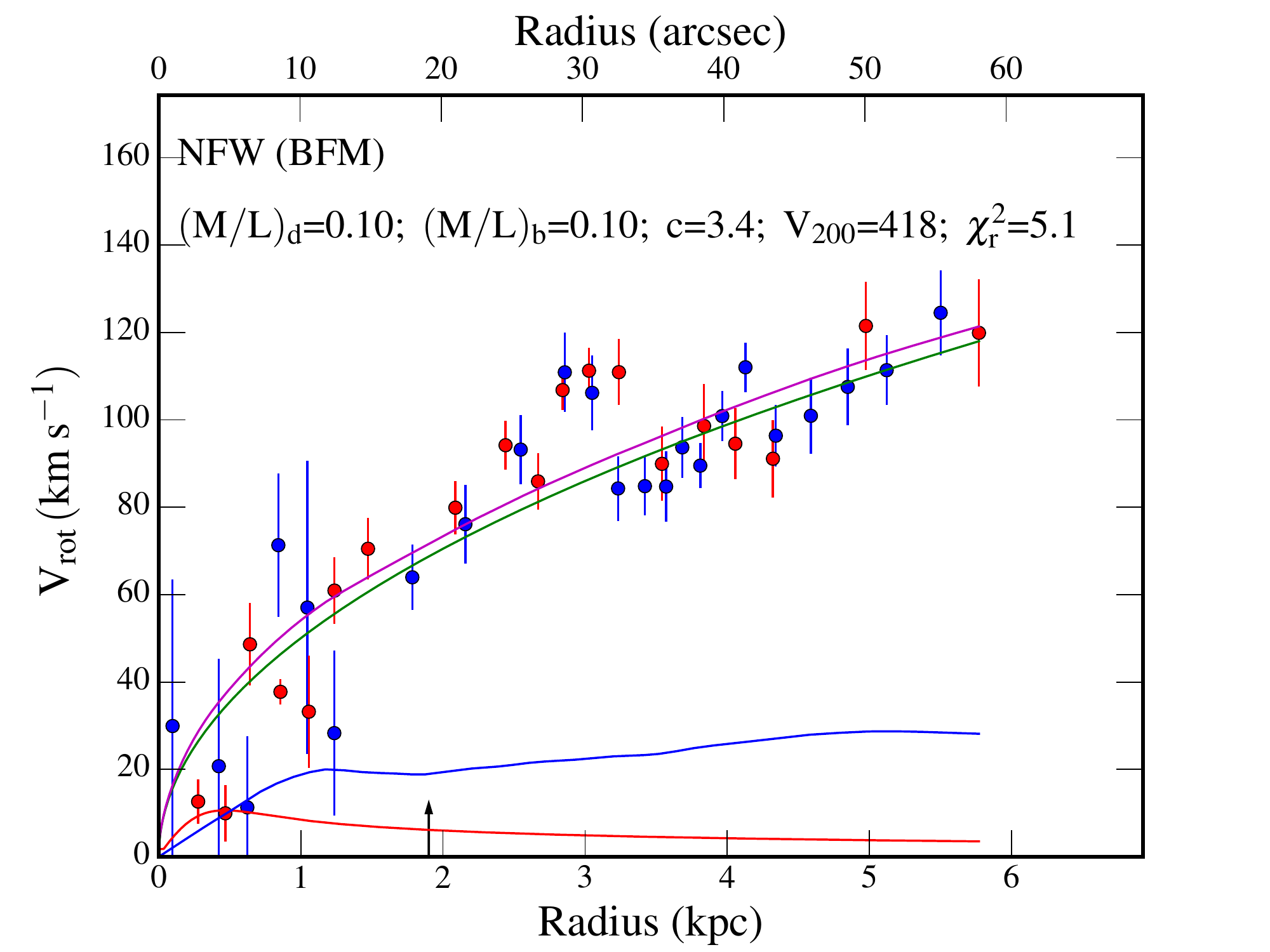}
\hspace*{-0.25cm} \vspace{-1.25cm} \includegraphics[width=0.31\textwidth]{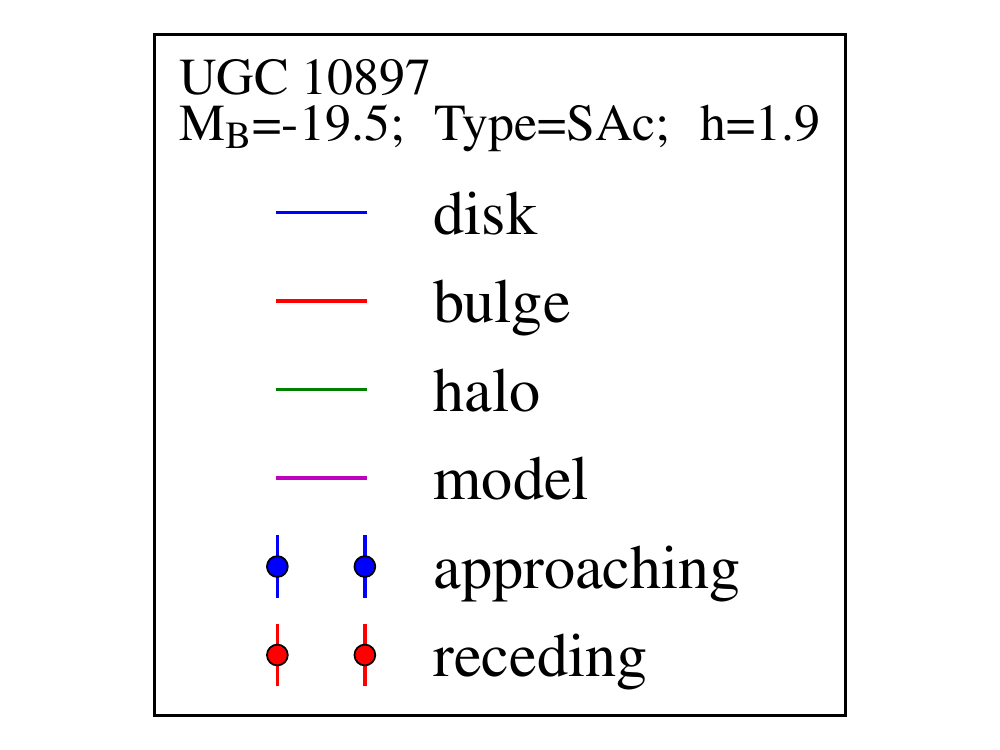} \vspace{1.25cm} \hspace*{-0.5cm}
\hspace*{-0.00cm} \includegraphics[width=0.35\textwidth]{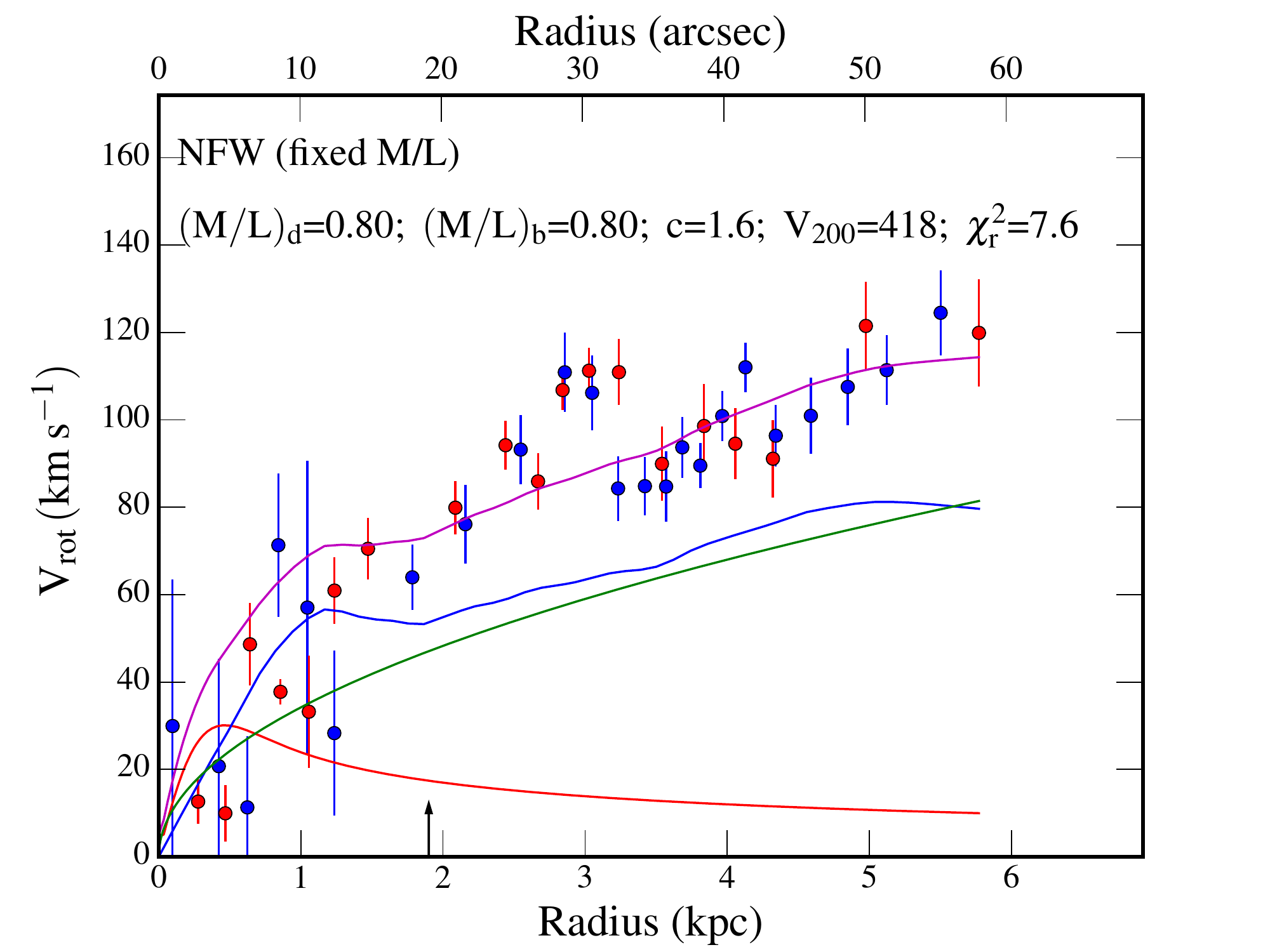}\\
\caption{Example of mass models for the galaxy UGC 10897. Lines and colors are same as in Fig. \ref{massmodel1ap}.}
\label{massmodel2}
\end{figure*}


\bsp	
\label{lastpage}
\end{document}

%% file: latex_table_paper.tex
\newgeometry{left=3cm,right=3cm,top=3cm,bottom=3cm}{\small}

\captionsetup{width=1.3\textwidth}				
\begin{table*}
\caption[Global properties]{Global properties: (1) Name of the galaxy in the UGC catalogue. (2): Morphological type taken from the RC3 catalogue (except for galaxies UGC 3521, 3708, 3915, 4393, 10652 for which the morphological types are taken from Epinat et al. (2008a)); (3): Absolute B-magnitude from Epinat et al. (2008a); (4): (B-V) colors corrected for galactic and internal extinction from RC3 catalogue when available. For galaxies with no (B-V) colour available in the RC3 catalogue, the value has been computed as explained in Section 3.2, those galaxies are marked with an asterisk (*). (5): Central surface brightness from the observed data in mag arcsec$^{-2}$; (6): Isophotal radius at the limiting surface brightness of 25 mag arcsec$^{-2}$ normalised by the disc scale length; (7): The last radius of the rotation curve normalised by the disc scale length; (8): Central surface brightness of the disc in mag arcsec$^{-2}$; (9): Disc scale length of the disc component in kpc; (10): Luminosity of the disc in unit of 10$^8$\ L$\odot$ calculated at the isophotal radius; (11): Surface brightness of the bulge at the effective radius in mag arcsec$^{-2}$; (12): Effective radius of the bulge in kpc; (13): S\'{e}rsic index of the bulge; (14): Luminosity of the bulge in units of 10$^8$\ L$\odot$, derived at the isophotal radius; (15): Classification flag of the rotation curves: 1 and 2 correspond respectively to very high and high quality rotation curves, while 3 represents poor quality rotation curves.}
\label{tab:photometry}
\begin{tabular}{c c c c c c c c c c c c c c c}

\hline
\multicolumn{1}{c }{\textbf{Galaxy}} & \multicolumn{6}{c}{\textbf{}} & \multicolumn{3}{c}{\textbf{Disc}} & \multicolumn{4}{c}{\textbf{Bulge}} & \multicolumn{1}{c}{\textbf{RC}} \\
\cmidrule(lr){1-1}
\cmidrule(lr){2-7}
\cmidrule(ll){8-10}
\cmidrule(ll){11-14}
\cmidrule(ll){15-15}
\multicolumn{1}{c}{\textbf{$\rm UGC$}} & \textbf{$\rm type$} & \textbf{$\rm Mag$} & \textbf{$\rm (B-V)$} & \textbf{$\mu_0$$_{obs}$} & \textbf{$\rm R_{25}/h$} & \textbf{$\rm R_{last}/h$} & \textbf{$\mu_0$}  & \textbf{$\rm h$} & \textbf{$L_D$} & \textbf{$\mu_e$} & \textbf{$r_e$ } & \textbf{$\rm n$} &\textbf{$L_B$} & \textbf{flag} \\ 
\multicolumn{1}{c}{(1)} & {(2)} & {(3)} & {(4)} & {(5)} & {(6)} & {(7)} & {(8)} & {(9)} & {(10)} & {(11)} & {(12)}& {(13)} &  {(14)} &  {(15)} \\ 
\cmidrule(lr){1-1}
\cmidrule(lr){2-7}
\cmidrule(ll){8-10}
\cmidrule(ll){11-14}
\cmidrule(ll){15-15}

00089 & $\rm SBa$ & $-21.5$ & $0.69$ & $16.4$ & $2.4$ & $1.6$ & $19.6$ & $6.0$ & $547.3$ & $16.73$ & $0.71$ & $1.0$ & $331.8$ & $1$ \\ 
00094* & $\rm SAab$ & $-20.4$ & $0.64$ & $18.7$ & $2.6$ & $2.8$ & $19.7$ & $4.0$ & $232.7$ & $18.69$ & $0.26$ & $1.57$ & $10.2$ & $1$ \\ 
00508 & $\rm SB(r)ab$ & $-21.8$ & $0.82$ & $16.6$ & $3.5$ & $3.0$ & $19.4$ & $7.5$ & $1378.5$ & $17.75$ & $0.6$ & $2.16$ & $235.13$ & $1$ \\ 
00528 & $\rm SABb$ & $-19.6$ & $0.45$ & $16.2$ & $4.2$ & $2.0$ & $18.0$ & $1.0$ & $90.3$ & $20.13$ & $0.71$ & $3.22$ & $12.88$ & $1$ \\ 
00763 & $\rm SABm$ & $-18.9$ & $0.43$ & $19.6$ & $3.4$ & $6.0$ & $19.9$ & $1.5$ & $33.4$ & $24.62$ & $0.77$ & $3.79$ & $3.55$ & $1$ \\ 
01256 & $\rm SBcd$ & $-18.9$ & $0.45$ & $19.4$ & $5.3$ & $6.1$ & $19.4$ & $1.4$ & $4.9$ & $-$ & $-$ & $-$ & $-$ & $3$ \\ 
01317 & $\rm SAB(r)c$ & $-21.5$ & $0.59$ & $18.1$ & $4.3$ & $4.5$ & $18.1$ & $5.4$ & $1023.7$ & $-$ & $-$ & $-$ & $-$ & $1$ \\ 
01437 & $\rm SABbc$ & $-21.8$ & $0.54$ & $17.6$ & $2.7$ & $5.0$ & $19.3$ & $5.2$ & $578.7$ & $20.59$ & $1.79$ & $3.24$ & $100.34$ & $1$ \\ 

\hline
\end{tabular}
\end{table*}

\setcounter{table}{1}

\begin{table*}

\caption[Parameters of mass models using the Best Fit Model (BFM) and fixed M/L techniques with the pseudo-isothermal (ISO) model]{Parameters of mass models using the Best Fit Model (BFM) and fixed M/L techniques with the pseudo-isothermal (ISO) model: (1) Name of the galaxy in the UGC catalogue; the columns (2) to (6) and (7) to (10) show respectively the BFM parameters, and the fixed M/L parameters for the ISO model. (2): M/L of the disc in  M$_{\odot}$/L$_{\odot}$; (3): M/L of the bulge in M$_{\odot}$/L$_{\odot}$. (4) \& (8): the core radius of the DM halo in kpc; (5) \& (9): the Central density of the DM halo in $10^{-3}$ M$_{\odot}$/pc$^3$; (6) \& (10): the reduced $\chi^2$; (7): M/L derived using the B - V color in units of M$_{\odot}$/L$_{\odot}$.}

\label{tab:iso}
\begin{tabular}{c c c c l c c c c l}
\hline
\multicolumn{1}{c }{\textbf{Galaxy}} & \multicolumn{5}{c}{\textbf{ISO (BFM)}} & \multicolumn{4}{c}{\textbf{ISO with fixed M/L}} \\  
\cmidrule(lr){1-1}
\cmidrule(lr){2-6}
\cmidrule(ll){7-10}
\multicolumn{1}{c}{\textbf{$\rm UGC$}} & \textbf{$\rm M/L$ $\rm Disc$} & \textbf{$\rm M/L$ $\rm Bulge$} & \textbf{$r_0$} & \textbf{$\rho_0$} & \textbf{$\chi^2$} & \textbf{$\rm M/L$ } & \textbf{$r_0$} & \textbf{$\rho_0$} & \textbf{$\chi^2$} \\ 
\multicolumn{1}{c}{(1)} & {(2)} & {(3)} &{(4)} & {(5)} & {(6)} & {(7)} & {(8)} & {(9)} & {(10)} \\ 
\cmidrule(lr){1-1}
\cmidrule(lr){2-6}
\cmidrule(ll){7-10}

00089* & $1.84_{1.74}^{0.01}$ & $1.87_{1.77}^{0.97}$ & $>39.2$ & $27_{17}^{372}$ & $7.0$ & $1.52$ & $2.8_{2.3}^{2.3}$ & $196_{186}^{203}$ & $9.3$ \\ 
00094 & $0.81_{0.71}^{0.60}$ & $5.13_{5.03}^{2.87}$ & $1.0_{0.5}^{0.7}$ & $702_{652}^{47}$ & $3.7$ & $1.41$ & $1.0_{0.5}^{0.1}$ & $686_{636}^{63}$ & $7.5$ \\ 
00508 & $2.81_{2.71}^{0.69}$ & $3.02_{2.92}^{0.98}$ & $7.3_{6.8}^{6.0}$ & $102_{102}^{397}$ & $1.4$ & $2.20$ & $4.9_{4.4}^{0.4}$ & $215_{215}^{284}$ & $1.5$ \\ 
00528* & $0.10_{0.01}^{0.05}$ & $0.28_{0.18}^{0.18}$ & $0.5_{0.1}^{7.5}$ & $203_{203}^{36}$ & $2.5$ & $0.78$ & $0.5_{0.1}^{7.5}$ & $0_{1}^{1}$ & $177.7$ \\ 
00763 & $0.32_{0.22}^{0.22}$ & $2.94_{2.84}^{1.06}$ & $2.0_{1.5}^{0.5}$ & $62_{52}^{43}$ & $2.2$ & $0.73$ & $2.7_{2.2}^{0.2}$ & $38_{28}^{67}$ & $2.4$ \\ 
01317* & $0.43_{0.33}^{0.05}$ & $-$ & $1.0_{0.5}^{0.2}$ & $750_{730}^{1}$ & $4.2$ & $1.15$ & $0.8_{0.3}^{0.1}$ & $750_{730}^{1}$ & $4.4$ \\ 
01437 & $1.08_{0.98}^{0.27}$ & $2.14_{2.04}^{1.86}$ & $3.4_{2.9}^{3.1}$ & $26_{16}^{723}$ & $2.4$ & $1.00$ & $0.7_{0.2}^{0.1}$ & $750_{740}^{1}$ & $6.2$ \\ 

\hline
\end{tabular}
\end{table*}

\setcounter{table}{2}

\begin{table*}
\caption[Parameters of mass models using the maximum disc model technique with the pseudo-isothermal (ISO)]{Parameters of mass models using the maximum disc model (MDM) technique with the pseudo-isothermal (ISO) model: (1) Name of the galaxy in the UGC catalogue; (2) M/L of the disc in  M$_{\odot}$/L$_{\odot}$; (3) M/L of the bulge in  M$_{\odot}$/L$_{\odot}$; (4) Core radius of the DM halo in kpc; (5) Central density of the DM halo in $10^{-3}$ M$_{\odot}$/pc$^3$; (6) The reduced $\chi^2$.}
\label{tab:maximumdisk}
\begin{tabular}{c c c c l c }
\hline
\multicolumn{1}{c }{\textbf{Galaxy}} & \multicolumn{5}{c}{\textbf{Maximum Disc Model}} \\  
\cmidrule(lr){1-1}
\cmidrule(lr){2-6}
\multicolumn{1}{c}{\textbf{$\rm UGC$}} & \textbf{$\rm M/L$ $\rm Disc$} & \textbf{$\rm M/L$ $\rm Bulge$} & \textbf{$r_0$} & \textbf{$\rho_0$} & \textbf{$\chi^2$} \\ 
\multicolumn{1}{c}{(1)} & {(2)} & {(3)} & {(4)} & {(5)} & {(6)} \\ 
\cmidrule(lr){1-1}
\cmidrule(lr){2-6}

00089 & $2.01_{1.91}^{0.01}$ & $2.05_{1.95}^{0.80}$ & $3.6_{3.1}^{4.4}$ & $36_{36}^{363}$ & $9.4$ \\ 
00094 & $2.20_{2.10}^{0.01}$ & $7.82_{7.72}^{0.18}$ & $2.4_{1.9}^{0.1}$ & $81_{31}^{668}$ & $4.8$ \\ 
00508 & $4.10_{4.00}^{0.01}$ & $4.11_{4.01}^{0.89}$ & $18.6_{18.1}^{0.1}$ & $23_{23}^{483}$ & $1.9$ \\ 
00528 & $0.18_{0.08}^{0.01}$ & $0.19_{0.09}^{0.27}$ & $0.5_{0.1}^{0.1}$ & $0_{1}^{239}$ & $3.2$ \\ 
00763 & $0.97_{0.87}^{0.01}$ & $0.99_{0.89}^{3.01}$ & $3.4_{2.9}^{0.1}$ & $23_{13}^{82}$ & $2.8$ \\ 
01317 & $3.10_{3.00}^{0.01}$ & $-$ & $0.5_{0.1}^{2.9}$ & $0_{1}^{106}$ & $16.0$ \\ 
01437 & $1.57_{1.47}^{0.01}$ & $1.96_{1.86}^{2.04}$ & $5.8_{5.3}^{2.2}$ & $2_{2}^{103}$ & $3.0$ \\ 

\hline
\end{tabular}
\end{table*}

\restoregeometry

\setcounter{table}{3}
\begin{table*}
\caption[Parameters of mass models using the Best Fit Model (BFM) and fixed M/L techniques with the Navarro-Frenk-White model (NFW)]{Parameters of mass models using the Best Fit Model (BFM) and fixed M/L techniques with the Navarro-Frenk-White model (NFW): (1) Name of the galaxy in the UGC catalogue; the columns (2) to (6) and (7) to (10) show respectively the BFM parameters, and the fixed M/L parameters for the ISO model. (2): M/L of the disc in  M$_{\odot}$/L$_{\odot}$; (3): M/L of the bulge in M$_{\odot}$/L$_{\odot}$. (4) \& (8): the central halo concentration index; (5) \& (9): the halo velocity in $\rm km\ s^{-1}$; (6) \& (10): the reduced $\chi^2$; (7): M/L derived using the B - V color in units of M$_{\odot}$/L$_{\odot}$.}
\label{tab:nfw}
\begin{tabular}{c c c c c c c c c c}

\hline
\multicolumn{1}{c }{\textbf{Galaxy}} & \multicolumn{5}{c}{\textbf{NFW (BFM)}} & \multicolumn{4}{c}{\textbf{NFW with fixed M/L}} \\  
\cmidrule(lr){1-1}
\cmidrule(lr){2-6}
\cmidrule(ll){7-10}
\multicolumn{1}{c}{\textbf{$\rm UGC$}} & \textbf{$\rm M/L$  $\rm Disc$} & \textbf{$\rm M/L$ $\rm Bulge$} & \textbf{$\rm c$} & \textbf{$\rm V_{200}$} & \textbf{$\chi^2$} & \textbf{$\rm M/L$ } & \textbf{$\rm c$} & \textbf{$\rm V_{200}$} & \textbf{$\chi^2$} \\ 
\multicolumn{1}{c}{(1)} & {(2)} & {(3)} & {(4)} & {(5)} & {(6)} & {(7)} & {(8)} & {(9)} & {(10)} \\ 
\cmidrule(lr){1-1}
\cmidrule(lr){2-6}
\cmidrule(ll){7-10}

00089 & $1.68_{1.58}^{0.05}$ & $1.70_{1.59}^{1.15}$ & $5.1_{4.1}^{6.5}$ & $500.0_{490.0}^{0.1}$ & $7.8$ & $1.52$ & $6.9_{5.9}^{5.3}$ & $455.5_{445.5}^{44.5}$ & $8.0$ \\ 
00094 & $0.37_{0.27}^{1.15}$ & $0.46_{0.36}^{7.54}$ & $45.2_{35.2}^{4.3}$ & $105.9_{85.9}^{94.1}$ & $3.8$ & $1.41$ & $43.1_{33.1}^{1.1}$ & $89.5_{69.5}^{110.5}$ & $4.0$ \\ 
00508 & $0.59_{0.49}^{1.91}$ & $2.14_{2.04}^{1.86}$ & $20.5_{12.5}^{7.9}$ & $390.9_{190.9}^{109.1}$ & $1.3$ & $2.20$ & $14.6_{6.6}^{2.0}$ & $397.7_{197.7}^{102.3}$ & $1.3$ \\ 
00528 & $0.10_{0.01}^{0.01}$ & $0.10_{0.01}^{0.35}$ & $58.8_{57.8}^{19.3}$ & $25.8_{24.8}^{74.2}$ & $2.3$ & $0.78$ & $1.0_{0.1}^{106.0}$ & $1.0_{0.1}^{0.1}$ & $177.7$ \\ 
00763 & $0.26_{0.16}^{0.16}$ & $0.28_{0.18}^{3.72}$ & $11.0_{10.0}^{1.2}$ & $89.1_{69.1}^{310.9}$ & $2.8$ & $0.73$ & $4.0_{3.0}^{0.6}$ & $175.5_{155.5}^{224.5}$ & $3.0$ \\ 
01317 & $0.10_{0.01}^{0.05}$ & $-$ & $28.0_{27.0}^{0.1}$ & $128.2_{108.2}^{171.8}$ & $4.1$ & $1.15$ & $28.0_{27.0}^{0.1}$ & $102.7_{82.7}^{197.3}$ & $5.4$ \\ 
01437 & $1.30_{1.20}^{0.11}$ & $2.05_{1.95}^{1.95}$ & $2.3_{1.3}^{53.5}$ & $159.1_{139.1}^{40.9}$ & $2.4$ & $1.00$ & $44.3_{43.3}^{1.3}$ & $77.3_{57.3}^{122.7}$ & $2.6$ \\ 

\hline
\end{tabular}
\end{table*}

\restoregeometry